# Near-Term Emission Targets Need Immediate Attention in the USA

## Author Information

Trevor Barnes[1*], Kamran Tehranchi[2], Brad Reinholz[3], Malcolm Metcalfe[3], Taco Niet[1]

[1]School of Sustainable Energy Engineering, Simon Fraser University. BC, Canada.

[2]Department of Civil & Environmental Engineering, Stanford University. California, USA

[3]Generac Power Systems. Wisconsin, USA

*Correspondence: trevor_barnes@sfu.ca

## Abstract

Given recent changes in federal climate policy, the United States is unlikely to meet its original 2030 Paris Agreement emission target of a 50-52% reduction from 2005 levels. However, rapid near-term abatement remains achievable through targeted multi-sector energy transitions. Extending the open-source energy system model, PyPSA-USA, to perform multi-sector analysis, we evaluate the primary drivers of USA energy costs and emissions though applying global sensitivity analysis. Our results suggest that fossil fuel price volatility is the dominant driver of marginal electricity and energy costs across most of the nation, however, uncoordinated state-level renewable mandates can induce localized cost spikes due to regional bottlenecks. We find that system climate impact (CO2e) is overwhelming sensitive to fugitive methane leakage rates and global warming potential assumptions. Addressing upstream methane leaks will play a crucial role in abating climate-related damages. Finally, demand-side electrification, specifically light-duty electric vehicles and service sector heating, can act as immediate levers for carbon abatement. The results of this work suggest that many of the Inflation Reduction Act's clean energy initiatives, that have since been repealed, are effective near-term solutions to reduce exposure to fossil fuel price and mitigate future financial penalties associated with the rising social cost of carbon.

## Introduction

As of January 2025, the USA notified the United Nations (UN) of their exit from the Paris Agreement [1]. With this exit, they leave behind their original commitment to achieving a 50% - 52% reduction in system level emissions from 2005 levels by 2030 [2]. This exit has been accompanied by the One Big Beautiful Bill Act (OBBBA), which scales back many of the Inflation Reduction Act (IRA) clean energy initiatives [3]. Critics of the OBBBA raise concerns that household and business energy expenditures will increase, in addition to Greenhouse Gas (GHG) emission reduction trends softening by 2030 [3], [4]. This has led to considerable concern as the USA is currently well off its original 50-52% reduction target, as shown in Figure 1. This study uses the Energy System Optimization Model (ESOM), PyPSA-USA Sector, to quantify the key uncertainties impacting energy prices and emissions in a 2030 USA. This helps policy makers understand what actions should be taken in the near-term given the current policy landscape.

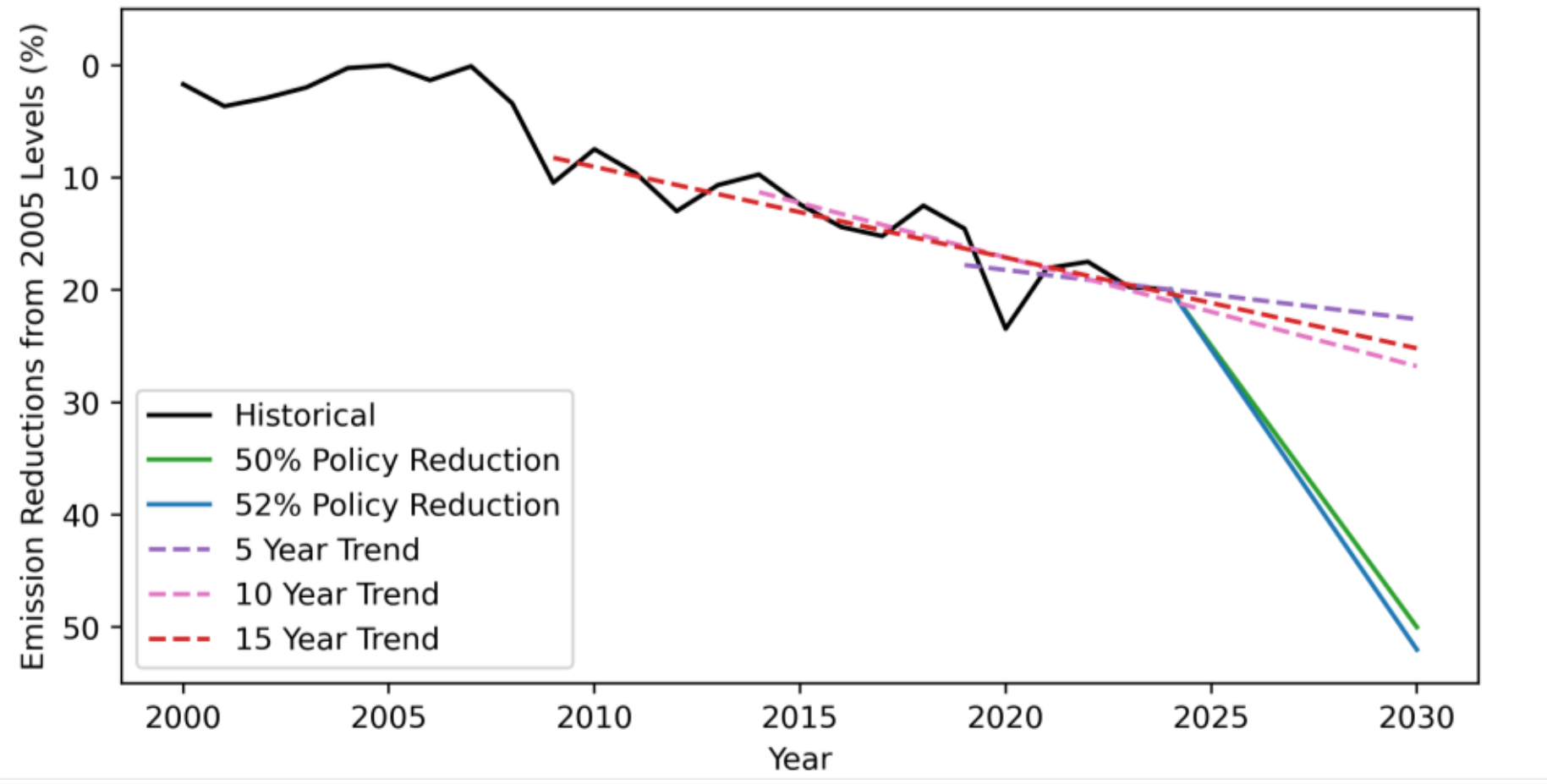

**Figure 1: CO2eq emission reduction change from 2005 levels in the USA.**

*Historical emissions are retrieved from the Our World in Data [5]. The yearly reduction in emissions to hit the 2030 USA targets are shown starting from 2024. Linear trends following the reductions over the last 5, 10, and 15 year time horizons highlight the significant difference between targets and reality. While emission reductions are not expected to follow linear curves, this illustrative example highlights the scale of the problem.*

Current energy planning literature predominantly focuses on 2050 decarbonization pathways. However, the quick implementations of both the IRA and OBBBA show that the energy policy landscape can shift quickly and have profound implications in the near-term [6]. Therefore, studying what decisions can be implemented in a time constrained framework offers value for immediate decarbonization pathways. The most significant study on near-term impacts is from the International Panel on Climate Change (IPCC) sixth assessment report [7], ranking carbon mitigation potential of different technologies in relation to 2030 targets (see Figure SPM.7 in the report). While thorough, these results provide global rankings and lack USA specific context. Bistline et al. [8] address this and perform an excellent model comparison study for a 2030 USA to assess actions that must be taken to meet emission targets. This national level study emphasizes what 2030 actions are similar across all national level models. Moreover, Iyer et al. [9] perform a similar analysis on the USA over a 2035 timeline, while Jenkins et al. [10] and Zhao et al. [11] study actions that can be undertaken to reduce emissions by 2030 and 2035 respectively.

The critical takeaway from these studies is that large scale infrastructure projects are not solutions to 2030 targets. This is not due to their importance in low-carbon system, but rather a function of deployment timelines. Instead, immediate reductions can be made through fuel-switching, multi-sector solutions, and demand-side solutions. While studies do highlight how geography in the USA will impact a regions decarbonization potential in the near-term [8], [12], they fail to rank the importance of these different solutions by geographic region. This study addresses this gap by quantifying the impact different parametric uncertainties have on near-term marginal costs and emission results in a high temporal and spatial resolution model.

We extend PyPSA-USA [13], an open-source electricity planning model of the USA, into a multi-sector model. We explicitly represent the residential, commercial, industrial, and transport sectors, including granular end-use demands for electricity, heating, and cooling. Additionally, we integrate state-level natural gas representation to capture the often-overlooked impact of methane leaks and trade flow constraints. Utilizing PyPSA-USA, we systematically perform a Global Sensitivity Analysis (GSA) across the Continental United States (CONUS) to identify the primary drivers of five key metrics; objective cost, average marginal cost of electricity and energy separately, Carbon Dioxide (CO2) emissions, and total emissions (CO2 plus methane (CH4)). We characterize 342 individual uncertainties to identify the most sensitive drivers in the model. We then propagated these uncertainties using a Monte Carlo analysis. This approach allows us to quantify what drives key metrics, and how these metrics are expected to respond in a 2030 USA.

We contribute to the larger energy planning field in three ways. First, we present an open-source and highly configurable ESOM of USA that represents the full energy system. Second, to the best of the authors knowledge, we present the first application of GSA as applied to marginal energy costs and emissions, including methane, within a capacity expansion framework. These metrics are critical, as marginal costs give indications of geographically specific wholesale electricity prices [14], [15], and have been shown to have interactions with carbon emissions in integrated grid studies [16]. Finally, from our results, we identify specific immediate strategies to decrease emission by 2030 and compare these solutions against existing literature.

To ensure transparency and facilitate stakeholder communication, all results are provided openly on GitHub [17] and with a public viewing dashboard available at the repository. All result images in this manuscript come directly from the public dashboard for maximum reproducibility.

## Results

We identify three key findings from our study. These being that marginal costs are overwhelmingly sensitive to fossil fuel price uncertainty, the extremely low likelihood of the USA meeting 2030 emission targets, and the importance of demand-side solutions for near-term carbon abatement.

### Marginal costs can be kept low but are sensitive to fossil fuel prices and portfolio standards.

In the near-term, the available solutions to achieve significant decarbonization through supply side investments are limited. This is a direct result of the 2030 planning horizon. Consequently, the USA will remain reliant on fossil fuel-based energy supply chains to meet immediate energy demands; particularly, natural gas in the power, service, and industrial sectors, and petrol in the transportation sector. This reliance makes the full energy system highly sensitive to fossil fuel price shocks, as shown in Figure 2. Natural gas export costs (ie. how much a state makes from selling gas to a neighboring state), natural gas import costs (ie. how much a state pays for natural gas from a neighboring state) and petrol costs emerged as a top three sensitivity in 43, 32, and 24 of the 48 modeled states, respectively. As a globally traded commodity with large price uncertainty and significant USA energy supply chain penetration, this is not necessarily surprising to see the high impact natural gas prices have on marginal costs. However, for energy security in the near term, the USA must minimize exposure to fossil fuel price volatility without major supply side changes.

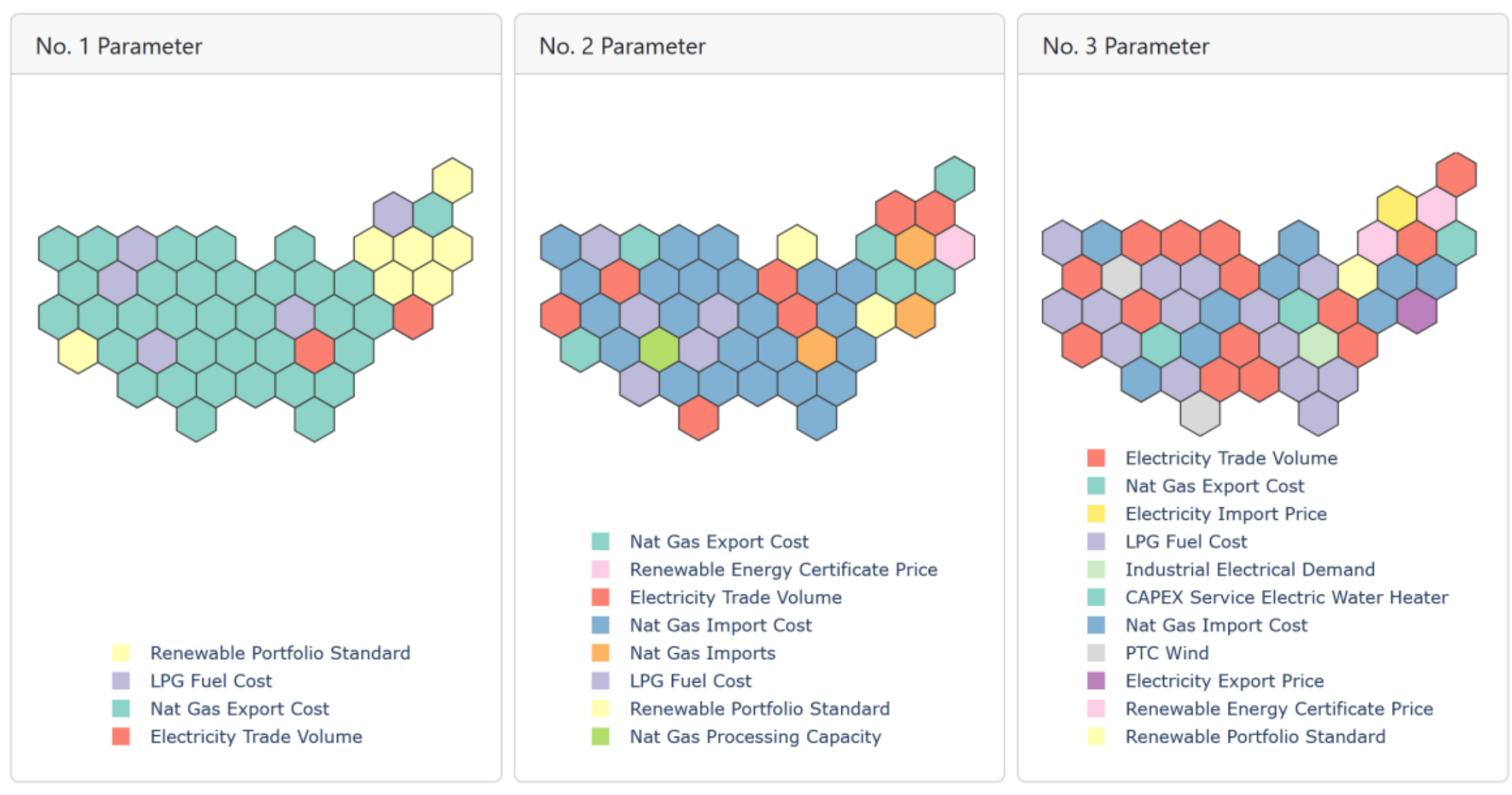

**Figure 2: Average marginal costs of energy sensitivity.**

*The state level sensitivity to average marginal costs of energy per state. The import and export costs represent the cost the state pay/earns from trading with the neighboring state. Electricity trade volume represents allowable yearly trade volume (MWh) to/from neighboring states.*

Propagating the influential uncertainties in Figure 2 highlights the role of geography in determining marginal prices factors. While fossil fuel price volatility is the most impactful parameter across most of CONUS, states with aggressive decarbonization mandates, in the form of Renewable Portfolio Standards (RPS), exhibit the largest spread in average energy costs. Figure 3 shows California (CA), Maine (ME), Massachusetts (MA) and New York (NY) all having the largest variations in marginal energy price outcomes. RPS outranks fossil fuel prices as the primary sensitivity in all of these states, as shown in Figure 2.

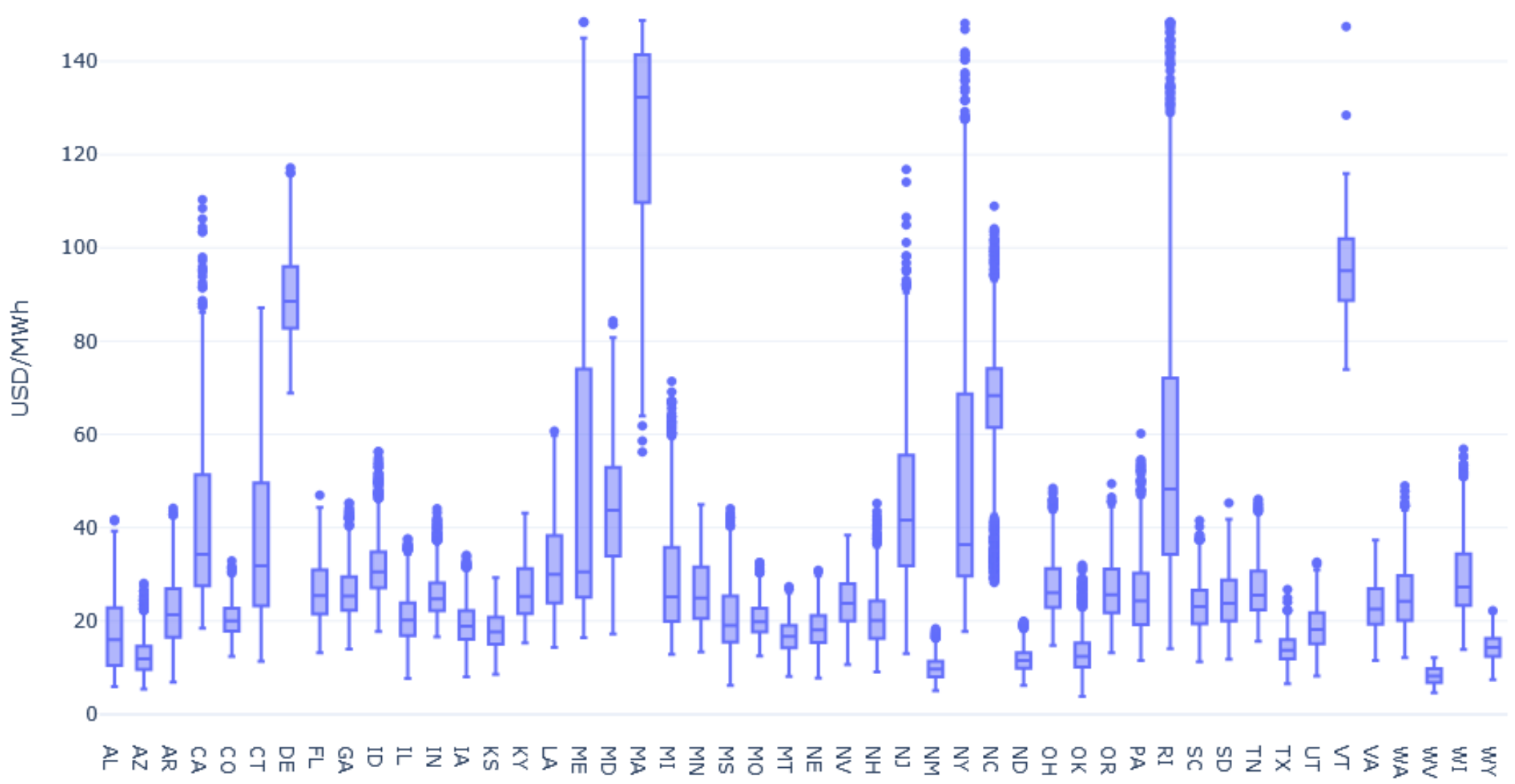

**Figure 3: Marginal costs of energy boxplot.**

*Boxplot of state level marginal costs of energy. Box plots are constructed by using GSA to find the uncertainty set of the 5 most impactful parameters impacting marginal energy costs, marginal electricity costs, objective costs, carbon emissions, and total emissions. These uncertainty sets, which are different per state, are propagated through the model to perform a Monte Carlo analysis. Note, this graph shows the 0-99% interval of data to remove one outlier result from Rhode Island to scale the axis.*

This RPS-induced volatility is a direct function of near-term infrastructure bottlenecks. While marginal costs remain stable under moderate RPS requirements (see Supplementary Information Figures 5-8), higher targets trigger extreme investment pathway volatility. For example, in Maine, a high RPS leads to maximum wind generation and minimum Combined-Cycle Gas Turbine (CCGT) generation, reducing reliance on firm capacity. Moreover, the system shifts heating loads from heat pumps back to fossil fuels (natural gas and heating oil furnaces) to conserve clean energy for electrical loads that cannot be shifted. These results suggest that when RPS requirements are set aggressively and without consideration of surrounding infrastructure, marginal costs can rise sharply. Energy planners must balance long-term policy goals with near-term emission targets; a concept well documented by Sepulveda et al. [18] in a USA context. It should be noted that one reason we see the strong sensitivity to this policy uncertainty is that we sample between 0% and 100% of the states 2030 RPS target, given the recent and rapid policy changes occurring in the USA.

A state's marginal cost profile in the near-term is governed by a trade-off between fossil fuel price volatility and policy-driven infrastructure constraints. While aggressive mandates can trigger price spikes, our analysis suggests that moderate RPS levels can be achieved while maintaining low system costs. It is not until extreme targets are set, coupled with unique regional bottlenecks, that price spikes occur. From a national energy security perspective, renewable heavy pathways show less volatility when compared to relying on global commodities for energy services. Natural gas and petrol supply chains in the USA are subject to unpredictable geopolitical induced market shocks [19], [20], driving unpredictable energy price trends. Moreover, as Liquified Natural Gas (LNG) terminal expansion continues, the commodity price domestically is subject to greater uncertainty, even in the near-term, as supply and demand markets must adjust accordingly [21]. Policy makers must balance the risk of fossil fuel price volatility and over prescription of renewable mandates to achieve cost optimal systems in the near-term.

## Carbon budgets will be exceeded

The 2030 climate targets are now so immediate that the USA is very unlikely to reduce emissions fast enough to meet its original Paris Agreement commitments. Specifically, no individual state achieved a 50% reduction in carbon emission (CO2 only) from 2005 levels by 2030. However, the Monte Carlo results demonstrate that substantial reductions are possible across a range of scenarios. What's more is that carbon reduction findings are likely conservative estimates due to the lack of regional cooperation represented. While we model Renewable Energy Certificates (RECs), our state level scope does not capture the benefits of multi-regional efforts that can accelerate decarbonization without sacrificing energy price [22], [23].

Despite missing the primary emission targets, significant opportunities for near-term impact remain. As will be discussed, the transportation and service sectors emerge as having the most potential to make decarbonization impacts quickly. This is due to policy that can assist in the deployment of existing commercially viable solutions, such as Electric Vehicles (EVs) and heat pumps. Nevertheless, as highlighted in Figure 4, carbon emissions remain highly uncertain when propagating the influential uncertainties through the system. Crucially, these figures represent a best-case scenario for total atmospheric impact, as Figure 4 only accounts for Co2 and does not incorporate the warming contributions of methane leakage in the near-term.

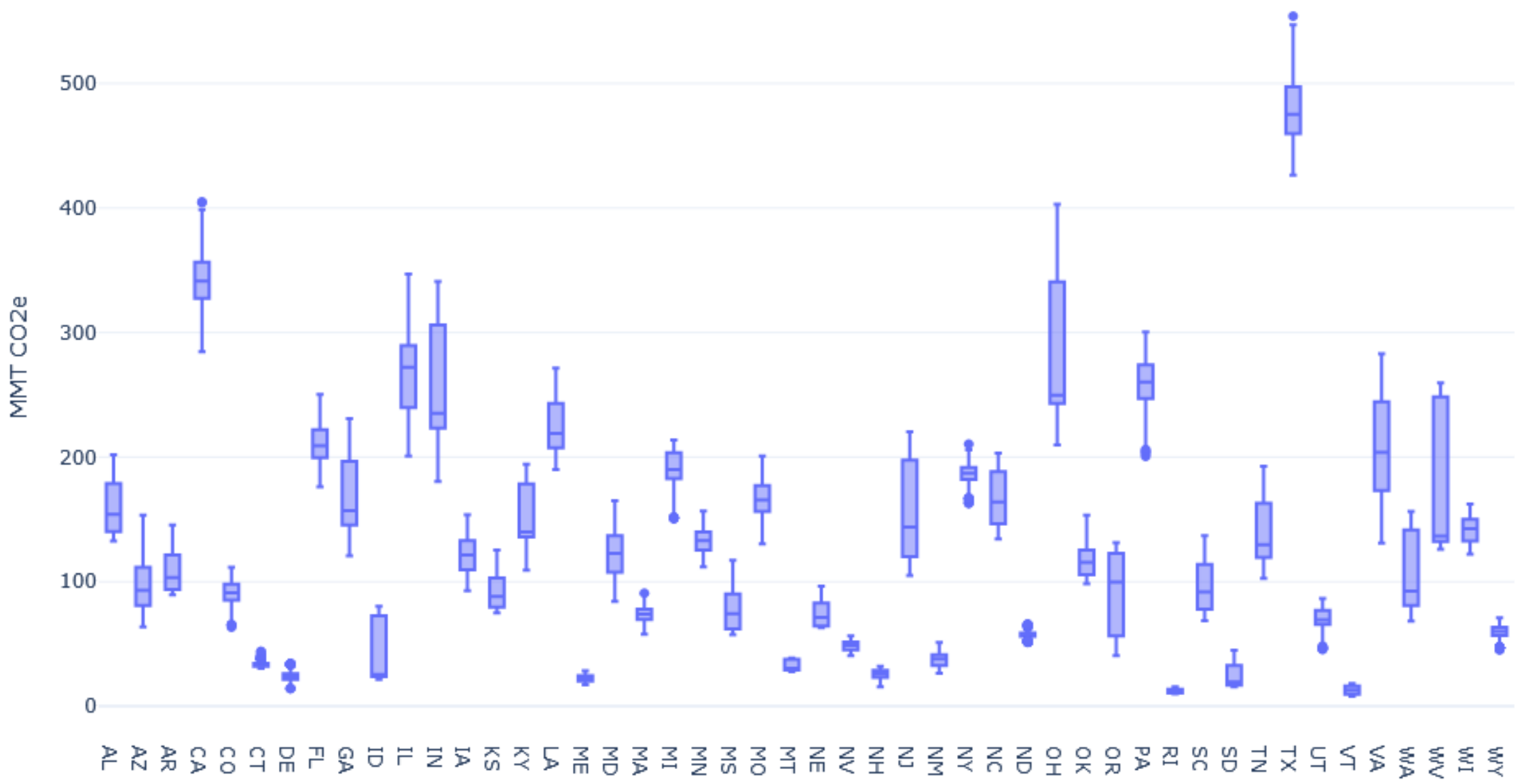

**Figure 4: Carbon emissions boxplot.**

*Boxplot of state level carbon emissions. Box plots are constructed by using GSA to find the uncertainty set of the 5 most impactful parameters impacting marginal energy costs, marginal electricity costs, objective costs, carbon emissions, and total emissions. These uncertainty sets, which are different per state, are then propagated through the model performing a Monte Carlo analysis.*

Integrating methane (CH4) into emission accounting shifts the sensitivity results and greatly expands the uncertainty ranges. Our GSA reveals that total emissions (CO2e) are overwhelmingly susceptible to changes in Global Warming Potential (GWP). GWP ranks as the most influential parameter in 38 of the 48 states, followed by downstream and upstream leakage rates, respectively. To model methane in a planning model, the modeler is responsible for selecting the correct GWP to apply [24] and must make decisions on how to manage the large leakage range uncertainties cited in literature [25]. These topics are both discussed later; however, the important takeaway is that simply ignoring fugitive emissions severely underestimates the energy system's true atmospheric impact. Given the continued heavy usage and reliance of natural gas in the USA, accounting for fugitive emissions is essential to correctly capture atmospheric warming impacts.

## Demand-side strategies offer viable pathways to near-term carbon abatement

While supply-side infrastructure projects are often constrained by years-long lead times, demand side transitions, particularly in the transportation and service sectors, emerge as the most valuable levers for immediate emission reductions. Our analysis identifies the transportation sector as having the highest potential for rapid decarbonization. Unlike industrial decarbonization pathways, which will require supply chain adjustments for large scale fuel switching opportunities [26], [27], EVs and heat pumps are cost competitive today and have support infrastructure in place to facilitate deployment, such as existing charging networks.

Across 18 states, EV adoption limits are identified as a top five sensitive parameter for carbon emissions. This limit highlights that policy intervention focused on EVs stock deployment, such as through manufacturing incentives or consumer subsidies, can yield measurable impacts in the near-term. This finding is consistent with literature [28], and provides future opportunities for large scale vehicle-to-grid policy. Moreover, initiatives that control fuel efficiency in internal combustion vehicles, such as through the Corporate Average Fuel Economy Standards [29], can have notable impacts. Across 20 states, the assumed efficiency of petrol cars is a top five sensitive parameter on carbon emissions. Importantly, this is in direct contrast to the 2026 Environmental Protection Agency's (EPA's) rescission of the motor vehicle GHG emission standards [30], which provided the legal framework to regulate vehicle emissions. The transportation sector offers significant opportunities for high-impact emission reductions in the near-term, but requires policy support for proper implementation.

Service sector heating sensitivity further reveals opportunities to quickly reduce emissions. While the cost of natural gas is generally the most impactful parameter on service sector heating generation, other uncertainties also have noticeable effects. For example, the capital cost of furnaces, Air-Source Heat Pumps (ASHPs) and Ground-Source Heat Pumps (GSHPs), the Coefficient of Performance (COP) of the heat pumps, estimated existing stock levels, and space-heat demand are all found as a top three sensitive parameters on heat generation. Moreover, while not explicitly modeled, uncertainty of heating demand in the service sector is a function of building performance, which has been found to be an immediate requirement for decarbonization pathways in the USA [31]. Policy prioritizing deployment of heat pumps, coupled with research grants to continue improving the cost competitiveness and efficiency of this technology, and incentives to reduce total heating demand can all have near-term decarbonization impacts.

## Discussion

Our analysis studying different 2030 scenario pathways yielded three main results. First, is that marginal costs are sensitive to fossil fuel uncertainty, second is that carbon budgets will be exceeded, and finally is the importance of considering demand side solutions. Here we discuss how to use these results to make positive changes in the energy transition.

## Results are regionally interdependent

A central finding of this research is that state-level energy outcomes are sensitive to the actions of neighboring jurisdictions. This geographic interdependence means that carbon emissions and marginal cost outcomes will be influenced by regional trade flows rather than internal policy and investment decisions alone. This finding is consistent with existing literature [22], however, for clarity we provide a concrete example using Ohio. We select Ohio as it is a major consumer and producer of both natural gas and electricity, making it a critical node in the USA energy network [32].

Our analysis suggests that Ohio's energy landscape is defined by high carbon emissions (5th highest median in Figure 4) and average marginal energy prices (20th highest median in Figure 3). Importantly, these outcomes are not results of Ohio's internal policy or investment decisions, but are driven by external boundary conditions. The GSA in Figure 5 highlights that Ohio's carbon emissions are most sensitive to electricity import prices and trade volume constraints. This suggests that neighboring states' behavior exerts significant pressure on Ohio's ability to decarbonize. This geographic interdependence is even more pronounced in the natural gas sector. Demonstrated in Figure 6 is how natural gas price shocks are felt across state lines. An increase in Ohio's natural gas export price is reflected as an increase in neighboring Indiana's natural gas import price. Ohio's marginal energy prices decrease with an increasing export price, as they make more from selling natural gas. However, Indiana's marginal prices increase as they pay more for the imported natural gas. This illustrates how state-level fossil fuel reliance will impact energy prices across state boundaries.

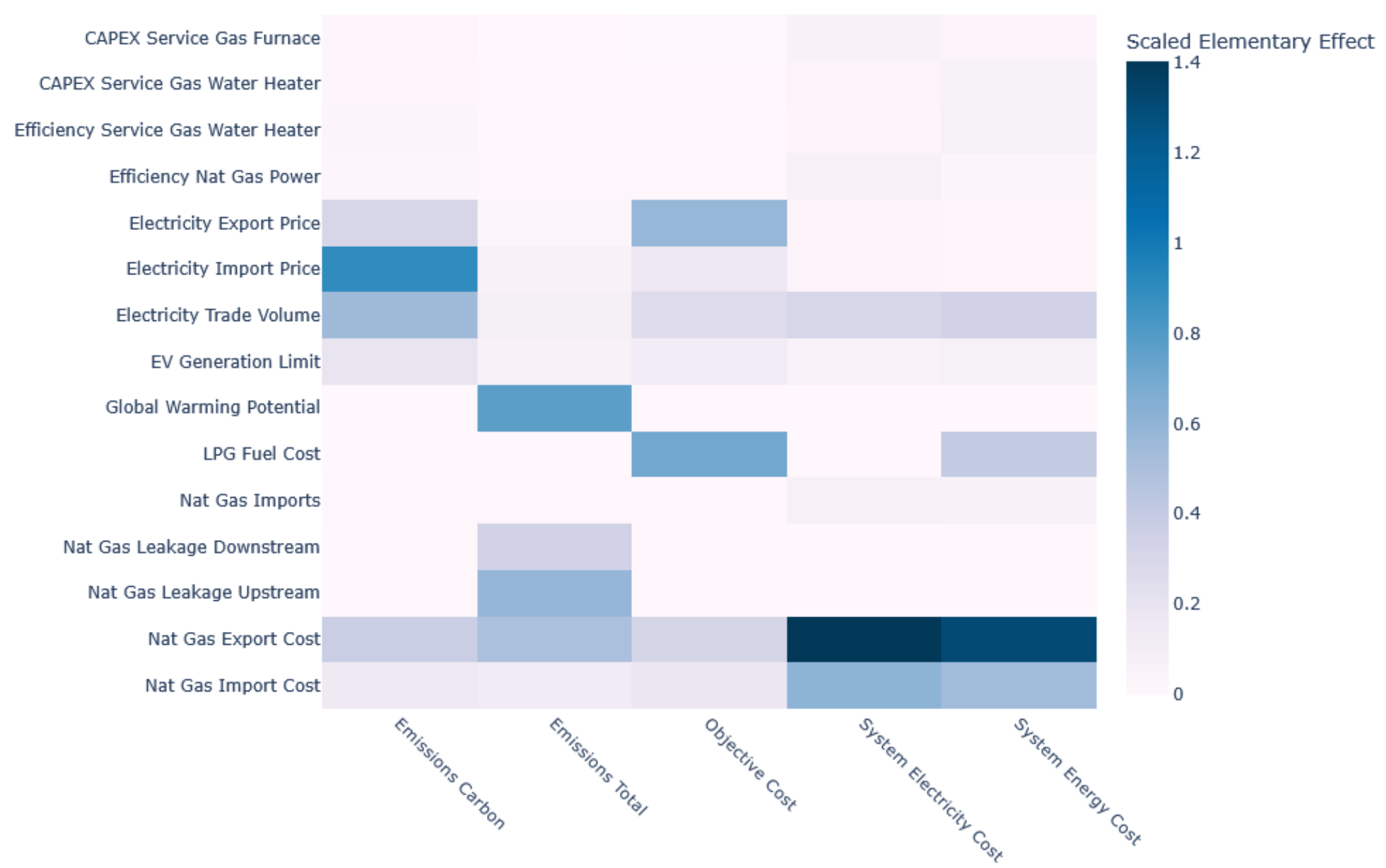

**Figure 5: Sensitivity Results from Ohio.**

*Heatmap showing the scaled elementary effect sensitivity results for Ohio. The values across the bottom horizontal are the five result variables the sensitivity analysis is run over. The values along the vertical are the union of the 5 most impactful parameters on each result parameter. The scaled elementary effect is shown as it is a unitless value that can be used to compare values across different units and scales.*

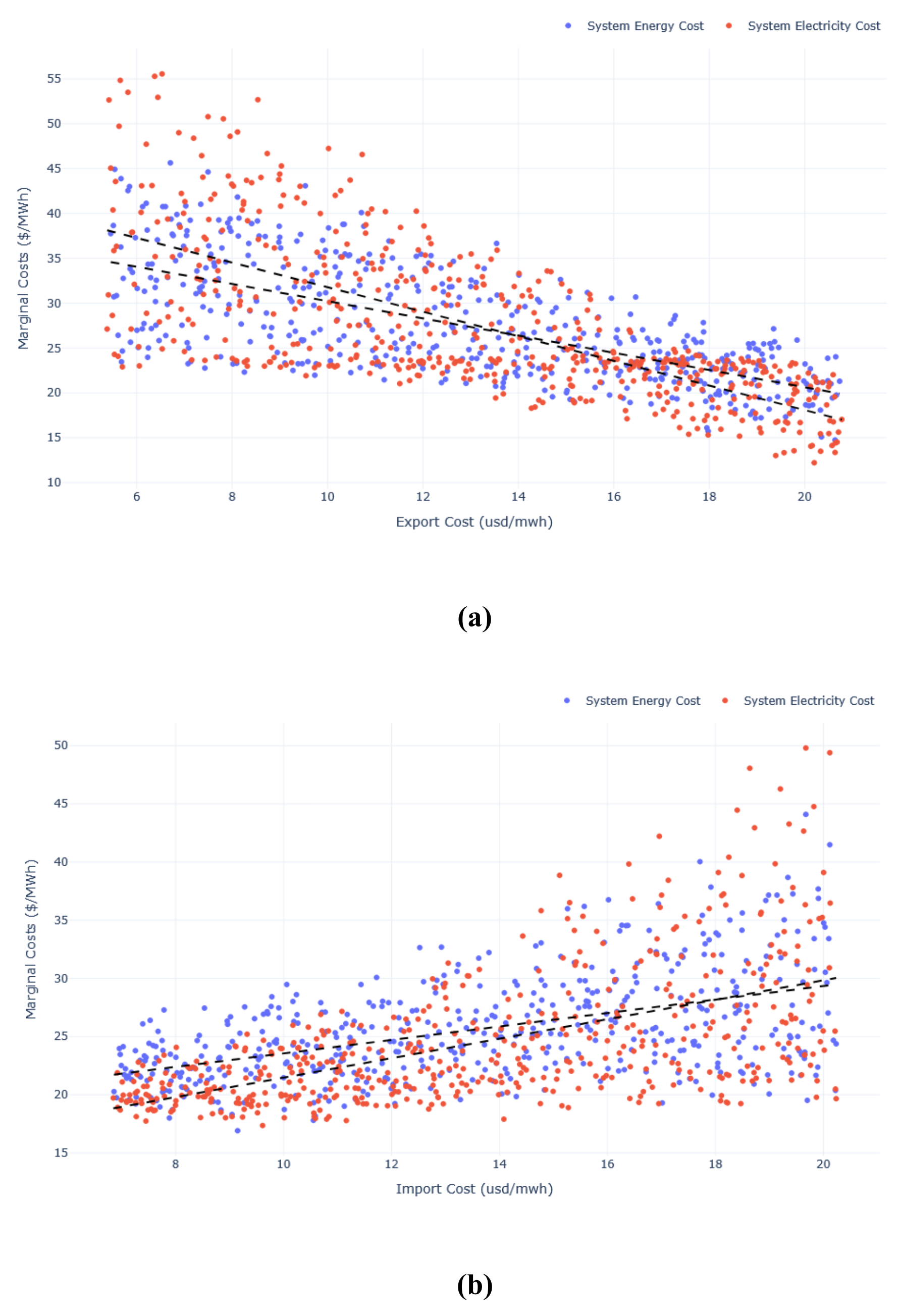

**Figure 6: Impact of Changing Natural Gas Prices.**

*The results of the Monte Carlo analysis are given. Black dashed lines represent the line of best fit for each parameter. (a) Shows how an increase in export costs of natural gas results in decreasing marginal energy costs in Ohio. (b) Shows how increasing import costs of natural gas results in increasing marginal energy costs in neighboring Indiana.*

The geographic sensitivities highlight two critical policy implications in the near-term. First, energy planners cannot effectively mitigate costs and emissions in isolation. A decision or policy in one state can inadvertently impact emissions and marginal prices in another. This sentiment is well highlighted by Jenkins et al. [10] who highlight from their Net-Zero America reports the need of the federal government to proactively employ stakeholder engagement. Second, to ensure policy robustness, energy system models need to move beyond simple heuristics to stress-test regional exogenous drivers. For example, non-obvious factors, such as electricity import prices, can emerge as dominant drivers impacting carbon footprints. While formal uncertainty handling has been identified as a critical modeling tool for many years [33], [34], many studies still do not employ it. This risks the improper exploration of the solution space and only reporting a subset of plausible futures.

## Strategic supply side investment

While immediate timeframe constraints likely place 2030 emission targets out of reach, accelerating low-carbon supply-side infrastructure investments remain a primary driver for low-cost energy systems. Our results reveal that many states are currently operating at the upper bounds of their Variable Renewable Energy (VRE) build rates. Improving the efficiency to permit and connect low-carbon infrastructure is therefore not only a regulatory goal [35], but a requirement for fast system resilience against fossil fuel price volatility. While debates regarding long-term supply-side solutions continue, such as the role of hydrogen [36], [37] and methanol [38], in the near-term the path is clear. Continued low-carbon expansion will positively impact marginal costs and emissions, as long as policy supports strategic planning.

Deployment of supply-side solutions must be strategically planned across the larger energy system. The high sensitivity to RPS in certain regions highlights that aggressive mandates can inadvertently cause price spikes, such as shown in Figure 7. This does not imply that renewables drive up energy prices, rather, it signals that parallel investments in supply side solutions, such as transmission expansion[23], resource diversification [39], and firm capacity [18], must be done in unison. Renewable mandates will inevitably expose system bottlenecks [40]; decision makers and energy modelers must work together to understand these interactions on a regional level and make responsible policy decisions to not sacrifice energy affordability.

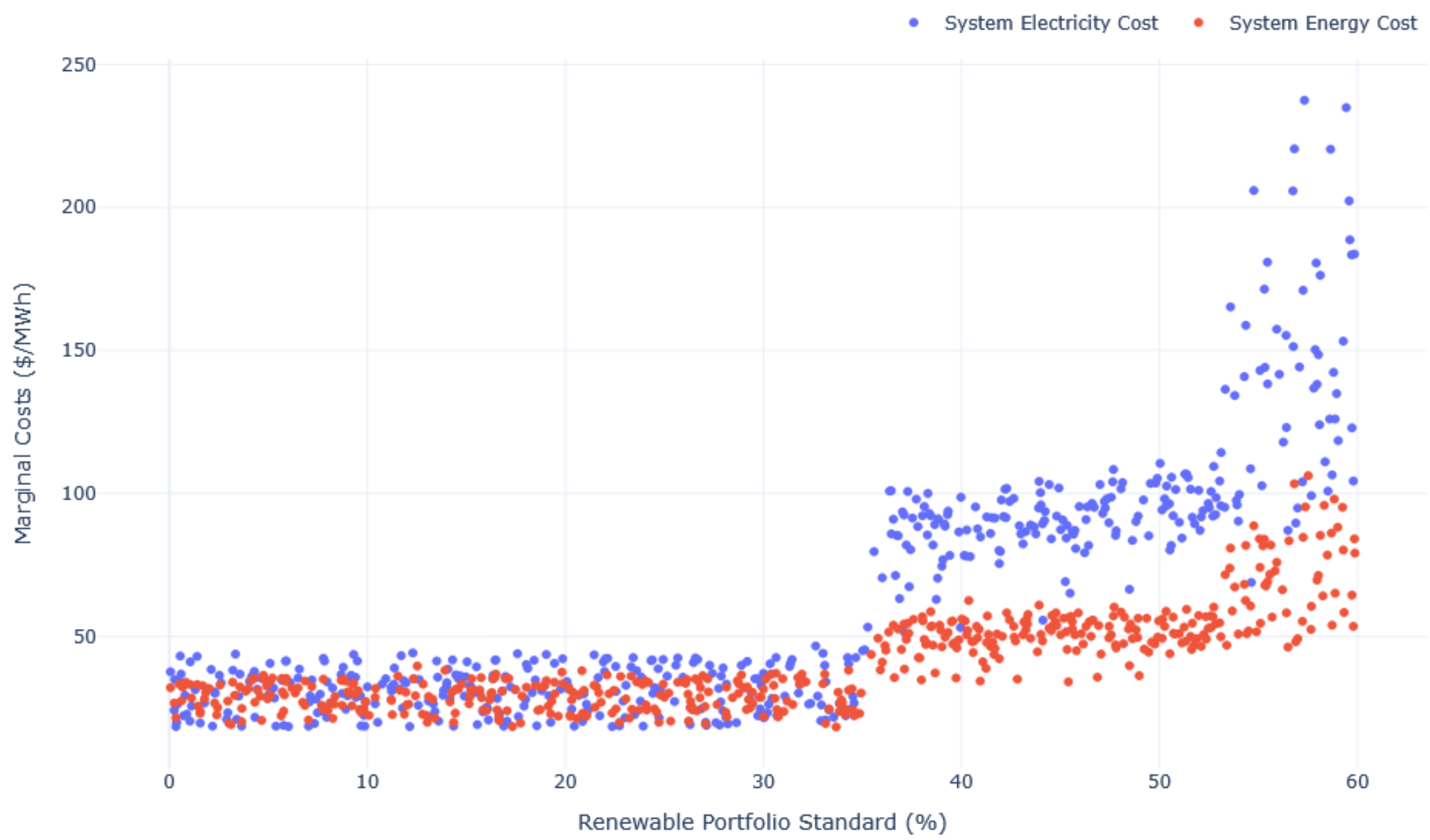

**Figure 7: California's Marginal Energy Costs vs. Renewable Portfolio Standard.**

*Scatter plot showing California's marginal electricity and energy costs against the renewable portfolio standard. The portfolio standard was identified as the most sensitive parameter for both marginal costs. This chart shows the results from the Monte Carlo analysis.*

## Leveraging the natural gas network

Rapid renewable deployment and fuel-switching remain important decarbonization levers in the near-term, however, reliance on the natural gas network will not diminish by 2030. Our model suggests that natural gas will be critical to providing both capacity and energy services across the full energy system. In the power sector, Open-Cycle Gas Turbines (OCGTs) provide essential capacity services, with median utilization rates below 15% in 39 states, while Combined-Cycle Gas Turbines (CCGTs) provide consistent energy services, with median utilization rates above 40% in 40 states. In the service sector, gas furnaces serve as a critical peaking resource providing cost-effective backup capacity during cold periods where heat pump COP degrades. This is evidenced by median utilization rates under 25% in 40 states. Additionally, due to the large legacy stock of gas furnaces, they continue to provide valuable energy services even with low utilization rates.

Continued reliance on natural gas presents both opportunities and risks for policy makers. While legacy gas infrastructure minimizes immediate capital expenditure and provides low-cost capacity, it is coupled to a volatile global commodity. This is a sentiment echoed by studies investigating the impact of infrastructure lock-in [41], [42]. Moreover, with the USA's heavy expansion of LNG terminals, this only increases the uncertainty of natural gas prices domestically [21]. Policy makers must balance the economic benefits of deferring new capital expenditures by relying on stable natural gas infrastructure, against the energy security benefits of decoupling the energy system from fossil fuel price uncertainty. Ignoring this price uncertainty can cause significant marginal price increases that will be passed onto the rate-payer.

## Control of Methane Leaks

As the USA maintains its reliance on natural gas through 2030 and beyond, addressing fugitive methane emissions is paramount to limit unintended atmospheric warming. Given methane's potent GWP, even marginal leakage rates can negate benefits of coal-to-gas fuel switching [41], [43]. This highlights a notable oversight as many planning studies do not account for fugitive emissions.

Our GSA demonstrates, as shown in Figure 8, that GWP is overwhelmingly the primary driver of total emissions (CO2e). When modelling methane, organizations like the EPA and IPCC will use a 100-year time horizon for their GWP estimates [7], [24] where methane is estimated at 27 to 30 CO2e. However, due to the natural decay of methane in the atmosphere, it is much more potent upon release, with a 20-year GWP value of 81-83 CO2e. Depending on the GWP value selected, the damage methane can cause to the atmosphere will drastically change. Clearly communicating this uncertainty to model users is required to fully understand the impact of climate change due to continued natural gas use. Following GWP, the second and third most impactful parameters on total emissions from Figure 8 are upstream and downstream methane leakage rates. Unlike GWP, opportunities exist to reduce leakage rates and limit the impact of methane across all time horizons.

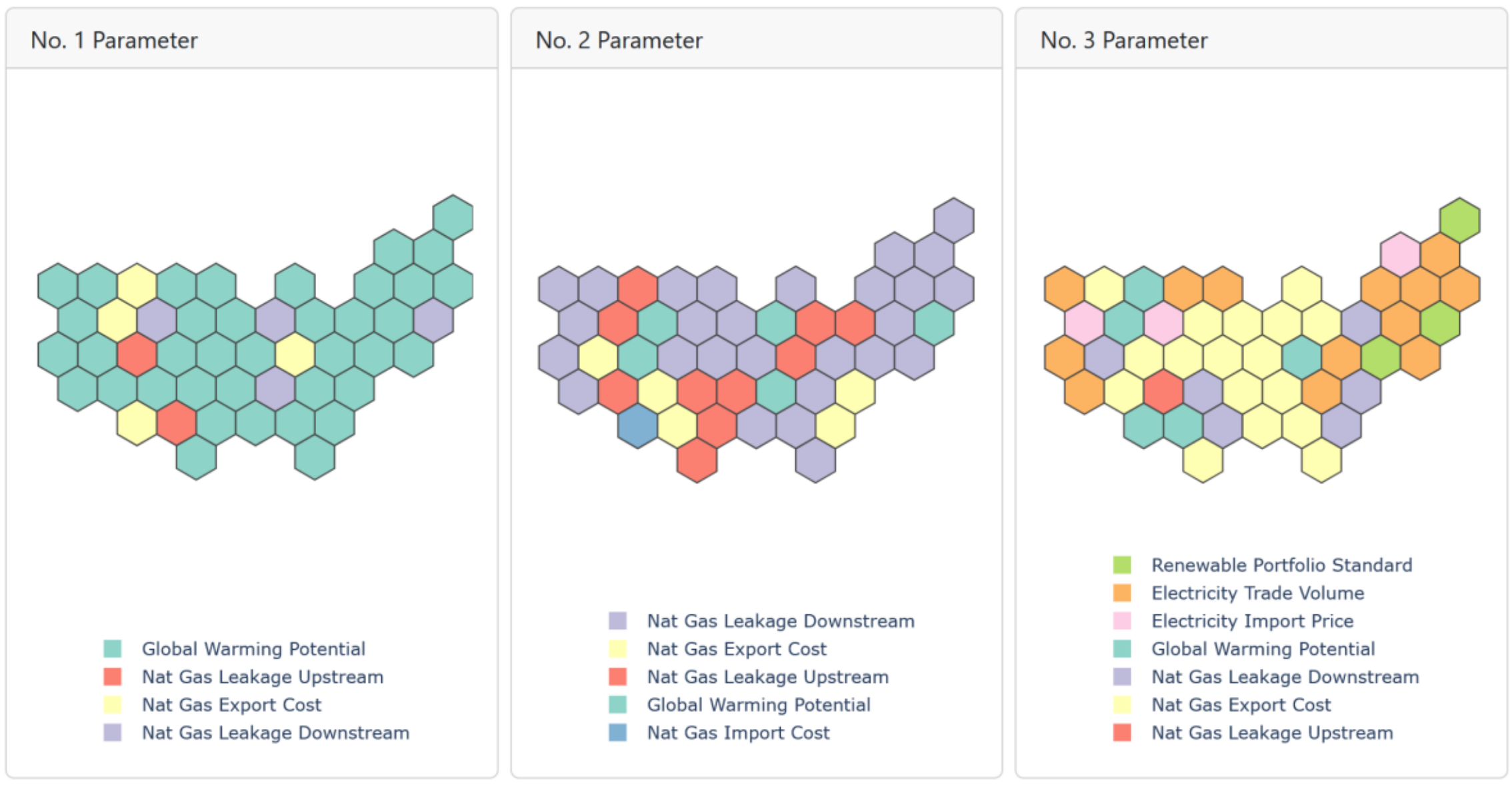

**Figure 8: Total Emissions (CO2 and Ch4) sensitivity**

*The state level sensitivity to total emissions (carbon and methane). Overwhelmingly, global warming potential and methane leakage rate (upstream and downstream) are the most impactful parameters.*

Literature has attempted to quantify and summarize methane leakage rates, however, estimates can vary quite significantly [25], [44], [45], [46]. Leakage rates are often represented as a percentage of total methane produced in natural gas system, and broken into upstream and downstream leaks. Upstream estimates can be from <1% to 20%, downstream estimate can be from <1% to 3.5%, while average total leakage rates from literature are estimated between 2.2% and 4.8%. The leakage rate becomes imperative when atmospheric warming is discussed. For example, estimates place the "break-even" point of coal-to-gas fuel switching between 0.2% and 10%, depending on time-horizon and if including other external factors [47], [48], [49]. While leveraging existing natural gas infrastructure is a cost-effective near-term solution, as we have shown, the leakage rates from this infrastructure cannot be ignored. Doing so significantly adds to total atmospheric warming impacts and can change damages associated with different transition pathways.

Solutions exist today that can help reduce leakage rates and aid in understanding plausible emission pathways. These solutions often target upstream leaks, as downstream leaks are associated with the distribution system and are costly to perform upgrades on. For example, studies have found that if industrial equipment is operating within specifications, upstream methane emission rates should not be as high as found in literature [50], [51]. Policies to enforce stricter methane reporting requirements, as fossil fuel companies have little financial incentive to reduce leaks, can help limit leakage in the near-term. Moreover, initiatives such as the Global Methane Pledge [52] have increased visibility of the fugitive emissions problem and aims to develop effective solution pathways. Modelers and policy makers must work together to properly account for fugitive emissions to make fully informed decisions on how to address the issue.

## The penalty of not acting immediately

In recent years literature has been published highlighting that the social cost of carbon, a metric estimating monetary damages of releasing one additional unit of carbon, has been steadily increasing [53], [54]. This literature suggests that, from not only an atmospheric perspective, but also a financial perspective, it is paramount to reduce emissions quickly. What's more is that the USA is one of the largest carbon emitters in the world, meaning their carbon emission can have a significant impact on global future damages [55]. These include future financial damages, but also damages in mortality rates [56]. When considering equity, if the social cost of carbon is weighted by income, meaning that dollar-valued impacts are adjusted based on the income of individuals, its value can increase by a factor of 8 [57]. As a wealthy nation, the USA has significant incentives, financial and otherwise, to reduce emissions in the near-term. Understanding the tradeoffs associated with the social cost of carbon and deferring immediate emission reductions must be investigated fully.

Although our findings suggest that 2030 targets are likely already out of reach, the importance of striving for as much decarbonization as possible in the near-term remains. Every year of deferred clean energy investment, especially when commercially viable solutions exist today, such as EVs and heat pumps, compound the economic penalty to be paid in the future. Moreover, as being one of the largest emitters in the world [5], the USA has opportunities to set a global precedence on how to make cost effective energy transitions. For example, as earlier discussed, implementing strategic supply side energy investment policy with energy mandates, focusing on demand side solutions for quick emission impacts, and enforcing strict methane accounting requirements. Prioritizing solutions that incentivize low-cost abatement, without sacrificing energy marginal costs, the USA can avoid substantial future damage premiums while showcasing effective methods that work at scale.

## Impact of climate policy retraction

Given the quick four-year political cycle in the USA, climate policy in the nation has seen numerous changes in direction over the last decade. For example, many of the clean-energy incentives that the IRA introduced in 2022 have since been stripped back under the OBBBA in 2025. From a climate and rate payer perspective, this has largely been seen as a negative by academics [3]. For example, studies show that this change in policy direction can lead to increased electricity and energy prices by 2030, due to continued reliance on fossil fuels [4], [12], [58], and force carbon targets further off-track, due to reduced clean energy investment [4], [59].

Our study adds to the growing evidence that the shift in USA policy, transitioning from the IRA to the OBBBA, does not benefit ratepayers or decarbonization progress. Our GSA suggests that managing marginal energy costs requires lessening the systems' reliance on fossil fuels; yet the OBBBA does the opposite [3]. For example, the OBBBA removes rebates and incentives aimed at service sector electrification and eliminates EV credits and charging centre incentives. EVs and service sector electrification are both identified in this work as critical solutions in the near-term to manage costs and emissions. Furthermore, under the OBBBA, clean energy capacity additions are expected to retract to pre-IRA estimates [3]. States implementing policy to drive decarbonization, such as through RPS, will likely face intensified bottlenecks as IRA financial support for clean energy generation is modified and/or rescinded. In these circumstances, our results suggest that the OBBBA does not only delay decarbonization, but it also removes the economic incentives aimed to protect the USA against commodity price volatility.

## Conclusions

This study is the first of its kind to investigate, through formal uncertainty analysis, the reality of near-term emission targets in the USA. Under current policy trajectories, the 50-52% reduction of emissions from 2005 levels proposed in the original Paris Agreement is likely out of reach. While this is not a novel finding, it is important to highlight the consequences of relaxing decarbonization targets. In a time where the social cost of carbon is increasing, repealing many IRA clean energy incentives under the OBBBA suggest larger future financial damages will need to be managed.

Our sensitivity analysis identifies fossil fuel price volatility as the primary driver of electricity and energy cost instability across much of the USA. While decarbonization policies, such as RPS, are essential to transition to equitable, secure, and clean energy, they must be deployed strategically. If parallel supply side clean energy investments (such as were incentivized under the IRA) do not occur, localized marginal price spikes caused by system bottlenecks can arise. To remedy this, demand side solutions, and in particular heat pumps and EVs, offer the most impactful decarbonization solutions in the near-term. Alternatively, as many states reached build limits on their VRE capacity, implementing policies to expedite supply side investment can manage exposure to fossil fuel price uncertainty.

Finally, this research highlights the immediacy of addressing the problem of fugitive emissions. Methods to report the impacts of carbon and the impacts of atmospheric warning within energy planning studies are uncertain and often accounted for by simply ignoring fugitive emissions. Our study shows that CO2e reporting is overwhelming sensitive to the values selected for GWP, upstream leakage rates, and downstream leakage rates. In particular, upstream leakage rates can be addressed with regulations ensuring the correct operation of equipment. Focusing efforts to correctly report atmospheric warming values, in addition to carbon emissions, will provide decision makers with more clarity on where to focus climate change efforts.

## Limitations of study

Our study provides a comprehensive probabilistic assessment of the 2030 energy transition; however, limitations of the study must be discussed. Our model runs each state individually at roughly balancing level authority level and with 3-hour temporal aggregation to maintain computational tractability. These decisions have two impacts. First, the spatial resolution likely results in conservative values for emission reduction potential, as interregional cooperation can accelerate decarbonization timeframe [22]. Second, the 3-hour resolutions smooth demand peaks, potentially underrepresenting the value of capacity serving technologies, such as OCGT plants and demand response.

Additionally, our industrial load growth projections rely on estimates from the EIA Annual Energy Outlook[60]. While an exceedingly reputable source, forecasted data center load growth trends are changing month-to-month. Moreover, the unique point-load nature and potential for on-site generation for data centers are not captured in this model. Unforeseen industrial electrical demand shocks in certain regions could intensify the fossil fuel price sensitivities identified in our GSA. Further investigation into how near-term data center demand uncertainty interacts with the other uncertainties in the system can be studied.

## Comparison with other sensitivity analysis literature

ESOMs are abstract computer representations of complex socio-technical systems, characterized by a high degree of uncertainty. Despite this, formal sensitivity analysis is not often performed, even when confident conclusions are being presented [61]. While the benefits of GSA as applied to ESOMs have been well documented, its overall use in literature remains underrepresented [33], [34], [62].

Our focus in this study was using GSA as a method to address parametric uncertainty; that being uncertainty associated with model data inputs. Literature has used GSA to address ESOM parametric uncertainty in both a USA context[63] and in other geographies[64], [65], [66]. These studies show that only a handful of parameters, such as demand, fuel costs, discount rates, and technology costs, are impactful. Our study confirms the impact of only a select group of parameters causing most result movement, however, our study disagrees with what these parameters are. Generally, we show that the boundary conditions (costs to import and export commodities), methane specific parameters, and policy levers have the largest impact. These different findings are likely driven by the results of interest and the time frame of the study. Long-term studies need to deal with larger uncertainty ranges, which will inevitably cause larger model responses. For example, over the next five years, we know reasonably well what technologies costs will be compared to over the next 30. Therefore, it's not necessarily surprising that the model is less sensitive to this proportionally smaller uncertainty range compared to a fossil fuel price.

## Additional Information

**Lead Contact**

Further information and requests for resources should be directed to and will be fulfilled by the lead contact, Trevor Barnes (trevor_barnes@sfu.ca).

**Data and Code Availability**

All data is generated using v0.8.0 of PyPSA-USA, available at https://github.com/PyPSA/pypsa-usa. The sensitivity analysis and data reporting code is available at https://github.com/DeltaE/pypsa-gsa. All datasets generated during this study are available at https://doi.org/10.5281/zenodo.20348340. A public dashboard for viewing results is available at https://pypsa-usa-gsa.com.

**Model Description**

This study uses the open-source and open-data energy system optimization model, PyPSA-USA. The model represents the full energy system of the USA. We model 134 load zones at 3hr resolution of the 2030 planning horizon. Each load zone roughly corresponds to the size of a balancing authority, and each state is run independently to maintain computational tractability. Formulated as a linear cost optimization, PyPSA-USA minimizes net-present capital and investment costs. The supplemental information provides extensive details on model development, data sources, and study setup.

To perform the GSA, we use the most common technique found in ESOM literature[62], that being the Method of Morris [67], [68]. The Method of Morris is computationally efficient, a necessity to handle the high dimensionality of our input space. We acknowledge that emerging techniques, such as applying variance based GSA methods to surrogate ESOMs [69] or implementing GSA via optimal transport theory[70] exist, but were not applied here. These methods allow deeper model insights, such as investigating high-order interaction effects, but were not required for this specific research question. The supplemental information provides extensive details on the Method of Morris implementation.

**Supplemental Information**

Document S1. Model details and experimental procedures

Document S2. Additional results

**Declaration of AI Use**

The generative artificial intelligence tool Google Gemini was used to improve the clarity and grammar of the manuscript throughout. The generative artificial intelligence tools Google Gemini and Claude were used to assist in writing the Python data processing scripts. These tools were not used to generate the research objective, perform the literature review, perform data analysis, perform data interpretation, or to generate any novel scientific findings presented in this article. After using these tools, the authors reviewed and edited all content and take full responsibility for the content of the publication.

**Acknowledgments**

This project was sponsored by Mitacs and Generac under grant number IT33642. We would like to thank Dr. Mariana Resner (SFU), Dr. Erik Schang (SFU), Yakov Familiant (Generac) and Dr. Tod Tesch (Generac) for providing guidance on project direction. Finally, Dr. Will Usher (Open Energy Transition) for providing feedback on the global sensitivity analysis methodology.

**Author Contributions**

Conceptualization: T.B., B.R., M.M., T.N. | Methodology: T.B., K.T., B.R., M.M., T.N. | Software: T.B., K.T. | Validation: T.B. | Formal Analysis: T.B. | Investigation: T.B. | Data Curation: T.B., K.T. | Writing

– Original Draft: T.B. | Writing – Review and Editing: T.B., K.T., B.R., M.M., T.N. | Visualization: T.B. | Supervision: B.R., M.M., T.N. | Project Administration: B.R., T.N. | Funding Acquisition: T.B., T.N.

## Declaration of Interests

Generac Power Systems and Mitacs provided funding for this research. Bradley Reinholz is an employee of Generac Power Systems. Malcolm Metcalfe was an employee of Generac Power Systems during project development. The remaining authors declare no competing interests

# References

[1] The White House, “Putting America First In International Environmental Agreements,” The White House. Accessed: May 04, 2026. [Online]. Available: https://www.whitehouse.gov/presidential-actions/2025/01/putting-america-first-in-international-environmental-agreements/

[2] United Nations, “Paris Agreement.” 2015. Accessed: May 05, 2023. [Online]. Available: https://unfccc.int/files/essential_background/convention/application/pdf/english_paris_agreement.pdf

[3] J. E. T. Bistline et al., “Impacts of the Inflation Reduction Act and One Big Beautiful Bill Act on the US energy system,” Nat. Rev. Clean Technol., pp. 1–12, Apr. 2026, doi: 10.1038/s44359-026-00168-z.

[4] J. Jenkins, J. Farbes, and B. Haley, “Impacts of the One Big Beautiful Bill On The US Energy Transition – Summary Report,” REPEAT Project, Jul. 2025. doi: 10.5281/zenodo.15801701.

[5] H. Ritchie, P. Rosado, and M. Roser, “Energy,” Our World in Data, Jan. 2024, Accessed: Feb. 14, 2024. [Online]. Available: https://ourworldindata.org/energy

[6] J. Bistline et al., “Emissions and energy impacts of the Inflation Reduction Act,” Science, vol. 380, no. 6652, pp. 1324–1327, Jun. 2023, doi: 10.1126/science.adg3781.

[7] A. Mukherji et al., “Synthesis Report of the IPCC Sixth Assessment Report (AR6).” 2023. [Online]. Available: https://www.ipcc.ch/report/ar6/syr/downloads/report/IPCC_AR6_SYR_SPM.pdf

[8] J. Bistline et al., “Actions for reducing US emissions at least 50% by 2030,” Science, vol. 376, no. 6596, pp. 922–924, May 2022, doi: 10.1126/science.abn0661.

[9] G. Iyer et al., “A multi-model study to inform the United States’ 2035 NDC,” Nat Commun, vol. 16, no. 1, p. 643, Jan. 2025, doi: 10.1038/s41467-025-55858-2.

[10] J. D. Jenkins, E. N. Mayfield, E. D. Larson, S. W. Pacala, and C. Greig, “Mission net-zero America: The nation-building path to a prosperous, net-zero emissions economy,” Joule, vol. 5, no. 11, pp. 2755–2761, Nov. 2021, doi: 10.1016/j.joule.2021.10.016.

[11] A. Zhao et al., “High-ambition climate action in all sectors can achieve a 65% greenhouse gas emissions reduction in the United States by 2035,” npj Clim. Action, vol. 3, no. 1, p. 63, Jul. 2024, doi: 10.1038/s44168-024-00145-x.

[12] S. Tuladhar, J. M. Carey, B. Ramkrishnan, and J. Chen, “Economic Impacts of Repealing Technology Neutral Tax Credits,” NERA, 2026.

[13] K. Tehranchi, T. Barnes, M. Frysztacki, and I. Azevedo, “Pypsa-Usa: A Flexible Open-Source Energy System Model and Optimization Tool for the United States,” Nov. 21, 2024, Social Science Research Network, Rochester, NY: 5029120. doi: 10.2139/ssrn.5029120.

[14] F. Schweppe, M. Caramanis, R. Tabors, and R. Bohn, Spot Pricing of Electricity. in Power Electronics and Power Systems. Springer New York, NY, 1988. doi: https://doi.org/10.1007/978-1-4613-1683-1.

[15] S. Harvey and W. Hogan, “Locational Marginal Prices and Electricity Markets,” Harvard Kennedy School, Oct. 2022. Accessed: May 04, 2026. [Online]. Available: https://whogan.scholars.harvard.edu/papers

[16] T. Jiang, H. Deng, L. Bai, R. Zhang, X. Li, and H. Chen, “Optimal energy flow and nodal energy pricing in carbon emission-embedded integrated energy systems,” CSEE Journal of Power and Energy Systems, vol. 4, no. 2, pp. 179–187, Jun. 2018, doi: 10.17775/CSEEJPES.2018.00030.

[17] T. Barnes, DeltaE/pypsa-gsa. (2026). Jupyter Notebook. ΔE+. doi: https://doi.org/10.5281/zenodo.20349002.

[18] N. A. Sepulveda, J. D. Jenkins, F. J. de Sisternes, and R. K. Lester, “The Role of Firm Low-Carbon Electricity Resources in Deep Decarbonization of Power Generation,” Joule, vol. 2, no. 11, pp. 2403–2420, Nov. 2018, doi: 10.1016/j.joule.2018.08.006.

[19] S. Li, H. Chen, and G. Chen, “The US-China tension and fossil fuel energy price volatility relationship,” Finance Research Letters, vol. 74, p. 106707, Mar. 2025, doi: 10.1016/j.frl.2024.106707.

[20] P. Maneejuk, N. Kaewtathip, and W. Yamaka, “The influence of the Ukraine-Russia conflict on renewable and fossil energy price cycles,” Energy Economics, vol. 129, p. 107218, Jan. 2024, doi: 10.1016/j.eneco.2023.107218.

[21] Department of Energy, “2024 LNG Export Study: Energy, Economic, and Environmental Assessment of U.S. LNG Exports,” Dec. 2024. Accessed: Dec. 11, 2025. [Online]. Available: https://www.federalregister.gov/documents/2024/12/20/2024-30370/2024-lng-export-study-energy-economic-and-environmental-assessment-of-us-lng-exports

[22] P. R. Brown and A. Botterud, “The Value of Inter-Regional Coordination and Transmission in Decarbonizing the US Electricity System,” Joule, vol. 5, no. 1, pp. 115–134, Jan. 2021, doi: 10.1016/j.joule.2020.11.013.

[23] National Laboratory of the Rockies, “National Transmission Planning Study,” Apr. 2026. Accessed: May 04, 2026. [Online]. Available: https://www.energy.gov/oe/national-transmission-planning-study-0

[24] Environmental Protection Agency, “Understanding Global Warming Potentials.” Accessed: May 04, 2026. [Online]. Available: https://www.epa.gov/ghgemissions/understanding-global-warming-potentials

[25] R. W. Howarth, “Methane Emissions from the Production and Use of Natural Gas,” The Magazine for Environmental Managers, 2022.

[26] J. Cresko et al., “U.S. Department of Energy’s Industrial Decarbonization Roadmap,” DOE/EE--2635, 1961393, Sep. 2022. doi: 10.2172/1961393.

[27] Department of Energy, “Pathways to Commercial Liftoff: Industrial Decarbonization,” 2023. [Online]. Available: https://climateprogramportal.org/resource/pathways-to-commercial-liftoff/

[28] M. Muratori et al., “Trends and 2025 insights on the rise of electric vehicles in the USA,” Nat. Rev. Clean Technol., vol. 1, no. 12, pp. 827–845, Oct. 2025, doi: 10.1038/s44359-025-00108-3.

[29] National Highway Traffic Safety Administration, “Corporate Average Fuel Economy.” Accessed: May 04, 2026. [Online]. Available: https://www.nhtsa.gov/laws-regulations/corporate-average-fuel-economy

[30] Environmental Protection Agency, Rescission of the Greenhouse Gas Endangerment Finding and Motor Vehicle Greenhouse Gas Emission Standards Under the Clean Air Act. 2026. doi: 10.2760/953322.

[31] P. Berrill, E. J. H. Wilson, J. L. Reyna, A. D. Fontanini, and E. G. Hertwich, “Decarbonization pathways for the residential sector in the United States,” Nat. Clim. Chang., vol. 12, no. 8, pp. 712–718, Aug. 2022, doi: 10.1038/s41558-022-01429-y.
[32] EIA, “Homepage - U.S. Energy Information Administration (EIA).” Accessed: Dec. 28, 2021. [Online]. Available: https://www.eia.gov/index.php
[33] J. DeCarolis et al., “Formalizing best practice for energy system optimization modelling,” Applied Energy, vol. 194, pp. 184–198, May 2017, doi: 10.1016/j.apenergy.2017.03.001.
[34] X. Yue, S. Pye, J. DeCarolis, F. G. N. Li, F. Rogan, and B. Ó. Gallachóir, “A review of approaches to uncertainty assessment in energy system optimization models,” Energy Strategy Reviews, vol. 21, pp. 204–217, Aug. 2018, doi: 10.1016/j.esr.2018.06.003.
[35] D. Boff and S. Barrows, “Waiting in Queue: A Historical Evaluation of Interconnection Policy,” PNNL--34350, 1996548, Jun. 2023. doi: 10.2172/1996548.
[36] Department of Energy, “Pathways to Commercial Liftoff: Clean Hydrogen,” 2024. Accessed: Mar. 23, 2023. [Online]. Available: https://climateprogramportal.org/resource/pathways-to-commercial-liftoff/
[37] L. Schumm et al., “The impact of temporal hydrogen regulation on hydrogen exporters and their domestic energy transition,” Nat Commun, vol. 16, no. 1, p. 7486, Aug. 2025, doi: 10.1038/s41467-025-62873-w.
[38] P. Glaum, F. Neumann, M. Millinger, and T. Brown, “A minimal methanol backstop for high-electrification scenarios,” Joule, p. 102464, May 2026, doi: 10.1016/j.joule.2026.102464.
[39] N. Patankar, H. Eshraghi, A. R. de Queiroz, and J. F. DeCarolis, “Using robust optimization to inform US deep decarbonization planning,” Energy Strategy Reviews, vol. 42, p. 100892, Jul. 2022, doi: 10.1016/j.esr.2022.100892.
[40] D. Young and J. Bistline, “The costs and value of renewable portfolio standards in meeting decarbonization goals,” Energy Economics, vol. 73, pp. 337–351, Jun. 2018, doi: 10.1016/j.eneco.2018.04.017.
[41] C. Kemfert, F. Präger, I. Braunger, F. M. Hoffart, and H. Brauers, “The expansion of natural gas infrastructure puts energy transitions at risk,” Nat Energy, vol. 7, no. 7, pp. 582–587, Jul. 2022, doi: 10.1038/s41560-022-01060-3.
[42] P. Achakulwisut, P. Erickson, C. Guivarch, R. Schaeffer, E. Brutschin, and S. Pye, “Global fossil fuel reduction pathways under different climate mitigation strategies and ambitions,” Nat Commun, vol. 14, no. 1, p. 5425, Sep. 2023, doi: 10.1038/s41467-023-41105-z.
[43] S. Yang, S. Hastings-Simon, and A. P. Ravikumar, “Global liquefied natural gas expansion exceeds demand for coal-to-gas switching in paris compliant pathways,” Environ. Res. Lett., vol. 17, no. 6, p. 064048, Jun. 2022, doi: 10.1088/1748-9326/ac71ba.
[44] R. W. Howarth, “A bridge to nowhere: methane emissions and the greenhouse gas footprint of natural gas,” Energy Science & Engineering, vol. 2, no. 2, pp. 47–60, 2014, doi: 10.1002/ese3.35.
[45] R. W. Howarth, R. Santoro, and A. Ingraffea, “Methane and the greenhouse-gas footprint of natural gas from shale formations,” Climatic Change, vol. 106, no. 4, pp. 679–690, Jun. 2011, doi: 10.1007/s10584-011-0061-5.
[46] S. Schwietzke et al., “Upward revision of global fossil fuel methane emissions based on isotope database,” Nature, vol. 538, no. 7623, Art. no. 7623, Oct. 2016, doi: 10.1038/nature19797.
[47] R. A. Alvarez, S. W. Pacala, J. J. Winebrake, W. L. Chameides, and S. P. Hamburg, “Greater focus needed on methane leakage from natural gas infrastructure,” Proceedings of the National Academy of Sciences, vol. 109, no. 17, pp. 6435–6440, Apr. 2012, doi: 10.1073/pnas.1202407109.
[48] D. Gordon, F. Reuland, D. J. Jacob, J. R. Worden, D. Shindell, and M. Dyson, “Evaluating net life-cycle greenhouse gas emissions intensities from gas and coal at varying methane leakage rates,” Environ. Res. Lett., vol. 18, no. 8, p. 084008, Jul. 2023, doi: 10.1088/1748-9326/ace3db.
[49] S. Schwietzke, W. M. Griffin, H. S. Matthews, and L. M. P. Bruhwiler, “Natural Gas Fugitive Emissions Rates Constrained by Global Atmospheric Methane and Ethane,” Environ. Sci. Technol., vol. 48, no. 14, pp. 7714–7722, Jul. 2014, doi: 10.1021/es501204c.
[50] D. Zavala-Araiza et al., “Super-emitters in natural gas infrastructure are caused by abnormal process conditions,” Nat Commun, vol. 8, no. 1, Art. no. 1, Jan. 2017, doi: 10.1038/ncomms14012.
[51] D. R. Lyon, R. A. Alvarez, D. Zavala-Araiza, A. R. Brandt, R. B. Jackson, and S. P. Hamburg, “Aerial Surveys of Elevated Hydrocarbon Emissions from Oil and Gas Production Sites,” Environ. Sci. Technol., vol. 50, no. 9, pp. 4877–4886, May 2016, doi: 10.1021/acs.est.6b00705.
[52] Climate and Clean Air Coalition, “Global Methane Pledge.” 2023. Accessed: Jun. 24, 2023. [Online]. Available: https://www.ccacoalition.org/en/resources/global-methane-pledge
[53] K. Rennert et al., “Comprehensive evidence implies a higher social cost of CO2,” Nature, vol. 610, no. 7933, pp. 687–692, Oct. 2022, doi: 10.1038/s41586-022-05224-9.
[54] R. S. J. Tol, “Social cost of carbon estimates have increased over time,” Nat. Clim. Chang., vol. 13, no. 6, pp. 532–536, Jun. 2023, doi: 10.1038/s41558-023-01680-x.
[55] M. Burke, M. Zahid, N. S. Diffenbaugh, and S. Hsiang, “Quantifying climate loss and damage consistent with a social cost of carbon,” Nature, vol. 651, no. 8107, pp. 959–966, Mar. 2026, doi: 10.1038/s41586-026-10272-6.
[56] R. D. Bressler, “The mortality cost of carbon,” Nat Commun, vol. 12, no. 1, p. 4467, Jul. 2021, doi: 10.1038/s41467-021-24487-w.
[57] B. C. Prest, L. Rennels, F. Errickson, and D. Anthoff, “Equity weighting increases the social cost of carbon,” Science, vol. 385, no. 6710, pp. 715–717, Aug. 2024, doi: 10.1126/science.adn1488.
[58] Energy Innovation, “Impacts of the One Big Beautiful Bill on US Energy, Health, Costs, Jobs, and Emissions,” Energy Innovation, 2026. Accessed: May 09, 2026. [Online]. Available: https://energyinnovation.org/wp-content/uploads/Impacts-Of-The-One-Big-Beautiful-Bill-On-U.S.-Energy-Costs-Jobs-Health-And-Emissions_FINAL.pdf
[59] Rhodium Group and MIT Center for Energy and Environmental Policy Research, “Clean Investment Monitor: US,” Clean Investment Monitor. Accessed: May 09, 2026. [Online]. Available: https://www.cleaninvestmentmonitor.org/us
[60] Energy Information Administration, “Annual Energy Outlook 2025.” Accessed: May 04, 2026. [Online]. Available: https://www.eia.gov/outlooks/archive/aeo25/index.php
[61] A. Puy et al., “Socio-environmental modeling shows physics-like confidence with water modeling surpassing it in numerical claims,” iScience, vol. 28, no. 4, Apr. 2025, doi: 10.1016/j.isci.2025.112184.
[62] W. Usher, T. Barnes, N. Moksnes, and T. Niet, “Global sensitivity analysis to enhance the transparency and rigour of energy system optimisation modelling [version 1; peer review: 1 approved, 2 approved with reservations],” Open Res Europe, vol. 3, p. 30, Feb. 2023, doi: 10.12688/openreseurope.15461.1.

[63] H. Eshraghi, A. R. De Queiroz, and J. F. DeCarolis, “US Energy-Related Greenhouse Gas Emissions in the Absence of Federal Climate Policy,” Environ. Sci. Technol., vol. 52, no. 17, pp. 9595–9604, Sep. 2018, doi: 10.1021/acs.est.8b01586.
[64] T. Tröndle, J. Lilliestam, S. Marelli, and S. Pfenninger, “Trade-Offs between Geographic Scale, Cost, and Infrastructure Requirements for Fully Renewable Electricity in Europe,” Joule, vol. 4, no. 9, pp. 1929–1948, Sep. 2020, doi: 10.1016/j.joule.2020.07.018.
[65] S. Moret, V. Codina Gironès, M. Bierlaire, and F. Maréchal, “Characterization of input uncertainties in strategic energy planning models,” Applied Energy, vol. 202, pp. 597–617, Sep. 2017, doi: 10.1016/j.apenergy.2017.05.106.
[66] N. Moksnes and W. Usher, “The relative importance of uncertain parameters and structural formulation for electricity systems planning in Kenya and Benin,” iScience, vol. 28, no. 2, p. 111792, Feb. 2025, doi: 10.1016/j.isci.2025.111792.
[67] M. D. Morris, “Factorial Sampling Plans for Preliminary Computational Experiments,” Technometrics, vol. 33, no. 2, pp. 161–174, 1991, doi: 10.2307/1269043.
[68] F. Campolongo, J. Cariboni, and A. Saltelli, “An effective screening design for sensitivity analysis of large models,” Environmental Modelling & Software, vol. 22, no. 10, pp. 1509–1518, Oct. 2007, doi: 10.1016/j.envsoft.2006.10.004.
[69] F. Neumann and T. Brown, “Broad ranges of investment configurations for renewable power systems, robust to cost uncertainty and near-optimality,” iScience, vol. 26, no. 5, p. 106702, May 2023, doi: 10.1016/j.isci.2023.106702.
[70] M. Nicoli et al., “A framework for global sensitivity analysis in long-term energy systems planning using optimal transport,” Energy, vol. 338, p. 138788, Nov. 2025, doi: 10.1016/j.energy.2025.138788.

# Supplementary Experimental Procedures

## Acronyms

| | |
|---|---|
| AEO | Annual Energy Outlook |
| ASHP | Air-Source Heat Pump |
| ATB | Annual Technology Baseline |
| CCGT | Combined Cycle Gas Turbine |
| CCS | Carbon Capture and Storage |
| CCU | Carbon Capture and Utilization |
| CES | Clean Energy Standard |
| CLIU | County Level Industrial Use Dataset |
| CONUS | Contiguous United States |
| COP | Coefficient of Performance |
| CO2e | Carbon Dioxide Equivalent |
| DAC | Direct Air Capture |
| DOE | Department of Energy |
| EE | Elementary Effect |
| EIA | Energy Information Agency |
| EPA | Environmental Protection Agency |
| EPRI | Electric Power Resource Institute |
| ESOM | Energy System Optimization Model |
| EULP | End-Use Load Profile Dataset |
| EV | Electric Vehicle |
| GSA | Global Sensitivity Analysis |
| GSHP | Ground-Source Heat Pump |
| GWP | Global Warming Potential |
| ISO | Independent System Operator |
| MECS | Manufacturing Energy Consumption Survey |
| NAICS | North American Industry Classification System |
| NERC | North American Electric Reliability Corporation |
| NLR | National Laboratory of the Rockies |
| OCGT | Open Cycle Gas Turbine |
| PUDL | Public Utility Data Liberation |
| PyPSA | Python for Power System Analysis |
| REC | Renewable Energy Certificate |
| RECS | Residential Energy Consumption Survey |
| ReEDS | Regional Energy Deployment System |
| RES | Reference Energy System |
| RPS | Renewable Portfolio Standard |
| SEE | Scaled Elementary Effect |
| TCT | Technology Capacity Targets |
| USA | United States of America |
| VMT | Vehicle Miles Travelled |
| VRE | Variable Renewable Energy |

## S.1 Model Overview

Python for Power System Analysis (PyPSA) is a reputable open-source Energy System Optimization Modelling (ESOM) framework to perform capacity expansion and economic dispatch studies [1], [2]. PyPSA-USA is a highly configurable ESOM of the United States built on top of the PyPSA platform [3]. This work extends the existing power sector representation of PyPSA-USA [3] to include multiple energy sectors and energy carriers.

ESOMs are prescriptive analytic tools to help researchers study the impacts of energy policies given different futures. PyPSA-USA can model a variety of policies under different spatial, temporal, and technological assumptions. This flexibility allows PyPSA-USA to be repurposed for many studies; however, it also requires the modeler to limit the model scope to maintain tractability. This section describes the key assumptions made for this study.

### S1.1 What is Modeled

In this study, we model the power sector, natural gas sector, service sector, industrial sector, and transport sector for a 2030 future energy system. A high-level overview of the interconnections between the sectors and fuels are shown in Figure 1. Sections S.2 Power Sector through S.6 Transport Sector describe each sector in detail.

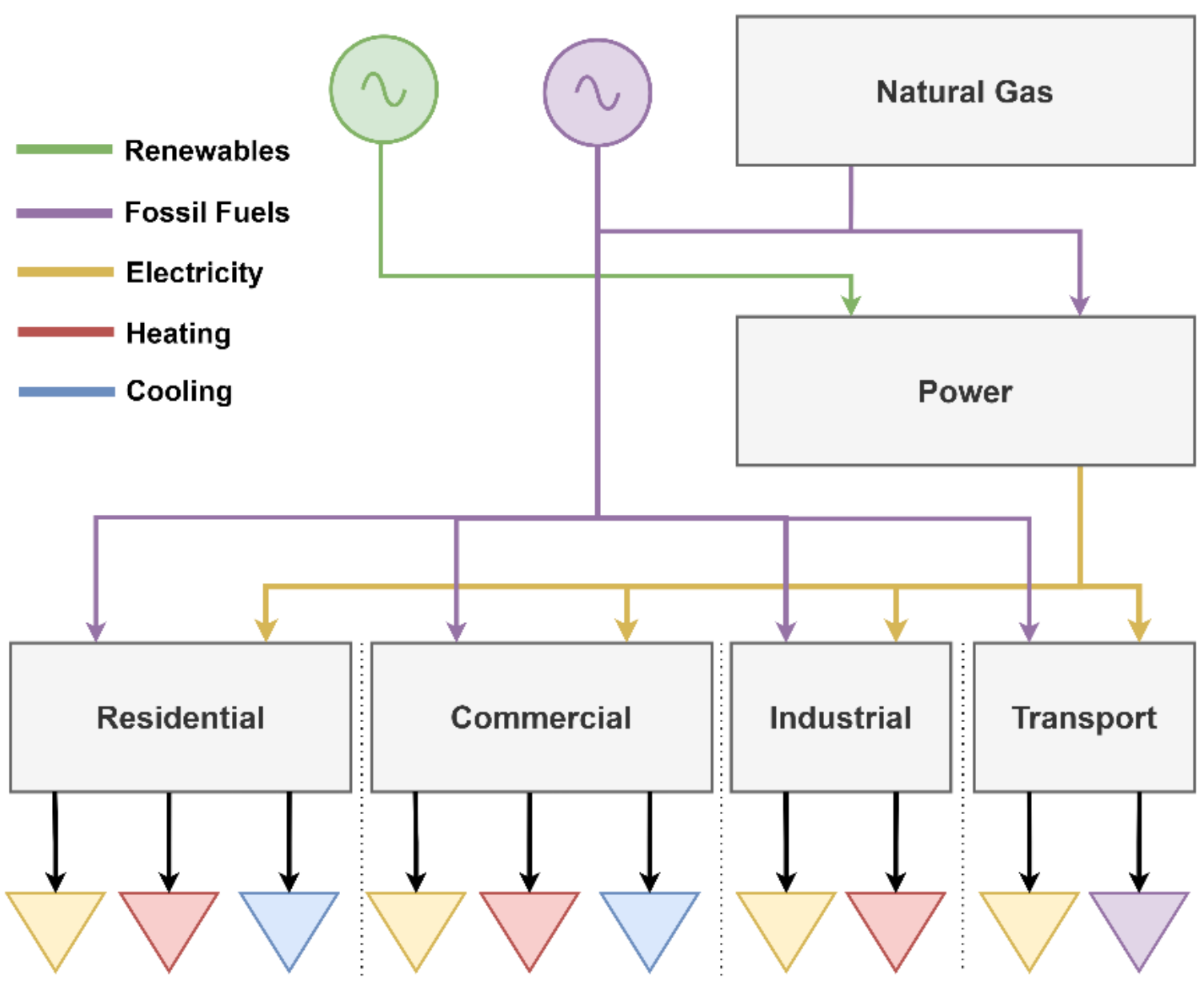

*Figure 1: High level reference energy system of PyPSA-USA Sector*

### S1.2 What is Not Modeled

The core research question in this study explores what uncertainties drive emissions and marginal costs in the USA in the near-term. As such, we restrict the solution space to eliminate decarbonization solutions that will be impactful after the 2030 horizon. These technologies will (likely) be impactful in long term timelines; however, they are yet to be commercialized on a scale to offer significant impact by 2030. Additionally, we do not model on-site generation technologies (for example, rooftop solar, service level batteries, and on-site industrial diesel generators). This is due to data restrictions that are elaborated in Sections S.2 through S.6.

The key technologies that we do not model, due to minimal commercialization potential by 2030, are:

- **Hydrogen**: Significant research shows that hydrogen will likely be a key solution for hard-to-decarbonize sectors [4]. However, over the 2030 timeline, long equipment lead times and infrastructure requirements position hydrogen as a 2035+ solution by the Department of Energy (DOE) [5][a].
- **Industrial Decarbonization:** Industrial electrification, and energy efficiency improvements are included in our model (see Section S.5), however, deep decarbonization solutions are not. These include solutions like low-carbon feedstock replacement (such as hydrogen or ammonia), capturing carbon from industrial processes, waste heat recovery investment, or alternative production methods. These are long term solutions, not near term.
- **Heavy Duty Transportation Decarbonization:** Road transportation can be electrified – including any of light-duty, medium-duty, heavy-duty and bus transport modes. However, air-travel, marine shipping, and rail transport demand cannot be met through any other means than fossil fuels due to timeframe restrictions.
- **Carbon Removal Technologies:** Reaching net-zero will likely require carbon removal technologies – such as Direct Air Capture (DAC) and Carbon Capture and Utilization (CCU) processes. However, significant infrastructure buildout needs to occur for this to be commercialized, likely occurring post-2030. We do model Cabon Capture and Storage (CCS) technologies in the power sector.

### S1.3 Spatial Resolution

Our spatial scope contains the Contiguous United States (CONUS), which includes the lower 48 states. We model each state individually to manage tractability. There are a total of 135 load zones modelled across the states – this spatial resolution matches the well-used National Laboratory of the Rockies (NLR) Regional Energy Deployment System (REeDS) model [6], [7]. Each state has between 1 and 8 load zones. The number of renewable potential zones is three times the number of load zones. For example, a state with two load zones will have 6 renewable potential zones. This allows us capture weather variability on the supply side and roughly matches the 356 resource zones in the REeDS model. Renewable potential zones are clustered following a k-medoids algorithm on the weather time-series data, with more details provided in the PyPSA-USA seminal paper [3]. The limitations of state level representation are expanded upon in the main manuscript.

### S1.4 Temporal Resolution

We model a future 2030 energy system of the USA. PyPSA-USA can model up to hourly resolution for the full year (ie. 8760 timesteps). To maintain tractability, we model at 3hr resolution. Temporal aggregation is done by averaging sequential timesteps (ie. all timesteps are weighted equally). The limitations of 3hr resolution are expanded upon in the main manuscript.

[a] Note that the Department of Energy has removed the "Pathway to Commercial Liftoff" reports from their domain. However, they have been archived independently.

### S1.5 Key Constraint Equations

The reader is referred to PyPSA's documentation site [8] for descriptions on all constraints in PyPSA. Here we describe the key constraints used in this study. Moreover, any constraints added represent individual sectors are described in Sections S.2 through S.6.

#### Variables

All variable definitions are given in Table 1, with a simple schematic of each bus in the network given in Figure 2.

*Table 1: List of Sets and Variables used in PyPSA-USA*

| Label | Description |
|---|---|
| $n \in N = \{0, 1, \ldots, N\}$ | Label of Bus, $n$ |
| $t \in T = \{0, 1, \ldots, T\}$ | Label of Snapshot, $t$ |
| $s \in S = \{0, 1, \ldots, S\}$ | Label of Component, $s$ |
| $l \in L = \{0,1, \ldots, L\}$ | Label of link, $l$ |
| $\omega_t$ | Weighting of time $t$ in the objective equation |
| $g_{n,s,t}$ | Dispatch of generator $s$ at bus $n$ at time $t$ |
| $G_{n,s}$ | Power capacity of component $s$ at bus $n$ |
| $\overline{g}_{n,s,t}$ | Upper bound availability of generator $s$ at bus $n$ at time $t$ |
| $\underline{g}_{n,s,t}$ | Lower bound availability of generator $s$ at bus $n$ at time $t$ |
| $\overline{G}_{n,s}$ | Upper bound power capacity of generator $s$ at bus $n$ |
| $\underline{G}_{n,s}$ | Lower bound power capacity of generator $s$ at bus $n$ |
| $h_{n,s,t}$ | Dispatch of store $s$ at bus $n$ at time $t$ |
| $H_{n,s}$ | Power capacity of store $s$ at bus $n$ |
| $e_{n,s,t}$ | Energy level of store $s$ at bus $n$ at time $t$ (ie. State of charge) |
| $\overline{e}_{n,s}$ | Maximum energy level of store $s$ at bus $n$ |
| $\underline{e}_{n,s}$ | Minimum energy level of store $s$ at bus $n$ |
| $f_{n,s,t}$ | Dispatch of link $s$ at bus $n$ at time $t$ |
| $F_{n,s}$ | Power capacity of link $s$ at bus $n$ |
| $\bar{f}_{n,s,t}$ | Availability of link $s$ at bus $n$ at time $t$ |
| $ri_{n,s}$ | Allowable increase in output for generator $s$ at bus $n$ |
| $rd_{n,s}$ | Allowable decrease in output for generator $s$ at bus $n$ |
| $\eta_{n,s}$ | Efficiency of component $s$ at bus $n$ |
| $d_{n,t}$ | Demand at bus $n$ at time $t$ |
| $c_{n,s}$ | Capital cost of extending component $s$ at bus $n$ by one unit of capacity |
| $o_{n,s}$ | Marginal cost to dispatch on unit of energy of component $s$ at bus $n$ |

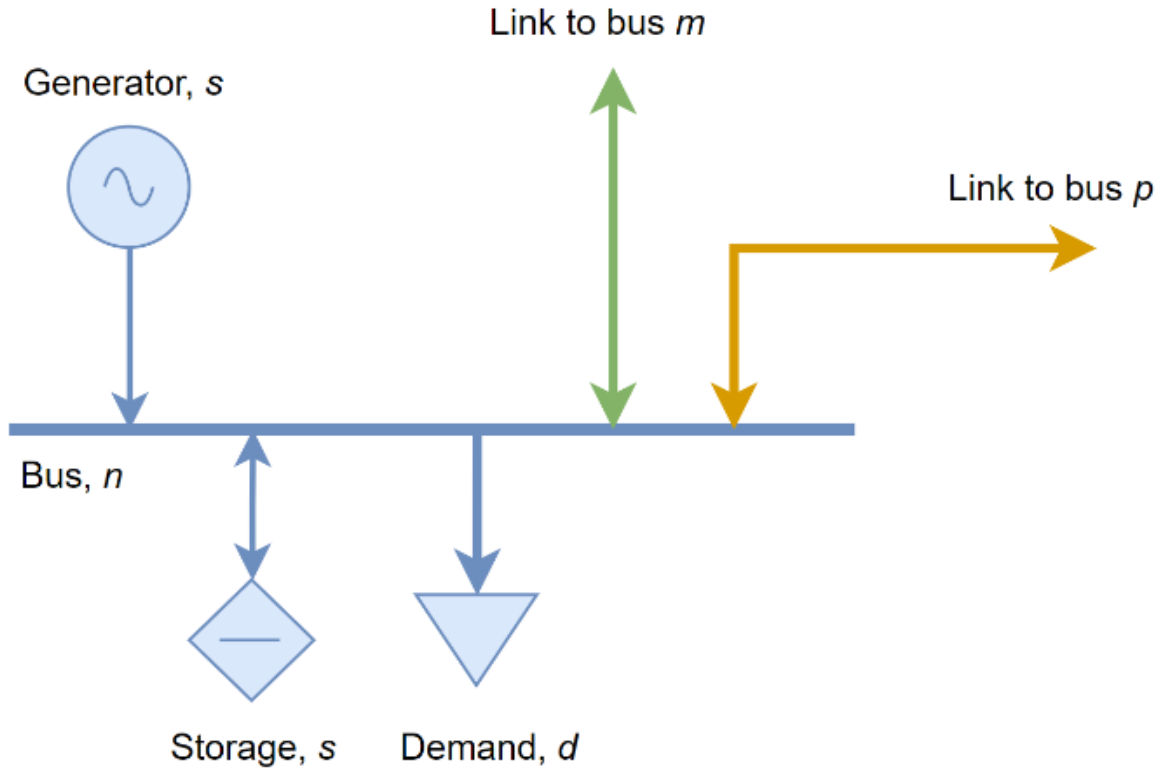

*Figure 2: Simple representation of energy flow at each bus*

**Objective Equation**

The objective function, Equation (1), states that the objective is to minimize the summed capital costs and operational costs for all components.

$$min\left(\sum_{n,s} c_{n,s} \cdot G_{n,s} + \sum_{n,s} c_{n,s} \cdot H_{n,s} + \sum_{l} c_{n,s} \cdot F_{n,s} + \sum_{t} \omega_t \left[\sum_{n,s} o_{n,s,t} \cdot g_{n,s,t} + \sum_{n,s} o_{n,s,t} \cdot h_{n,s,t} + \sum_{n,s} o_{n,s,t} \cdot f_{n,s,t}\right]\right) \quad (1)$$

**Energy Balance**

Every node in the network must satisfy the energy balance constraint, described in Equation (2). This equation guarantees that demand is met at every bus $n$ and snapshot $t$.

$$g_{n,s,t} + h_{n,s,t} + f_{n,s,t} = d_{n,s,t} \qquad \forall_{n,s,t} \quad (2)$$

**Dispatch Constraints**

The dispatch of each generator, storage, and link at time $t$ and bus $n$ must respect available capacity constraints. Using a generator as an example, Equation (3) states that the dispatch of generator $g_{n,s,t}$ must be between the installed capacity, $\mathrm{G}_{\mathrm{n,s}}$, multiplied by its availability profile. Ramping of generators is governed by Equation (4).

$$\underline{g}_{n,s,t} \cdot G_{n,s} \leq g_{n,s,t} \cdot \frac{1}{\eta_{n,s}} \leq \overline{g}_{n,s,t} \cdot G_{n,s} \qquad \forall_{n,s,t} \quad (3)$$

$$-rd_{n,s} \cdot \overline{g}_{n,s} \leq \left(g_{n,s,t} - g_{n,s,(t-1)}\right) \leq ri_{n,s} \cdot \overline{g}_{n,s} \qquad \forall_{n,s,t} \quad (4)$$

**Capacity Constraints**

The installed capacity must abide by maximum and minimum requirements, given in Equation (5). For example, wind and solar power are constrained by land use, while other technologies may have build rates.

$$\underline{G}_{n,s} \leq G_{n,s} \leq \overline{G}_{n,s} \qquad \forall_{n,s} \quad (5)$$

**Storage Constraints**

The energy level connected to the dispatch variables through Equation (6). Standing losses are given by $\eta_{n,s,t}^{\omega_t}$, while charging and discharging losses are given by $\eta_{n,s,t}^{+}$ and $\eta_{n,s,t}^{-}$ respectively.

$$e_{n,s,t} = e_{n,s,t-1} \cdot \eta_{n,s,t}^{\omega_t} + \omega_t \left( h_{n,s,t}^{+} \cdot \eta_{n,s}^{+} - h_{n,s,t}^{-} \frac{1}{\eta_{n,s}^{-}} \right) \qquad \forall_{n,s,t} \tag{6}$$

The storage inflow and outflow must respect charging and discharging limits in Equation (7), while the state of charge must respect minimum and maximum charge levels in Equation (8).

$$\underline{h}_{n,s,t} \cdot H_{n,s} \leq h_{n,s,t} \cdot \frac{1}{\eta_{n,s}} \leq \overline{h}_{n,s,t} \cdot H_{n,s} \qquad \forall_{n,s,t} \tag{7}$$

$$\underline{e}_{n,s} \leq e_{n,s,t} \leq \overline{e}_{n,s} \qquad \forall_{n,s,t} \tag{8}$$

**Policy Constraints**

Three main policy constraints are implemented in the model: a renewable portfolio standard (RPS) constraint[b], a Renewable Energy Certificates (RECs) system, and Technology Capacity Target (TCT) constraints. The RPS constraint given in equation (9) is applied to the power sector and specifies the minimum fraction of energy that must come from renewable (or clean) energy sources. This can be applied to an individual zone, groups of zones, or globally. The RPS value to apply is taken from the Lawrence Berkley National Lab [9]. In equation (9), let the regions to apply the policy constraint to be $Z \subseteq N$. Let the renewable power generator be $R = \{r \in S \mid s \text{ is renewable power generator}\}$, and all power generators be $A = \{a \in S \mid s \text{ is power generator}\}$.

$$\frac{\sum_{r,t}(\omega_t \cdot g_{z,r,t})}{\sum_{a,t}(\omega_t \cdot g_{z,a,t})} \geq RPS_z \qquad \forall_z \tag{9}$$

RECs are implemented following equation (10). In general, it states that the generation from renewables across the REC trading zone must be greater than the state level generation RPS targets. The REC trading zone is taken following the USA Environmental Protection Agency (EPA) definitions [10]. Let the REC trading zone be $U \subseteq N$.

$$\sum_{r,t}(\omega_t \cdot g_{u,r,t}) - \sum_{r,t}(\omega_t \cdot g_{z,r,t}) \times RPS_z \geq 0 \qquad \forall_z \tag{10}$$

In this study each state is run individually. As such, an additional constraint to capture REC pricing behavior is added. Each state connects to a neighboring state via capacity constrained electricity transmission lines (discussed later). An additional "dummy line" is run in parallel with any importing line within the REC Zone, $U$. Equation (11) states that the price of importing electricity on this "dummy line" is the REC value added on top of the nominal import value. Equation (12) states that the summed imports of the real line and dummy line must not exceed the capacity of the real line at any snapshot. The model will choose to use the more expensive imports to buy RECs. This is graphically given in Figure 3.

[b] Both renewable portfolio standards and clean energy targets are implemented. They follow the same constraint definition.

$$c_{dummy} = c_{real} + REC_u \qquad \forall_z \tag{11}$$

$$F_{n,s} \geq f_{n,s_{real},t} + f_{n,s_{dummy},t} \qquad \forall_{n,s,t} \tag{12}$$

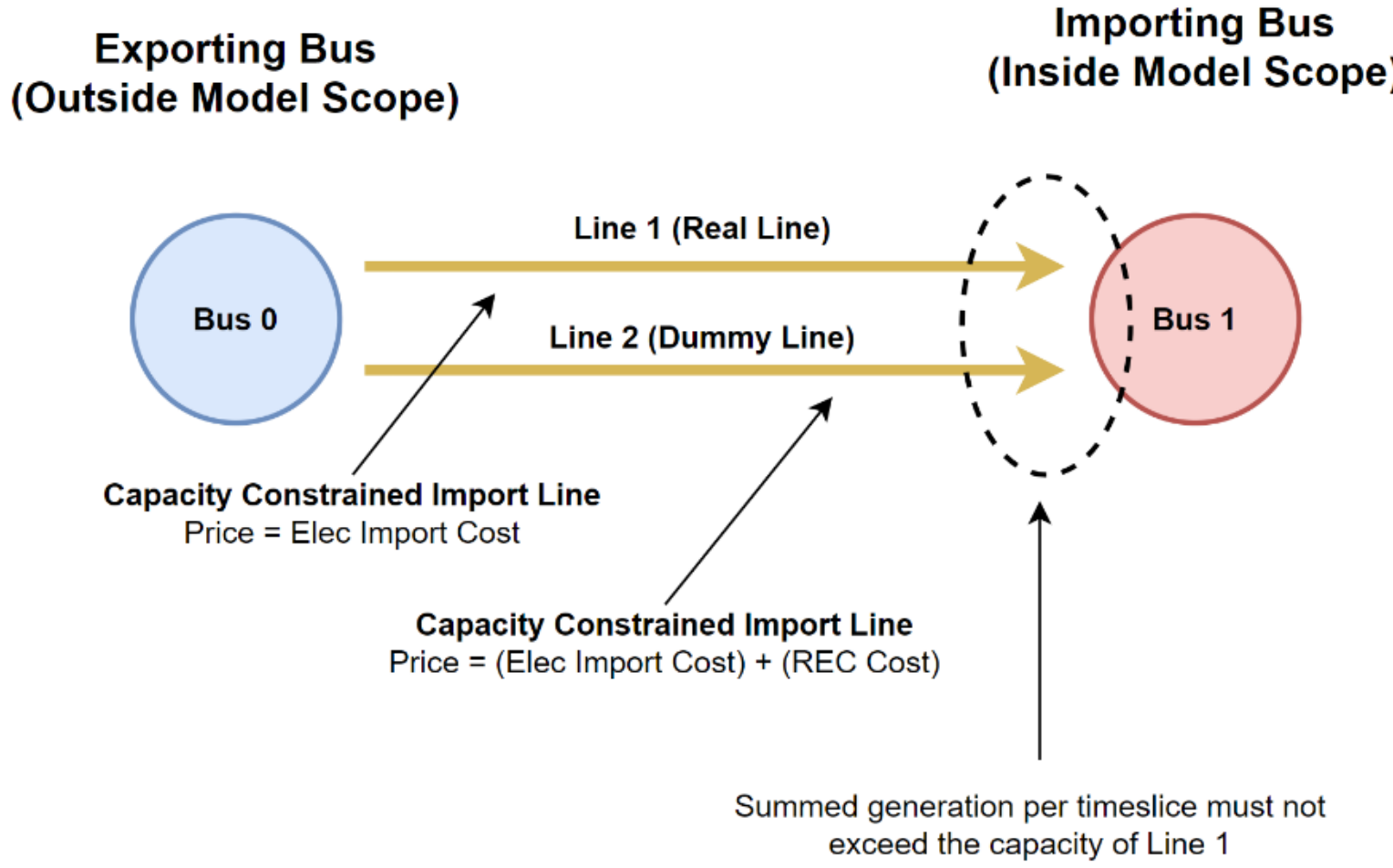

*Figure 3: Schematic showing how renewable energy certificates are modelled*

Finally, groups of technologies can have TCTs to restrict or require total installation values, described in equation (13). This is done to enforce build limits or to force buildouts of groups of technologies. Let the technologies to apply the TCT to be $B \subseteq S$.

$$TechLimit_{MIN_{z,b}} \leq \sum_{b} (G_{z,b} + H_{z,b} + F_{z,b}) \leq TechLimit_{MAX_{z,b}} \qquad \forall_{z,b} \tag{13}$$

### S.1.6 Cost Assumptions

Each component in PyPSA-USA is subject to an overnight capital cost, a fixed operation cost, and a fuel cost. Fuel costs are discussed below in their respective sections. Capital costs of new technologies are discounted to present value following equation (14), where $occ$ is the Overnight Capital Cost, $fom$ is the Fixed Operation and Maintenance cost (given in percentage of $occ$ per year), and $a$ is the annuity given in Equation (15), where $r$ is the discount rate and $n$ is the lifetime. Each power generation technology has a unique discount rate, while all other sectors have a single discount rate that is applied to all technologies within that sector.

$$capex = occ \cdot (a + fom) \tag{14}$$

$$a = \frac{1 - (1 + r)^{-n}}{r} \tag{15}$$

Utility scale powerplants can be economically retired before the end of their technical lifetime. The salvaged value of these technologies is discounted based on their start year, retirement year, and discount rate.

## S.2 Power Sector

In this section we give a high-level overview of the power sector representation. The reader is referred to the original PyPSA-USA paper [3] which exclusively details the power sector.

### S2.1 Demand

The power sector electrical demand is endogenously determined by the energy demands in the service, transport, and industrial sectors. There is no direct end-use electrical load attached to the power sector.

### S2.2 Technologies

The power generation technologies available in PyPSA-USA can broadly be broken into dispatchable generators, and Variable Renewable Energy (VRE) generators. The transmission system is discussed separately below.

#### Generators

PyPSA-USA ingests unit-level generator data and are then clustered by technology type. Fossil based generators are clustered to Combined-Cycle Gas Turbines (CCGT), Open-Cycle Gas Turbines (OCGT), CCGTs with Carbon Capture and Storage (CCS), coal turbines, oil generators, and waste powerplants. Clean dispatchable generators include large scale nuclear, biomass, and hydroelectric facilities. VRE generators include photovoltaic solar, onshore wind, and offshore wind.

Full details on all techno-economic parameters included in the power sector can be found in the original PyPSA-USA paper [3]. Solar and wind technology maximum capacities are limited based on resource potentials that consider land usage. Nameplate capacities, heat rates, and seasonal deratings are incorporated from the Energy Information Agency [11] (EIA) via the Public Utility Data Liberation [12] (PUDL), while financial parameters are taken from the NLR Annual Technology Baseline [13] (ATB). Capacity factor profiles for VRE sources are calculated via Atlite [14] using ERA5 data [15].

#### Electrical Transmission

We implement a zonal transmission network that is based on the NLR ReEDS model [6], [7]. This results in 134 unique load zones distributed throughout CONUS. Each zone is connected to adjacent regions through capacity constrained transmission lines. We model electrical trade as a transport model, but implement pre-calculated interface transmission limits [16] to better represent capacity limits. The original PyPSA-USA paper extensively discusses electrical transmission [3].

#### Demand Response

We use PyPSA-USA's price-based demand response system for electrical load [17]. Demand can be shifted, either forward or backwards, and stored for any period of time. For each hour energy is stored, a cost is incurred to the system at a per-unit level (ie. $/MWh). The graphical representation of this is given in Figure 4.

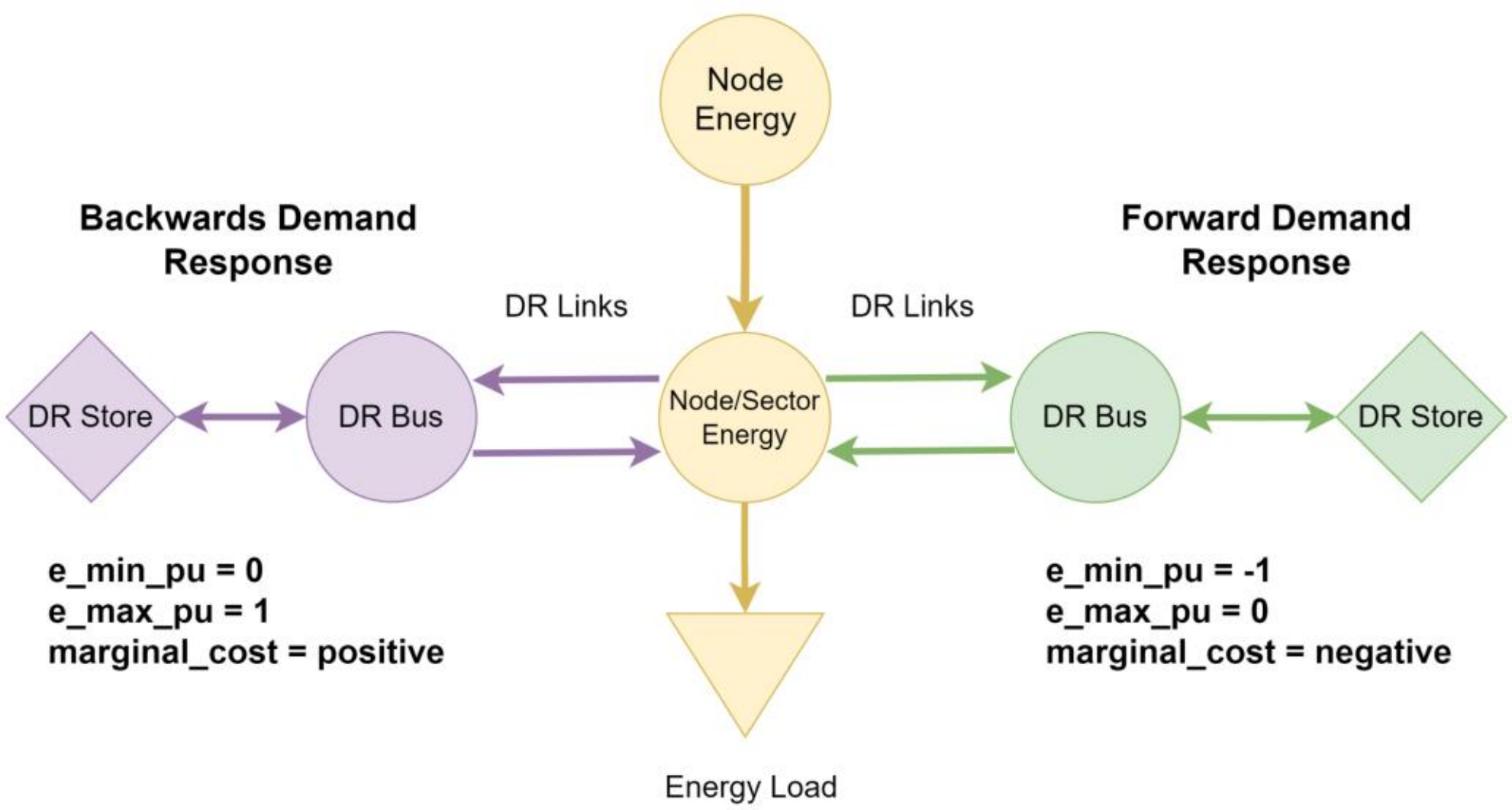

*Figure 4: Demand Response Implementation in PyPSA-USA*

## S.3 Natural Gas Sector

This section describes the components included in the natural gas sector.

### S3.1 Reference Energy System

Given in Figure 5 is the flow of energy in the natural gas sector. Natural gas is tracked at a state level due to data restrictions on more resolved data. The figure shows all technologies attached to the state-level accounting bus. Specifically, there are four components: pipelines attaching to neighboring states, production facilities to inject natural gas into the grid, underground storage facilities, and line pack storage facilities. Data sources for each component are included in Table 2.

*Table 2: Natural Gas Data Sources*

| Data | Source | Assumptions |
|---|---|---|
| State-to-State Pipeline Capacity | [11] | All currently installed pipelines will not be retired before 2050. |
| State Level Transmission Pipeline Volume | [18] archived at [19] | Includes both interstate and intrastate pipelines. Pipeline operating pressures. Pipeline diameters. |
| State Level Gas Processing | [11] | Marketed dry natural gas production. |
| State Level Underground Storage | [11] | Starting level state of charge. Includes salt cavers, depleted fields, and aquifers. Not extendable. |
| State Level Domestic Imports/Exports | [11] | |
| State Level International Imports/Exports | [11] | A state can only trade with one other province. If the state is connected to two provinces, the larger capacity connection is taken as the connection. |
| Gas Extraction Marginal Costs | [20] | Costs do not depend on well type |

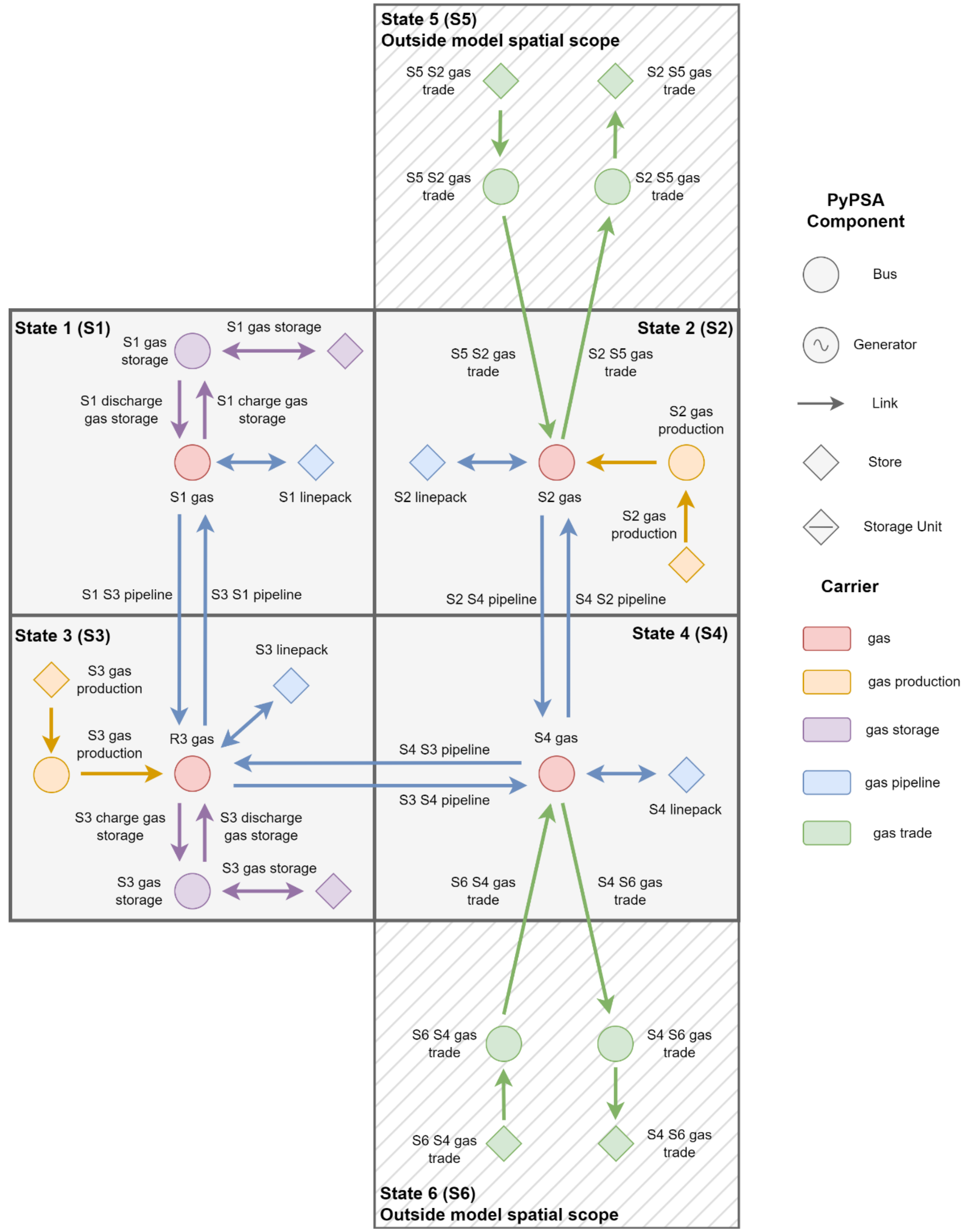

*Figure 5: Natural Gas Representation in PyPSA-USA*

Pipelines are modelled as links with linear efficiencies. As this study investigates near-term emission targets, it is assumed that natural gas pipeline expansion is not allowed. Underground storage facilities are aggregated to state level and respect minimum and maximum operating conditions. Both storage options (underground and line pack) operate to start and end the modelling period with the same storage level.

Line pack volumes ($V$) are calculated following equation (16), where the pipe pressure ($P$) and radius ($r$) are estimated [21], and pipeline lengths ($L$) are extracted from geolocated data [18], [19]. The internal energy ($E$) is calculated following the ideal gas law in equation (17), where $R$ is the universal gas constant $\left(8.314 \frac{J}{mol \cdot K}\right)$ and $C_v$ is the heat capacity at a constant value $\left(35.7 \frac{J}{mol \cdot K}\right)$ for methane.

$$V = L\pi r^2 \tag{16}$$

$$E = nC_vT = \frac{PV}{RT}C_vT = \frac{PV}{R}C_v \tag{17}$$

Pipe flows between neighboring states within the model scope are endogenously solved for. Pipe flows to neighboring regions outside the model scope (for example, between Washington and British Columbia) are modelled as an exogenous boundary condition. The imports/exports in this pipeline are constrained to meet the historical flows, balanced each month, following Equations (18) through (21); where $B = \{b \in S \mid s \text{ } is \text{ } a \text{ } natural \text{ } gas \text{ } pipeline\}$, and where $X$ and $Y$ are neighboring states, and $M$ is a user configurable multiplier.

$$\sum_t p_{b(X \to Y)} \leq (Monthly\ Historacal\ Flow_{X \to Y}) \cdot m_1 \tag{18}$$

$$\sum_t p_{b(X \to Y)} \geq (Monthly\ Historacal\ Flow_{X \to Y}) \cdot m_2 \tag{19}$$

$$\sum_t p_{b(Y \to X)} \leq (Monthly\ Historacal\ Flow_{Y \to X}) \cdot m_3 \tag{20}$$

$$\sum_t p_{b(Y \to X)} \geq (Monthly\ Historacal\ Flow_{Y \to X}) \cdot m_4 \tag{21}$$

Upstream and downstream methane leakage tracking is supported, as given in Figure 6. For every unit of natural gas injected into the system (upstream) or used by a technology (downstream), a user configurable value is specified to release methane. A Global Warming Potential (GWP) is specified to convert methane into $CO_2$ equivalent ($CO_{2_e}$) following Equation (22). Upstream methane leakage counts against the emission budget of the state where it is produced, while downstream leaks count against the emission budget of the technologies state. Leakage rates with storage and state level transmission are assumed to be zero.

$$CO2_e = GWP \cdot Leakage\ Rate \cdot NG_{production} \tag{22}$$

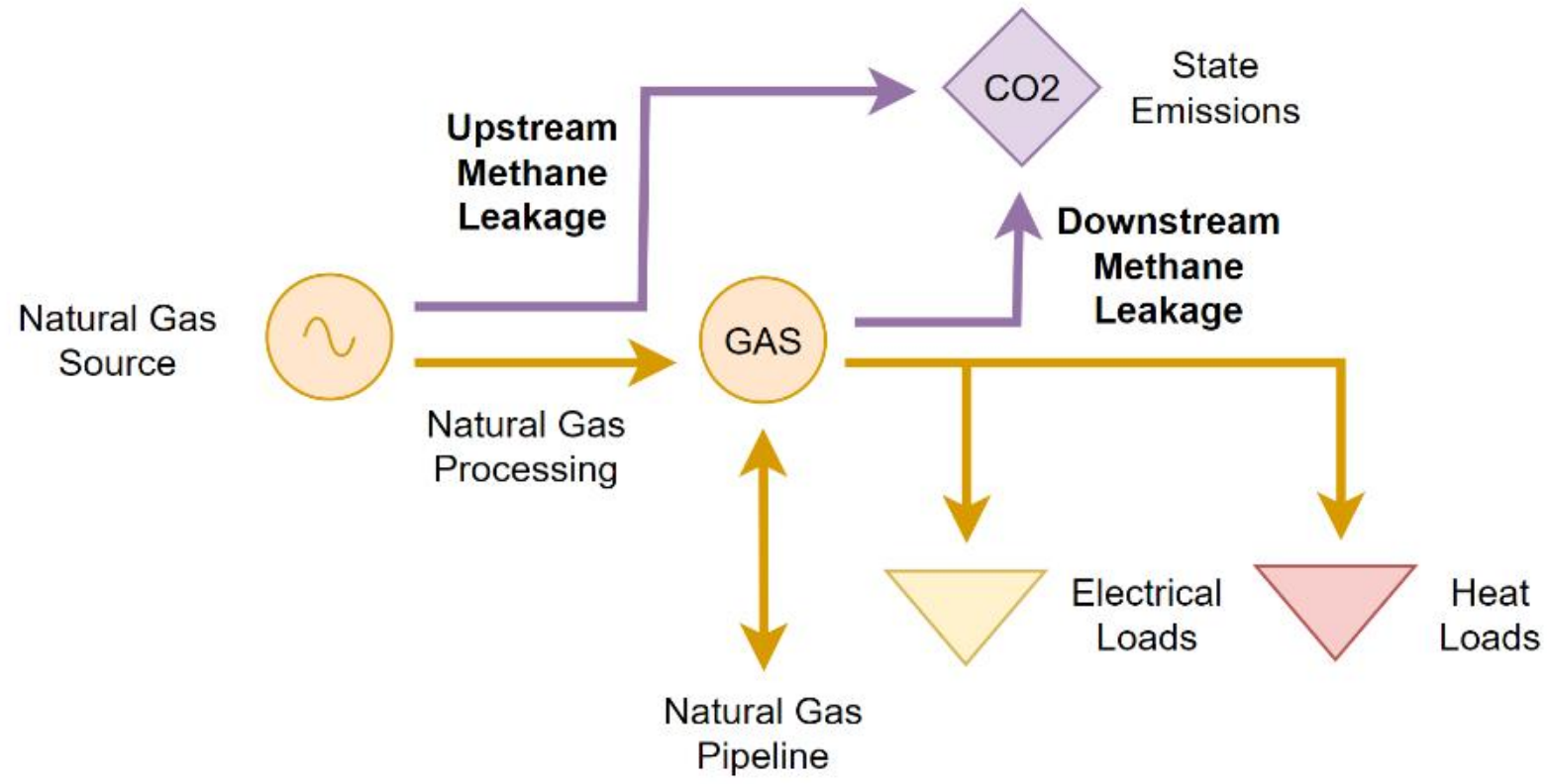

*Figure 6: Methane tracking schematic*

### S3.2 Demand

There is no direct natural gas demand within PyPSA-USA. Rather, natural gas is a primary resource that can be used for power generation, via CCGT and OCGT power plants, can be used as a heating source in the service sector and industrial sector, or can be traded. Its demand is endogenous.

## S.4 Service Sector

The service sector represents both the residential and commercial sectors. Each of these sectors are represented in the same way, but have different load shapes, profiles, and technology characteristics.

### S4.1 Demand

The demand module for the service sector in PyPSA-USA can broadly be described in three steps. End-use demand profile retrieval, demand disaggregation, and demand magnitude projecting. This section will describe these three steps.

#### Demand Profiles

The service sector represents demand for electricity, cooling, space heating, and water heating for the residential and commercial sector separately. These loads are retrieved at a state level from the NLR End-Use Load Profiles (EULP) dataset [22]. Loads are grouped based by end-use fuel and aggregated to hourly values. Heating loads can be modelled to include space-heating and water-heating separately or aggregated together into a single heat load. Figure 7 and Figure 8 show processed state level data for the residential and commercial sector in California.

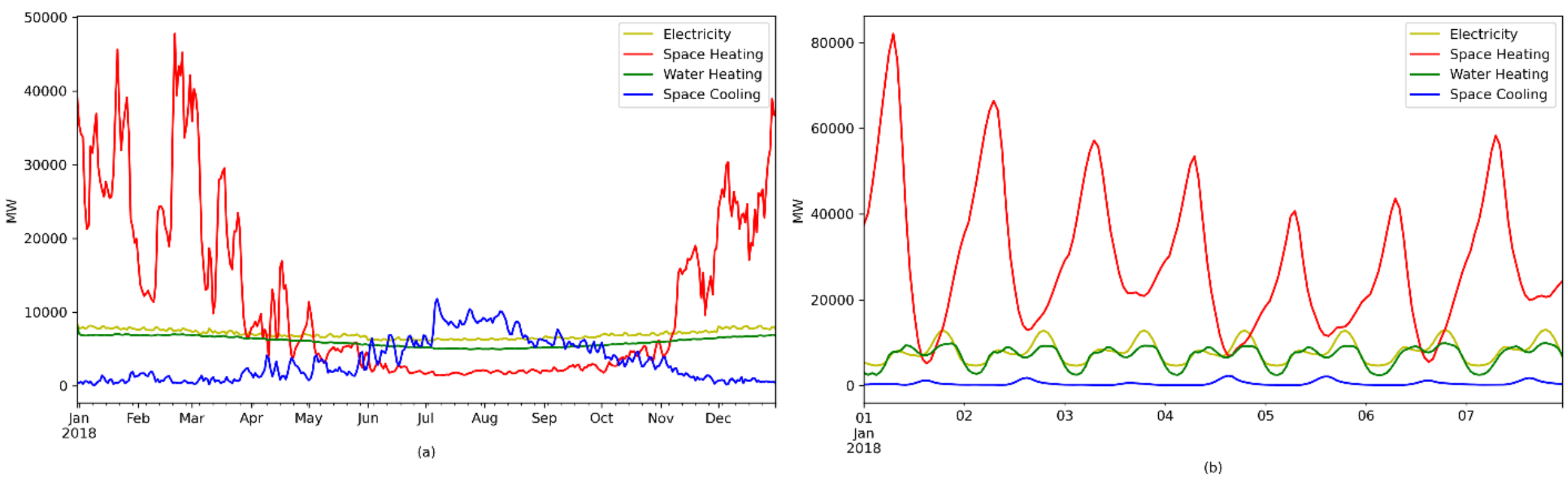

*Figure 7: Residential sector demand in California. (a) Yearly load profile resampled to daily average. (b) Actual hourly profile for the first week of the year read by PyPSA-USA*

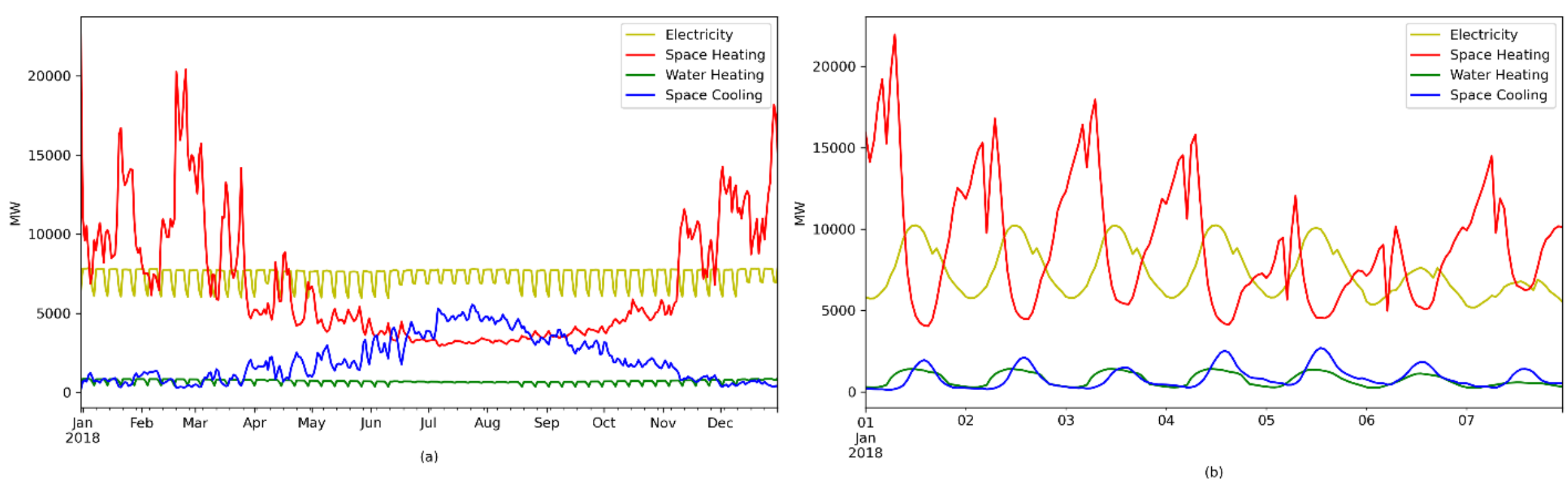

*Figure 8: Commercial sector demand in California. (a) Yearly load profile resampled to daily average. (b) Actual hourly profile for the first week of the year read by PyPSA-USA*

### Demand Disaggregation

The EULP dataset [22] provides state-level aggregated data. This load must be allocated to clustered regions, which are generated based on user configuration options. To do this, the state level load is disaggregated based on population metrics [23]. Figure 9 shows a sample of how each of the four end-use residential demands are allocated for a 4-zone network in California.

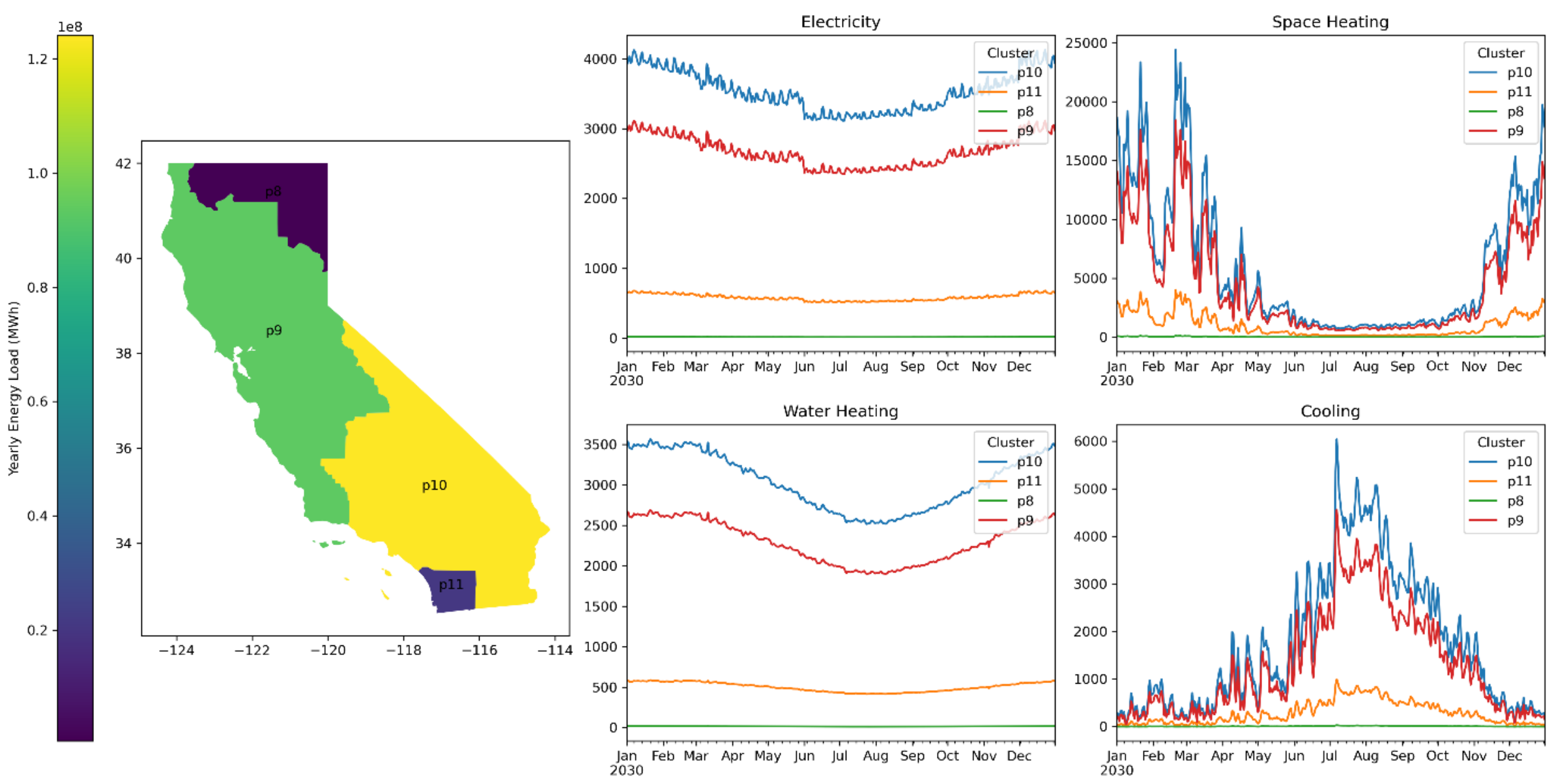

*Figure 9: Sample clustered residential sector demand in California*

## Demand Forecasting

The EULP dataset [22] is a synthetic dataset tied to the 2018 weather year and demand. For future years, this load is forecasted while retaining the same profile. The EIA Annual Energy Outlook [24] (AEO) provides end-use energy consumption projections by sector. Users can select different AEO scenarios to obtain different demand projections. The residential and commercial are scaled using national values due to state level data not being availability. Figure 10 and Figure 11 show the residential and commercial projections from the AEO for different scenarios.

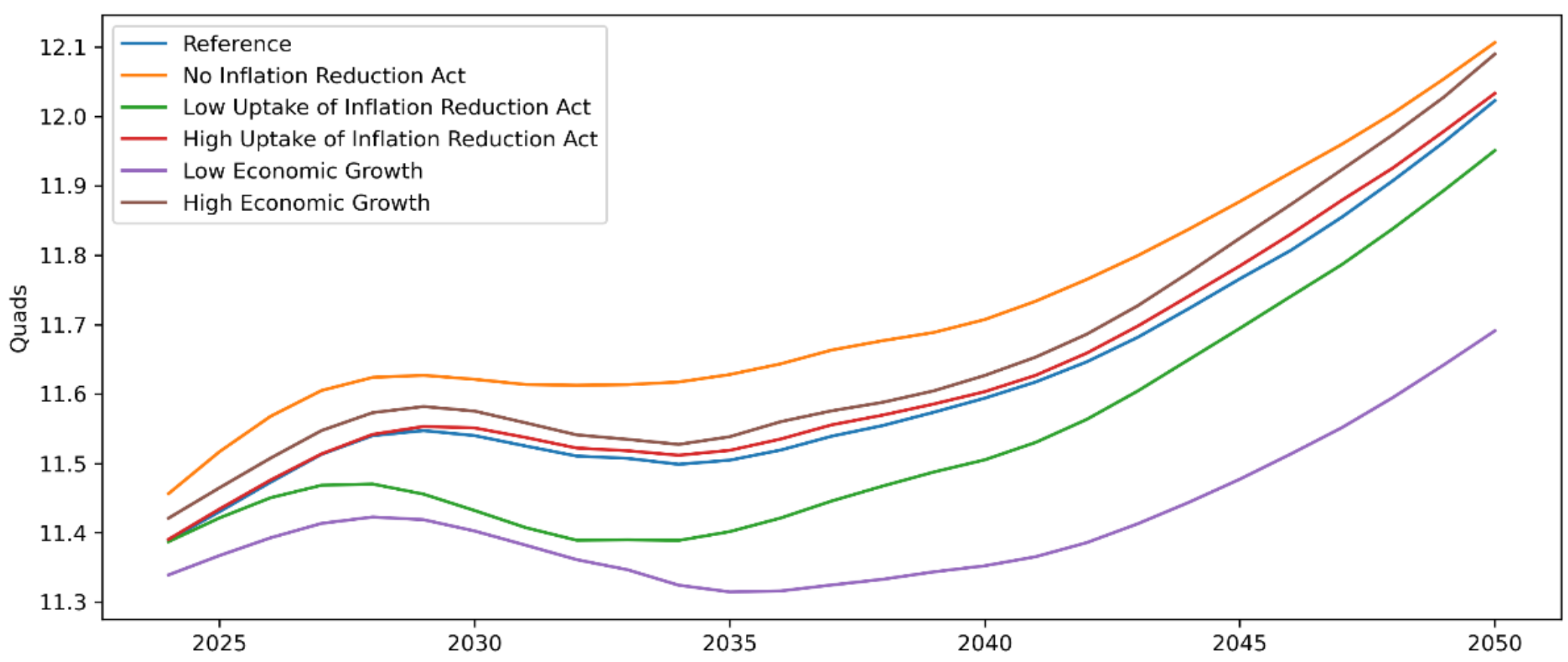

*Figure 10: Annual Energy Outlook residential end-use consumption projections, with select scenarios.*

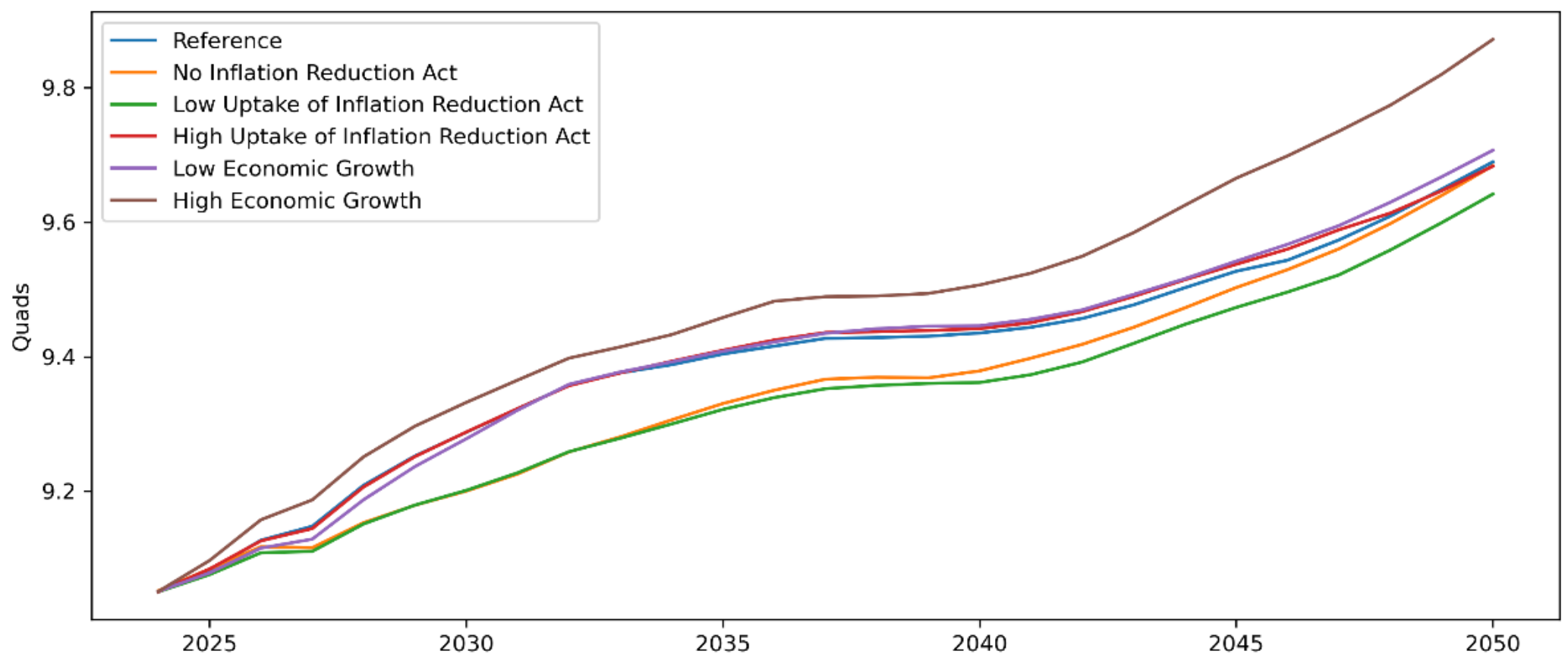

*Figure 11: Annual Energy Outlook commercial end-use consumption projections, with select scenarios.*

## S4.2 Technologies

There are 10 different technologies to meet end-use service sector demand. Base electrical load (which includes existing electrified loads, such as lights and washing machines) are met through electrical distribution. As PyPSA-USA does not ingest distribution level electrical network data, this is modelled as a simple lossy link. Space heat demand can be met through oil, gas, or electrical furnaces. Additionally, heat pumps can be used to meet both space heating and space cooling demand, the latter which can also met through air-conditioners. Finally, water heat demand can be met through oil, gas, or electric hot-water tanks. A Reference Energy System (RES) of the service section is shown in Figure 12.

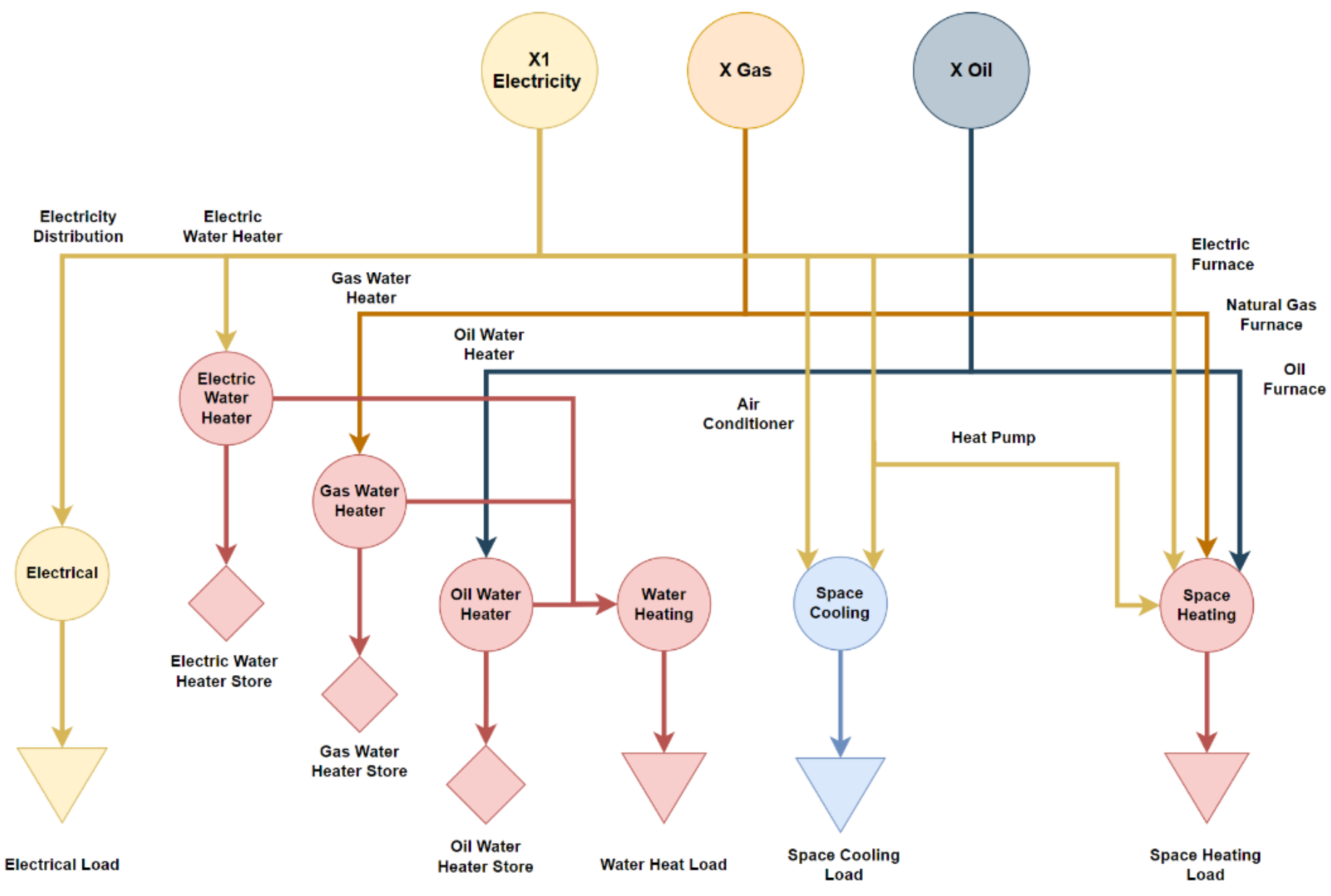

*Figure 12: Reference Energy System of the Service Sector*

All technologies have techno-economic parameters associated with them. These include capital costs, fixed costs, lifetimes, efficiencies, and existing stock values. All data sources are listed in Table 3. Details on heat pump implementation are given in next section. The existing stock values originate from state-level survey data. Wherever state level data was not collected, USA averages were assumed.

*Table 3: Service Sector Data Sources*

| Sector | Technology | Metric | Source |
|---|---|---|---|
| Residential \| Commercial | Furnaces | Capital Cost \| Lifetime \| Efficiency | [25] |
| Residential \| Commercial | Water Heaters | Capital Cost \| Lifetime \| Efficiency | [25] |
| Residential \| Commercial | Air Conditioner | Capital Cost \| Lifetime \| Efficiency | [25] |
| Residential \| Commercial | Heat Pumps | Capital Cost \| Lifetime | [25] |
| Residential \| Commercial | Heat Pumps | Efficiency | [15], [26] |
| Residential | Furnaces | Existing Stock | [27] |
| Residential | Water Heaters | Existing Stock | [27] |
| Residential | Air Conditioner | Existing Stock | [27] |
| Residential | Heat Pumps | Existing Stock | [27] |
| Commercial | Furnaces | Existing Stock | [28] |
| Commercial | Water Heaters | Existing Stock | [28] |
| Commercial | Air Conditioner | Existing Stock | [28] |
| Commercial | Heat Pumps | Existing Stock | [28] |
| Residential \| Commercial | Natural Gas | Fuel Cost | [11] |
| Residential \| Commercial | Heating Oil | Fuel Cost | [11] |
| Residential \| Commercial | - | Demand | [22], [24] |

**Heat Pumps**

Two different types of Heat Pumps (HP) are modelled; Air-Source Heat Pumps (ASHP) and Ground-Source Heat Pumps (GSHP). The HP coefficient of performances (COP) is exogenously calculated using the same weather data used to calculate wind and solar generation [15]. Based on a user configurable set-point, the COP is calculated following equations (23) and (24), originally presented by Staffel et. al. [26]. An example of the profiles is given in Figure 13.

$$COP_{ASHP} = 6.81 - 0.121\Delta T + 0.000630\Delta T^2, \qquad for\ 15 \leq \Delta T \leq 60 \tag{23}$$
$$COP_{GSHP} = 8.77 - 0.160\Delta T + 0.000734\Delta T^2, \qquad for\ 20 \leq \Delta T \leq 60 \tag{24}$$

Where:

$$\Delta T = (T_{sink} - T_{source})$$

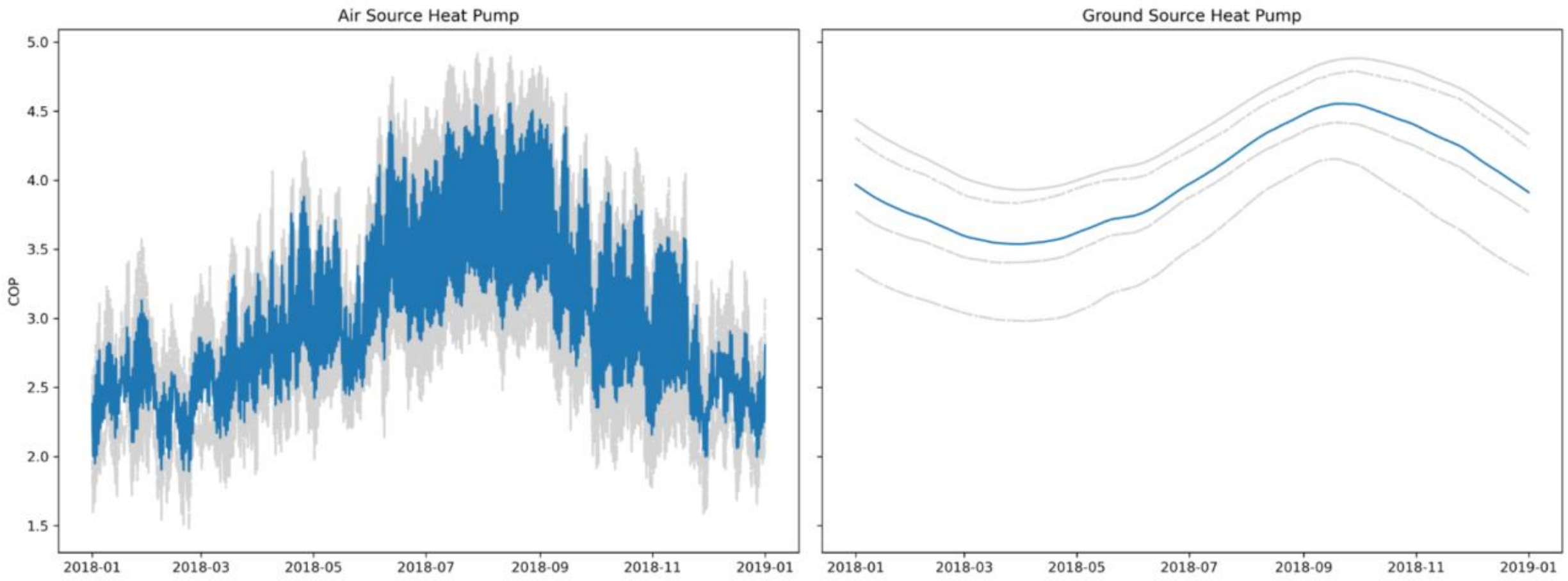

*Figure 13: Example COP profiles for air-source heat pumps (left) and ground source heat pumps (right). Grey lines show the profile for each zone in California (Figure 9), while the blue shows the average.*

We limit the expansion of GSHP as land requirements will restrict deployment in urban areas. We use county level urbanization ratios [23] to apply deployment constraints. Specifically, the ratio of GSHP to ASHP must not exceed the ratio of rural population to urban population for the cluster, given in equation (25). An example of the urbanization data ingested for California is given in Figure 14.

$$ASHP_{capacity} - \frac{urban}{rural} \times GSHP_{capacity} \leq 0 \tag{25}$$

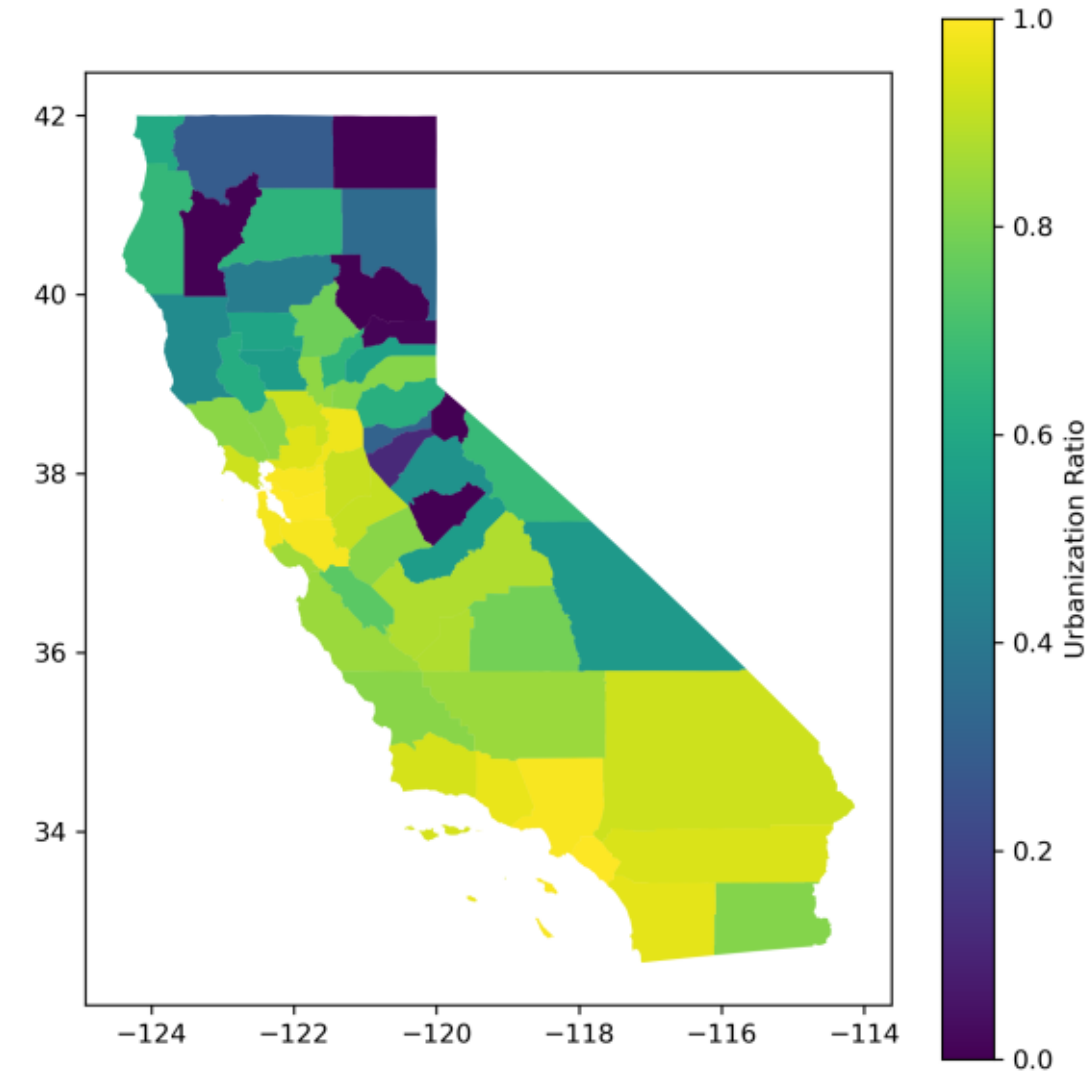

*Figure 14: Urbanization Ratios for California. Extracted from [23],*

## S.5 Industrial Sector

The industrial sector captures the flow of energy associated with all heavy industry. This module aggregates all industry together and does not differentiate between different industry class.

### S5.1 Demand

The industrial demand module takes inspiration from NLR's Demand Side Grid model [29]. End-use demand profiles, yearly demand magnitudes, demand projections, and industrial demand disaggregation strategy are next discussed. Only electrical and process heating loads are tracked in the industrial sector.

**Demand Profiles**

Demand profiles are taken from the Electric Power Resource Institute (EPRI) Load Shape Library [30]. Average profiles for on-peak and off-peak seasons at a North American Electric Reliability Corporation (NERC) region level are extracted. We manually group EPRI profiles into electrical and heating applications; with the average daily profile for each fuel given in Figure 15.

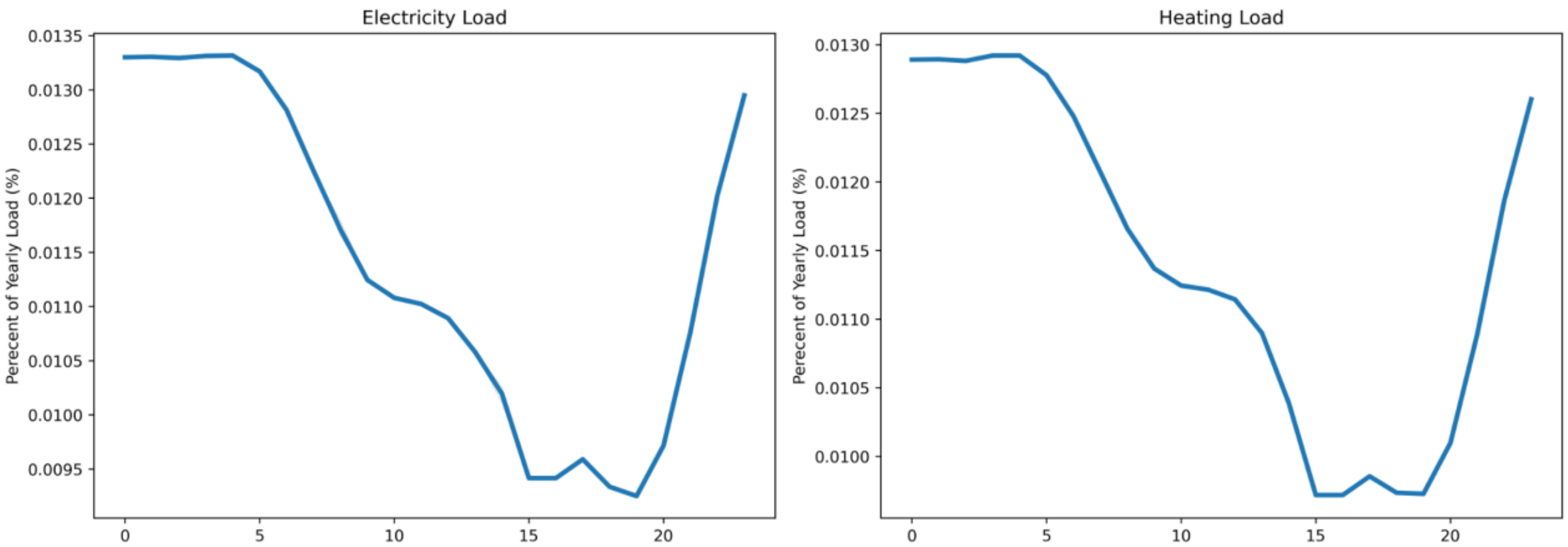

*Figure 15: Average daily industrial load profiles for electrical and heating applications. Data taken from EPRI [30].*

**Demand Projections**

Annual demand magnitudes are taken from NLR County Level Industry Use (CLIU) dataset [31] and scaled according to the EIA Manufacturing Energy Consumption Survey (MECS) [32]. The CLIU dataset gives estimates of total end-use industrial fuel demand by county in 2014. As this dataset is now quite old, and derived from emissions rather than end-use processes, it is scaled against the MECS. The MECS gives regional (9 regions for the USA) reported end-use energy consumption by North American Industry Classification System (NAICS) code. Therefore, the CLIU fuel demand is scaled to match regional energy demand from MECS. This process follows a similar process to pervious NLR studies [33]. Daily load results of overlaying the EPRI demand profiles on the CLIU/MECS data is shown in the figure below Figure 16.

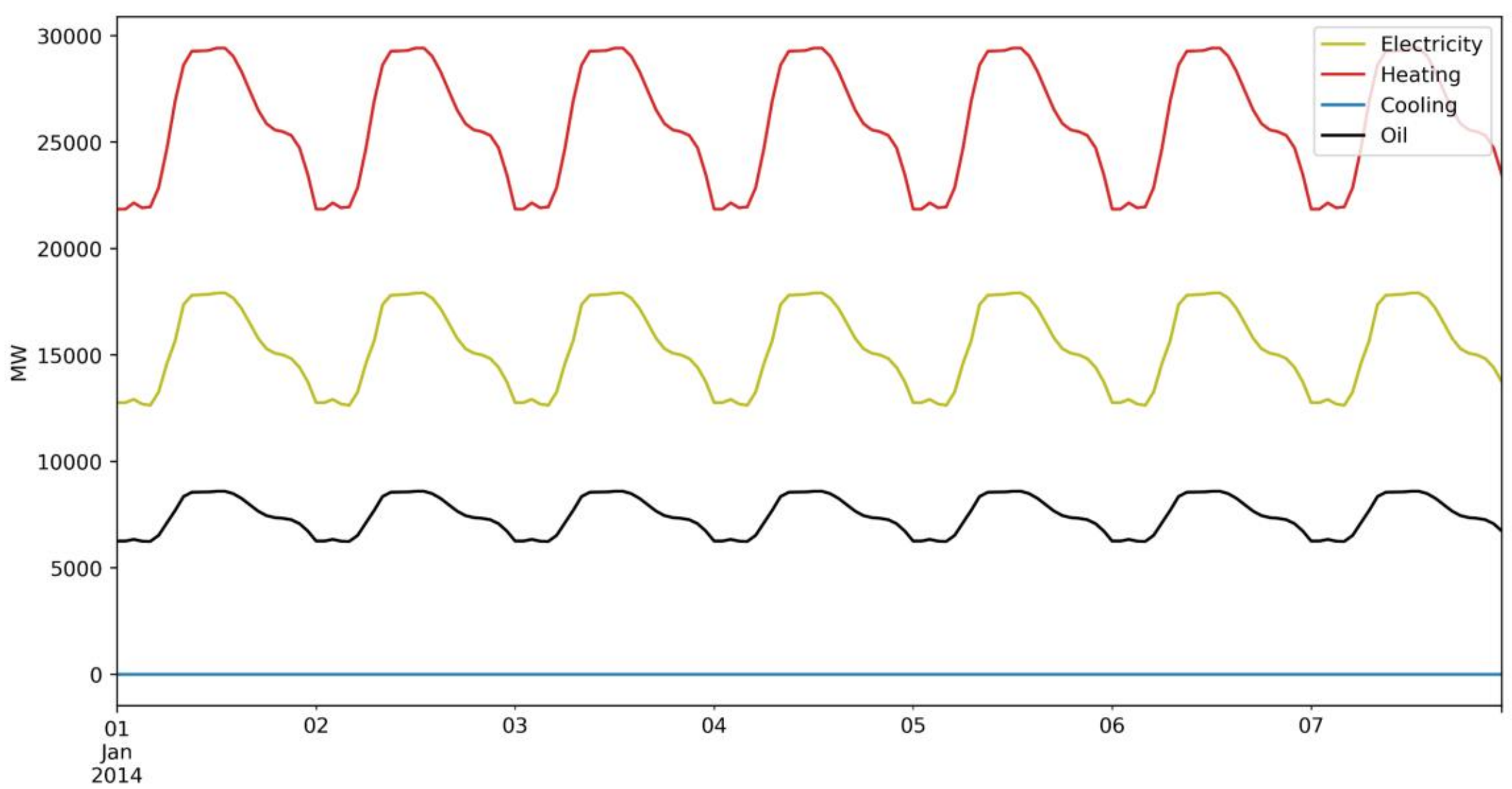

*Figure 16: Industrial profiles applied to annual demand taken from the County Level Industry Use dataset.*

**Demand Projections**

Following the same method used to project service level demand, industrial demand is projected following EIA AEO projections [24]. Figure 17 shows the industrial demand projections for select AEO scenarios. We assume heating and electrical load proportions do not change. The AEO does give NAICS specific (ie. food, paper, mining, ect.) energy consumption projections, therefore, total industrial energy is used to scale demand.

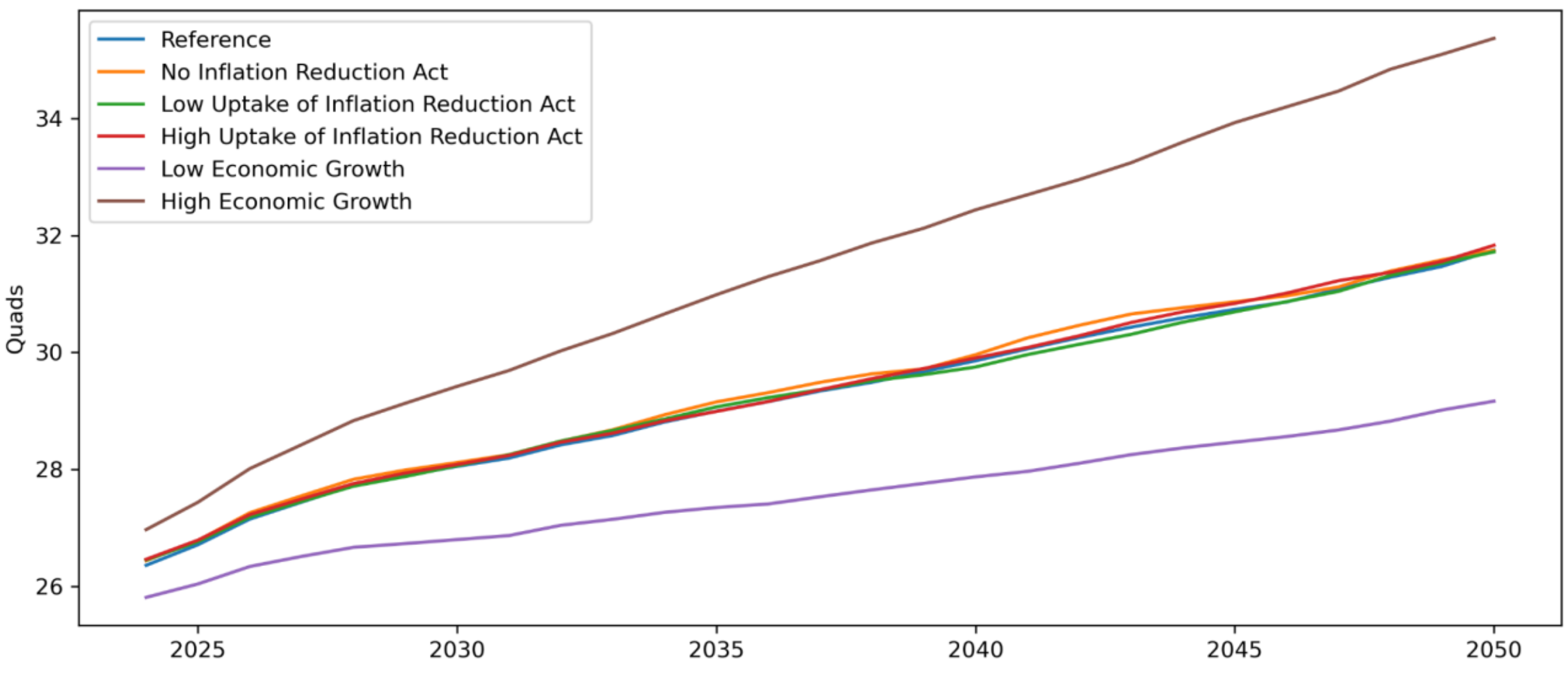

*Figure 17: Annual Energy Outlook industrial end-use consumption projections, with select scenarios.*

**Demand Disaggregation**

Industrial demand disaggregation follows NLR's CLIU dataset [31]. This dataset gives industrial end-use demand by fuel type at a county level. Notably, this disaggregation strategy is different than the strategies used for the service and transport sector, which are based on population distributions. Industrial demand often does not follow population distributions, and disaggregating against CLUI captures this. Figure 18 gives an example of the industrial demand disaggregation for four select states.

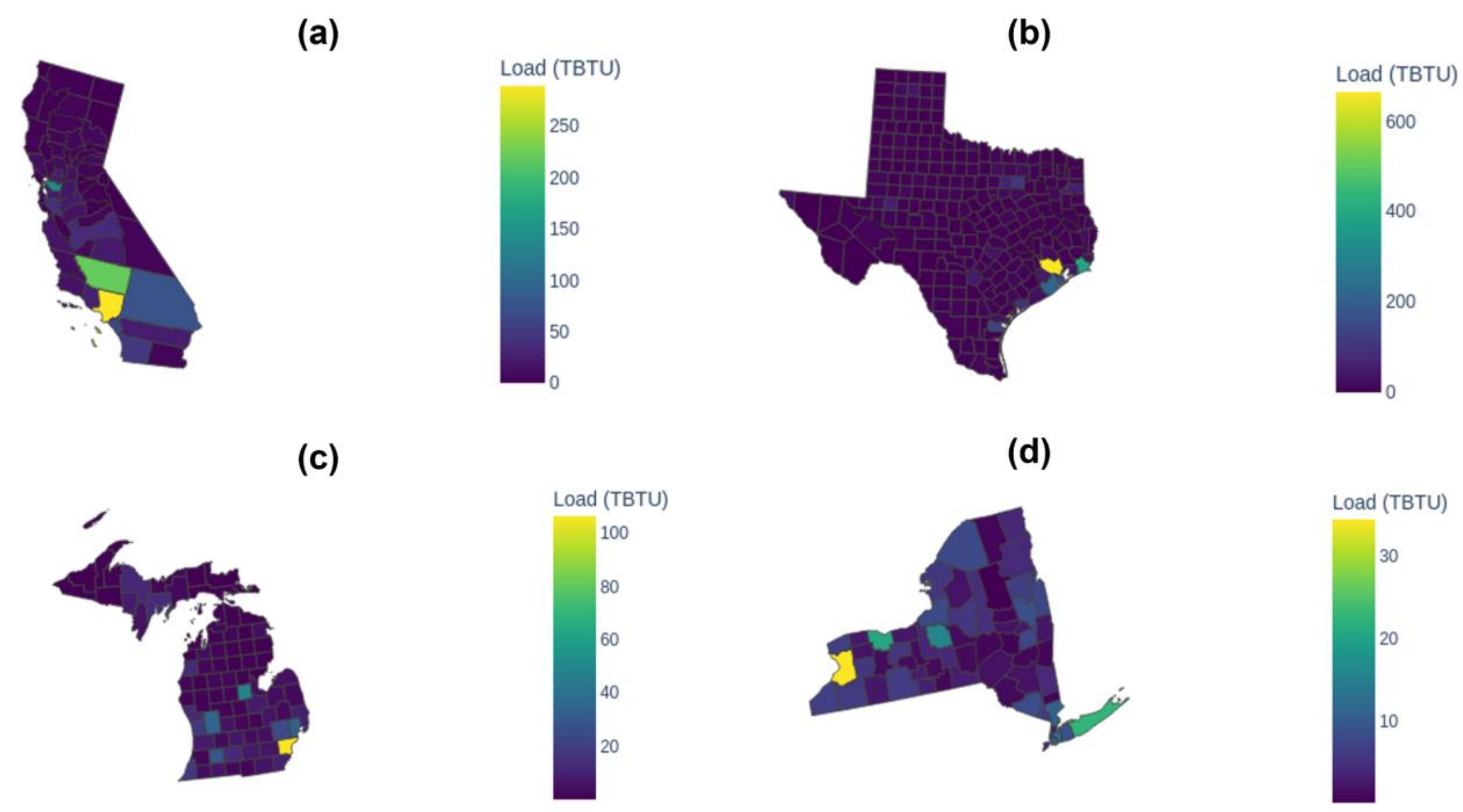

*Figure 18: Industrial demand disaggregation examples for (a) California. (b) Texas. (c) Michigan. (d) New York.*

## S5.3 Technologies

There are 4 different technologies to meet end-use industrial demand of heat and electricity. Base electrical loads (which includes loads such as lighting or machine drives) are met through electrical distribution. As PyPSA-USA does not ingest distribution level electrical network data, this is modelled as a simple lossy link. Heat demand can be met through gas, coal, or electrical furnaces, or through heat-pumps. The graphical representation of the industrial sector is given in Figure 19. As noted above, we do not differentiate industrial load via NAICS code; moreover, we do not differentiate between different heat-level requirements. This will miss capturing nuanced interaction in industrial decarbonization pathways (for example, heat-pumps cannot necessarily meet high temperature requirements for some applications). As this paper focuses on near-term targets, which will likely not rely on industrial decarbonization, we include build rates to limit expansion of technologies. Moreover, we include a configurable parameter to ensure industrial heating is met via a certain share of fossil fuels, to represent current heating practices. This is represented in equation (33), where $K = \{k \in S \mid k\ is\ fossil\ fuel\ heat\ source\ in\ industry\}$.

$$\frac{\sum_{n,k,t}(g_{n,k,t})}{\sum_{n,t}(d_{ind-heat_{n,t}})} \geq FossilGenerationRatio \tag{26}$$

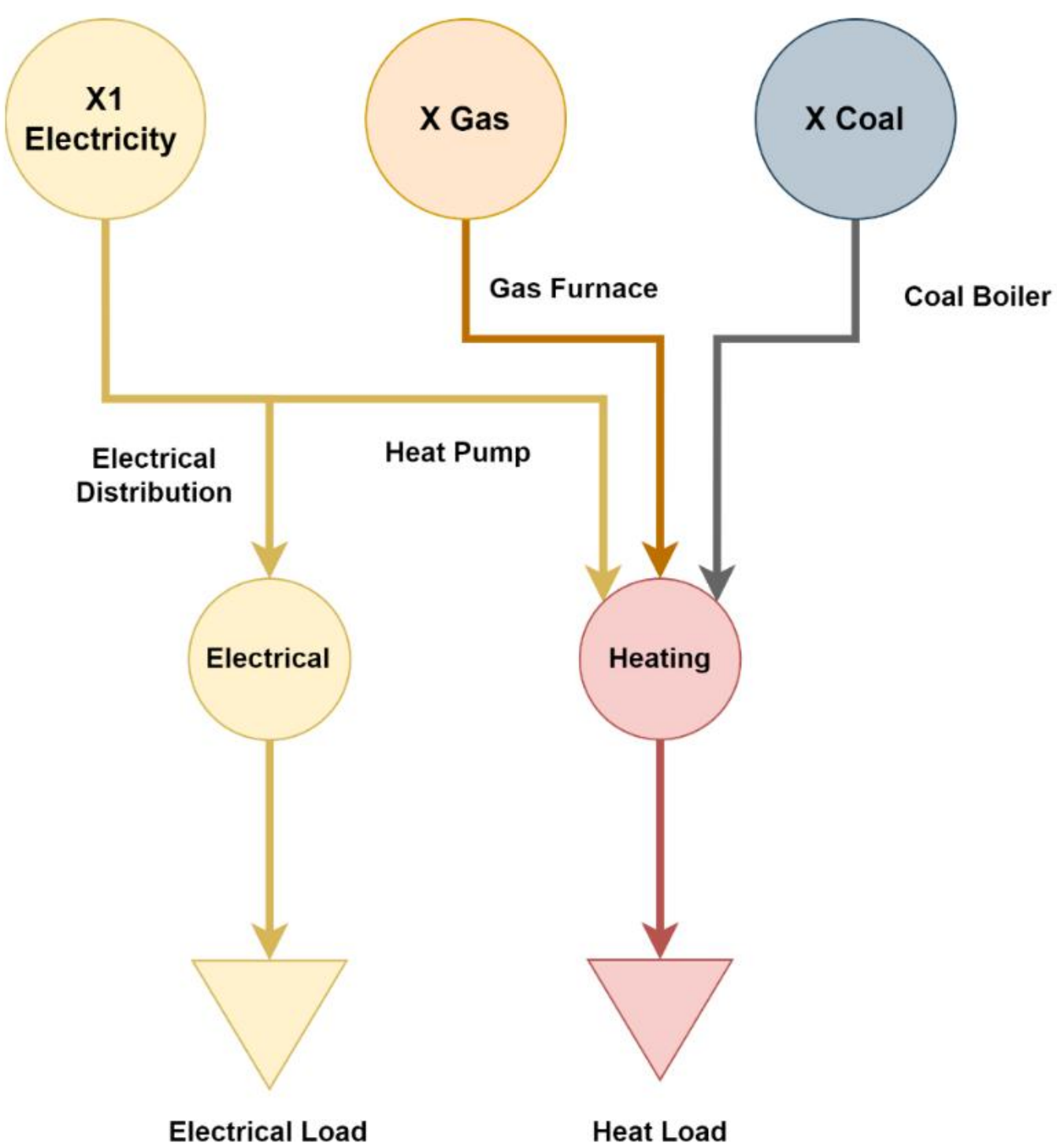

*Figure 19: Reference Energy System of the Industrial Sector*

All technologies have techno-economic parameters associated with them. These include capital costs, fixed costs, lifetimes, efficiencies, and existing stock values. All data sources are listed in Table 4. Heat pump COP are calculated following the same process described for service level heat pumps described in Figure 13.

*Table 4: Industrial Sector Data Sources*

| Technology | Metric | Source |
|---|---|---|
| Furnaces | Capital Cost \| Lifetime \| Efficiency | [34] |
| Heat Pumps | Capital Cost \| Lifetime | [34] |
| Heat Pumps | Efficiency | [15], [26] |
| Furnaces \| Heat Pumps | Existing Stock | [32] |
| Natural Gas | Fuel Price | [11] |
| Coal | Fuel Price | [11] |
| - | Demand | [24], [30], [31], [32], [33] |

## S.6 Transport Sector

The transportation module of PyPSA-USA is split into two categories; Road transportation, and non-road transportation. Each category has numerous vehicle types. This section will discuss each category separately.

### S6.1 Technologies

#### Road Vehicles

Four road vehicle classes are modelled in PyPSA-USA; light-duty vehicles, medium-duty vehicles, heavy-duty vehicles, and buses. Demand for each of these vehicles can be met through electricity or petrol. Distinctions are not made between petrol levels (ie. all road vehicles operate using the same petrol gasoline). Biogas, natural gas, hydrogen, and other transport fuels are not represented. The reference energy system for road transportation is given in Figure 20.

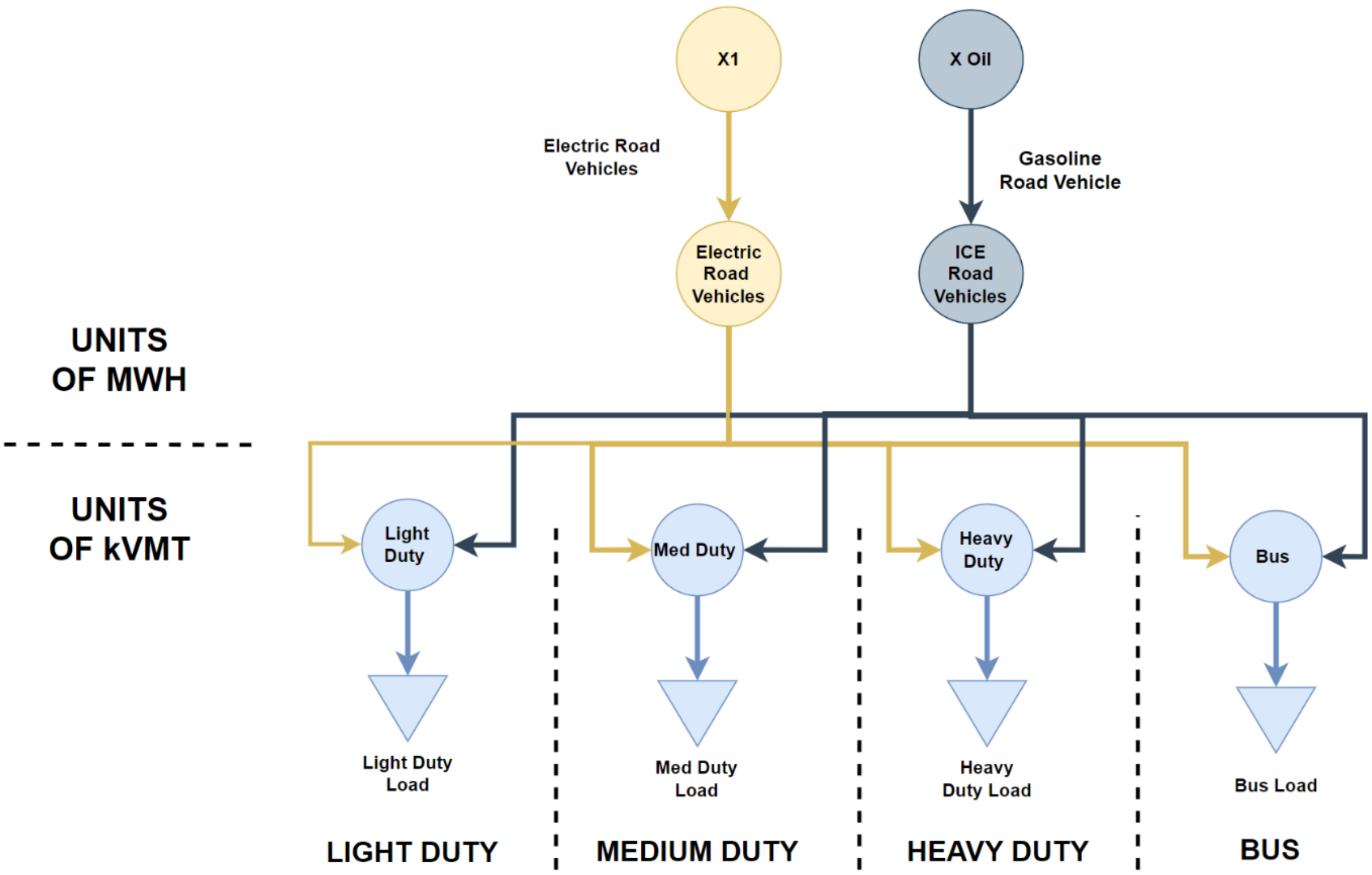

*Figure 20: Schematic of the road transportation sector in PyPSA-USA*

#### Non-Road Vehicles

Four non-road vehicle classes are modelled in PyPSA-USA; passenger air-travel, passenger rail-travel, shipping rail, and domestic marine shipping. Currently, these travel modes are included for emission accounting purposes only, and energy trade-off questions cannot be answered with them. This is because heavy duty transportation (ie. non-road vehicles) are generally not seen as a near-term emission target. To reduce model complexity, we remove decarbonization options from this sector. The reference energy system for non-road transportation is given in Figure 21.

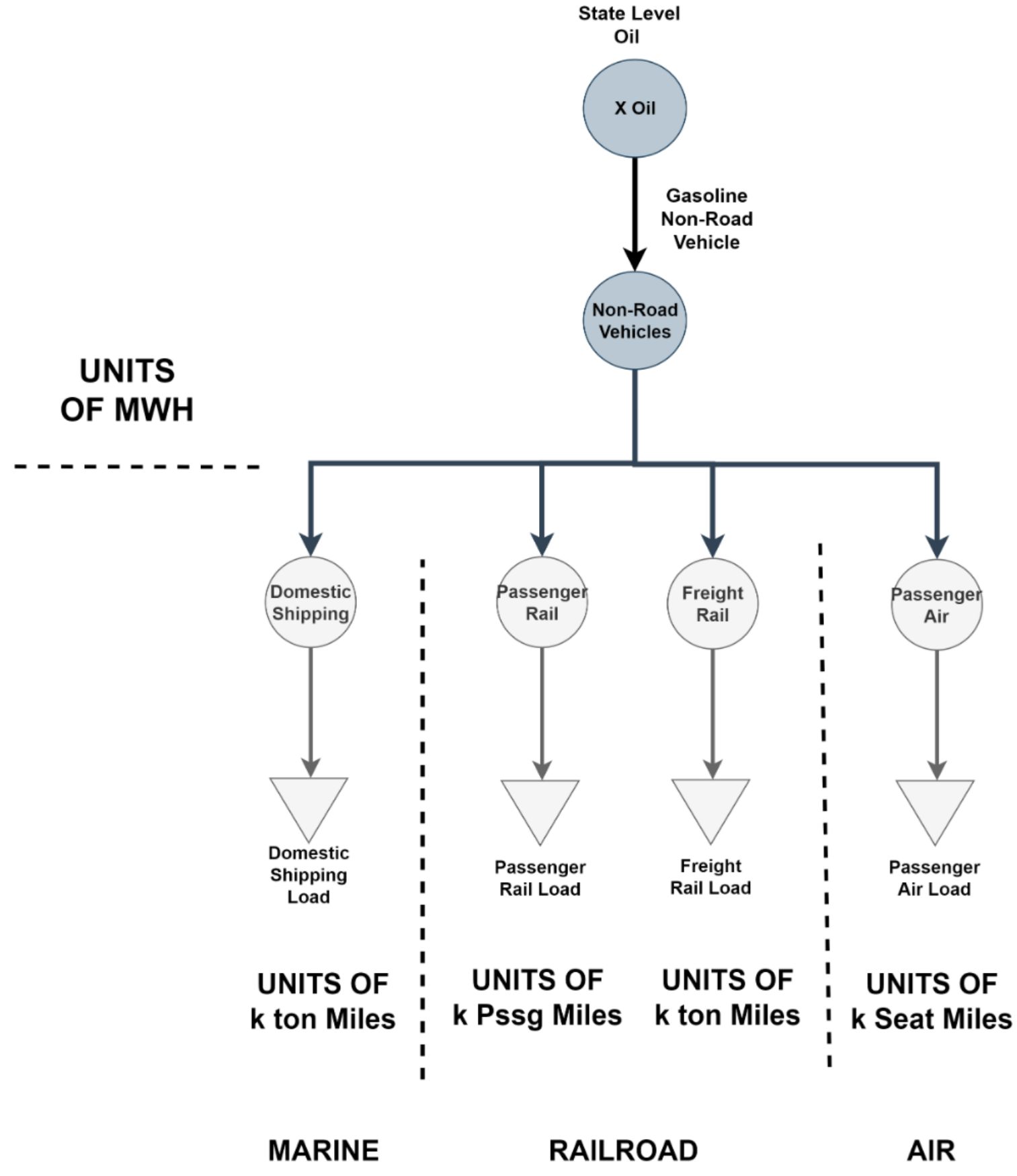

*Figure 21: Schematic of non-road vehicle transportation sector in PyPSA-USA*

**Transportation Sector Parameters**

The techno-economic parameters and sources for both road and non-road transportation is given in Table 5. As non-road transportation is included for emission accounting purposes only, no existing stock and investment costs are included.

*Table 5: Data Sources for the transportation sector*

| Technology | Metric | Source |
|---|---|---|
| Road Vehicles | Capital Costs \| Efficiency | [35] |
| Road Vehicles | Lifetime | [24] |
| Road Vehicles | Demand | [24], [36] |
| Road Vehicles | Existing Stock | [35] |
| Road Vehicles | Fuel Price | [11] |
| Non-Road Vehicles | Efficiency | [24] |
| Non-Road Vehicles | Demand | [24] |
| Non-Road Vehicles | Fuel Price | [11] |

## S6.2 Demand

Both road and non-road transportation sectors model demand in end-use units, for example, Vehicle Miles Travelled (VMT) and passenger miles, rather than energy (ie. MWh). This reduces the number of assumptions needed when working with our primary data sources, and to allow for mode shifting studies in the future.

### Road Vehicles

Road vehicle demand is created by overlaying load profiles on top of annual demand values. Electric Vehicle (EV) charging profiles for each of the vehicle classes (light-duty, medium-duty, heavy-duty and buses) are taken from the NLR Electrification Futures Study (EFS) study [36]; selecting the EFS year closest to the PyPSA-USA planning year. EFS gives hourly charging profiles per vehicle type and per state. An example of these profiles is given in Figure 22.

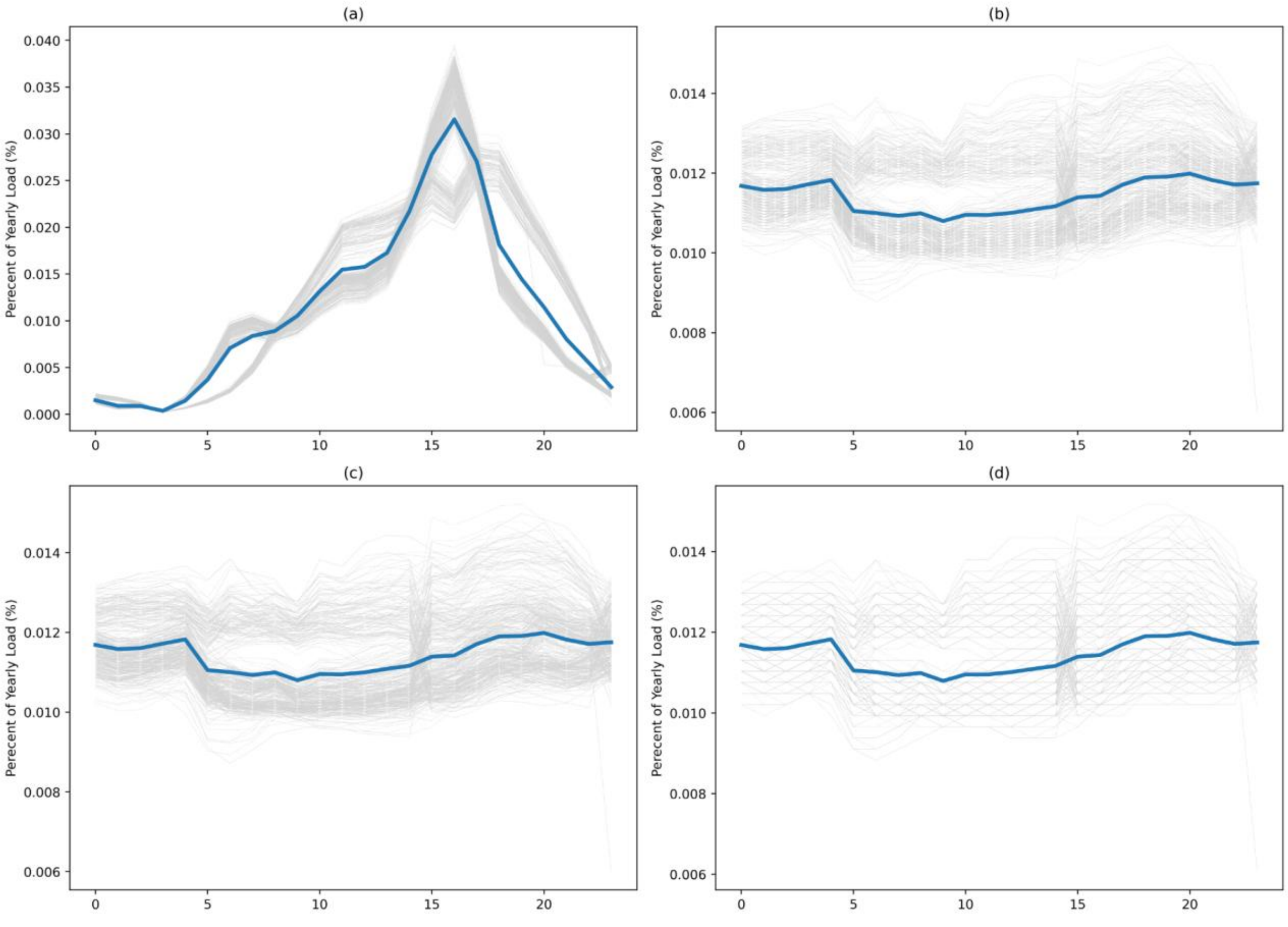

*Figure 22: Electrical Vehicle load profiles for 2050 taken from NLR EFS [36]. Gray lines are hourly profiles over the year. The blue line is the average profile over the year. (a) Light-Duty Vehicles. (b) Medium-Duty Vehicles. (c) Heavy-Duty Vehicles. (d) Buses*

Historical and projected annual load values are extracted from the EIA AEO [24]; Figure 23 thru Figure 26. The annual loads are overlaid with the charging profiles to obtain state level hourly charging profiles for a future year. The load is disaggregated following the same population-based method described for the service sector.

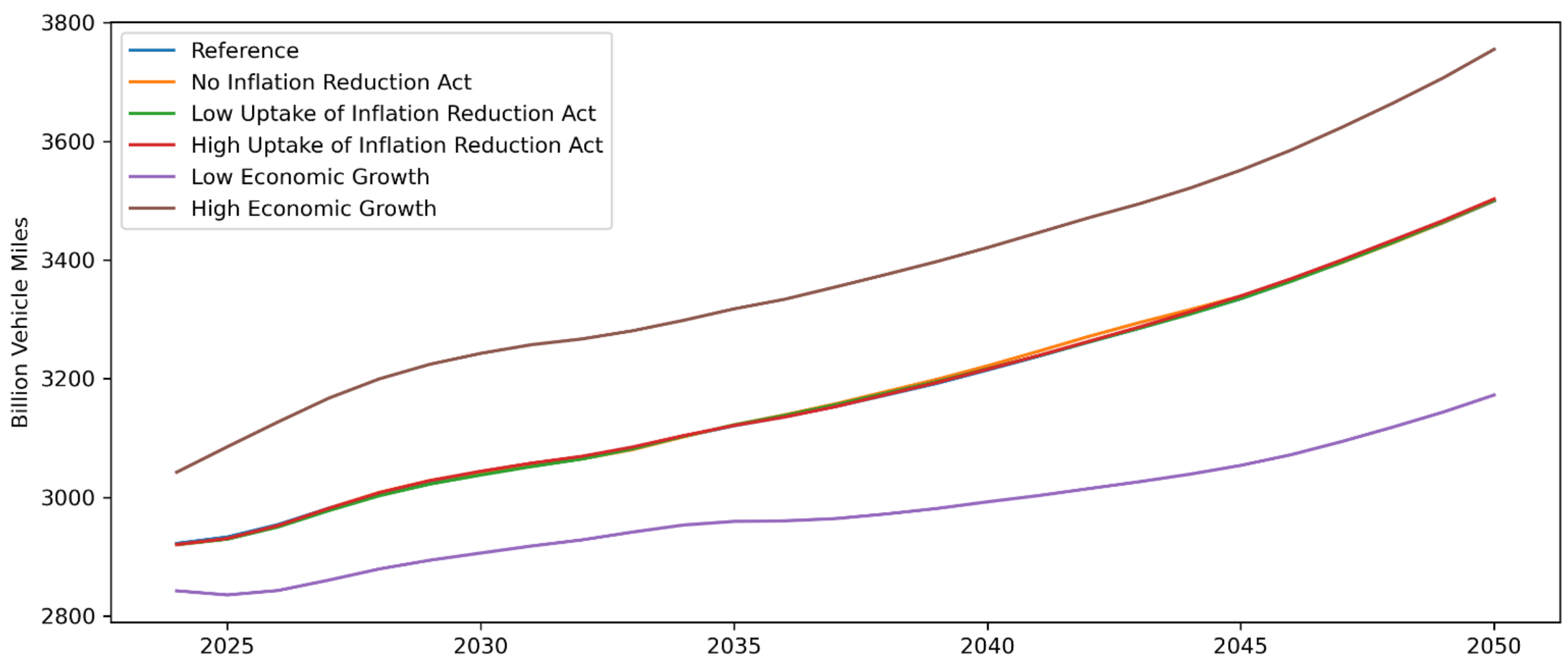

*Figure 23: EIA AEO Light-Duty Projected Vehicle Demand*

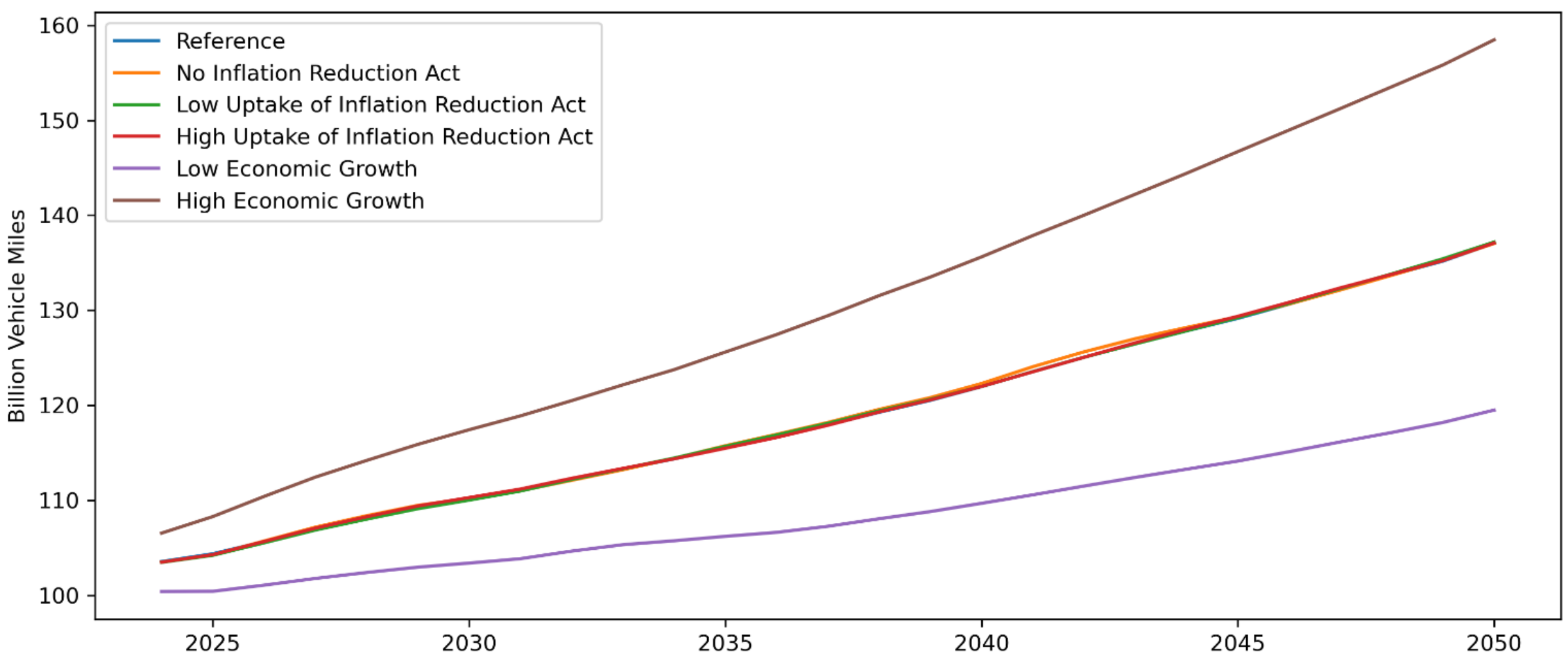

*Figure 24: EIA AEO Medium-Duty Projected Vehicle Demand*

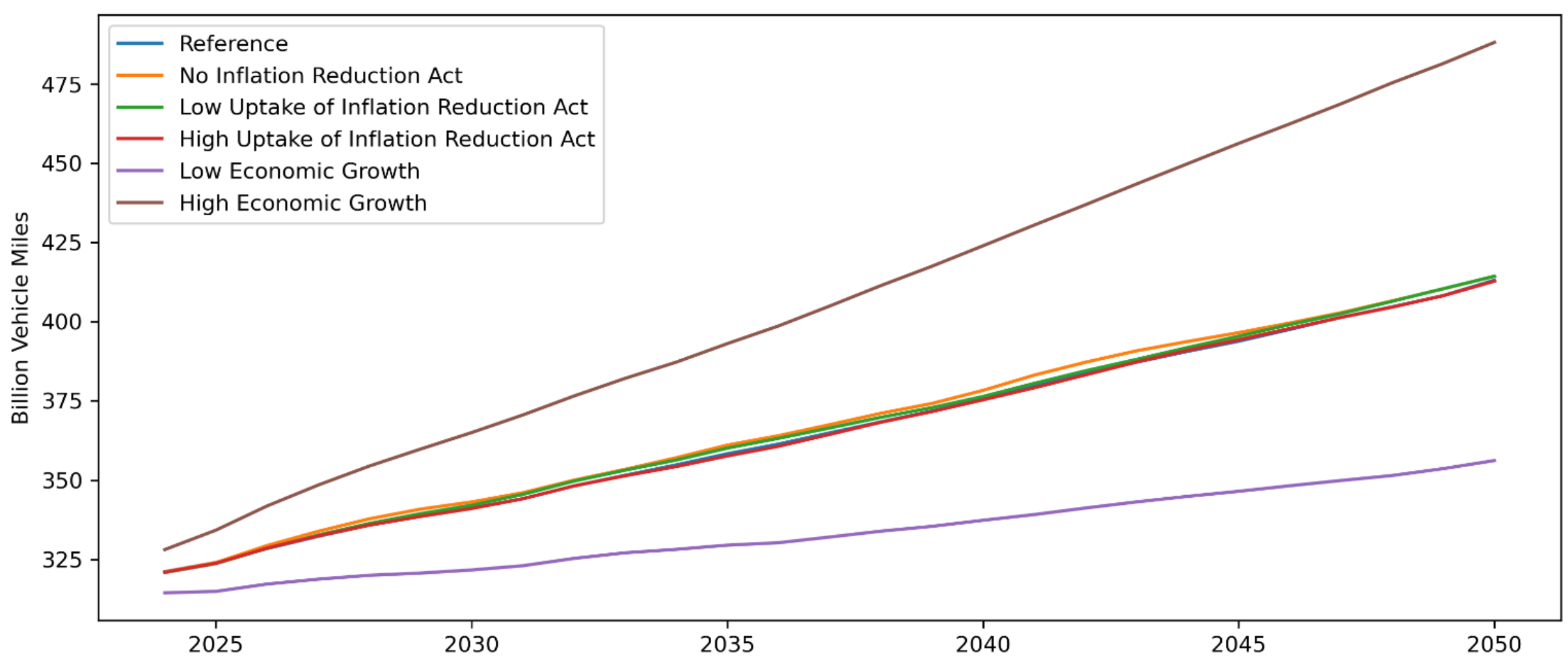

*Figure 25: EIA AEO Heavy-Duty Projected Vehicle Demand*

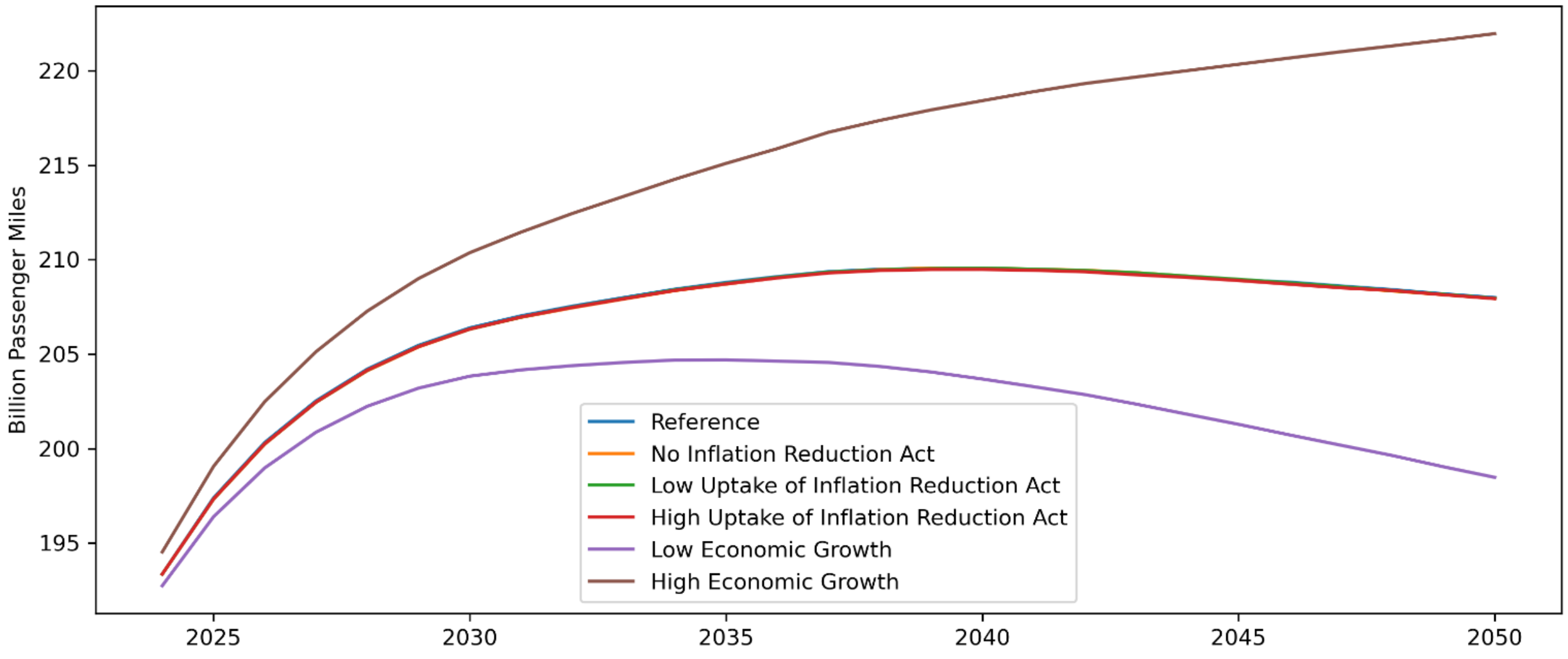

*Figure 26: EIA AEO Bus Projected Vehicle Demand*

A challenge with endogenous investment decisions is modifying the electrical load to account for a greater share of EVs in the system. Load is exogenous to the system, meaning it must be defined by the user. However, if more EVs are invested in, then the magnitude of the electrical load profile should grow. To account for this, we set all EVs to a “must-run” condition. The EV links minimum and maximum operating limits are set to match the load profile. The petrol links are free to operate as little or as much as they like. This setup forces EVs to always be run, with petrol vehicles filling in remaining load. A schematic of this is shown in Figure 27.

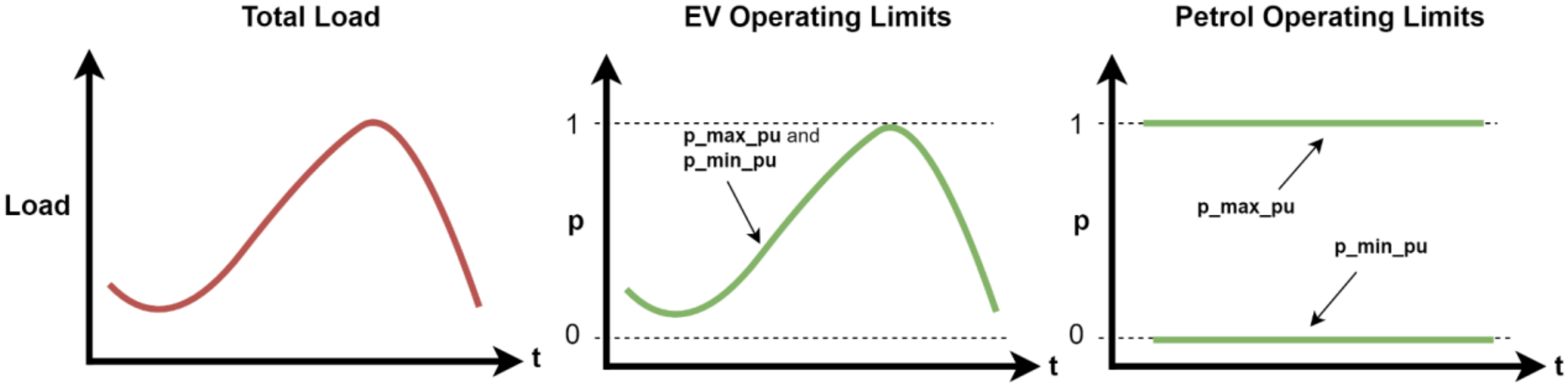

*Figure 27: Schematic of Must-Run EV configuration for the transportation sector*

### Non-Road Vehicles

Non-road transport demands do not have load profiles. Rather, yearly load values are uniformly distributed evenly over the full year. The annual demand projections are taken from taken from the EIA AEO [24], following the same data source as road demand; shown in Figure 28 through Figure 31. As previously mentioned, non-road transportation is not open to decarbonization due to it being a sector that will likely be decarbonized over a longer time horizon. It is included for emission counting purposes only. Therefore, all non-road demand must be met with petrol.

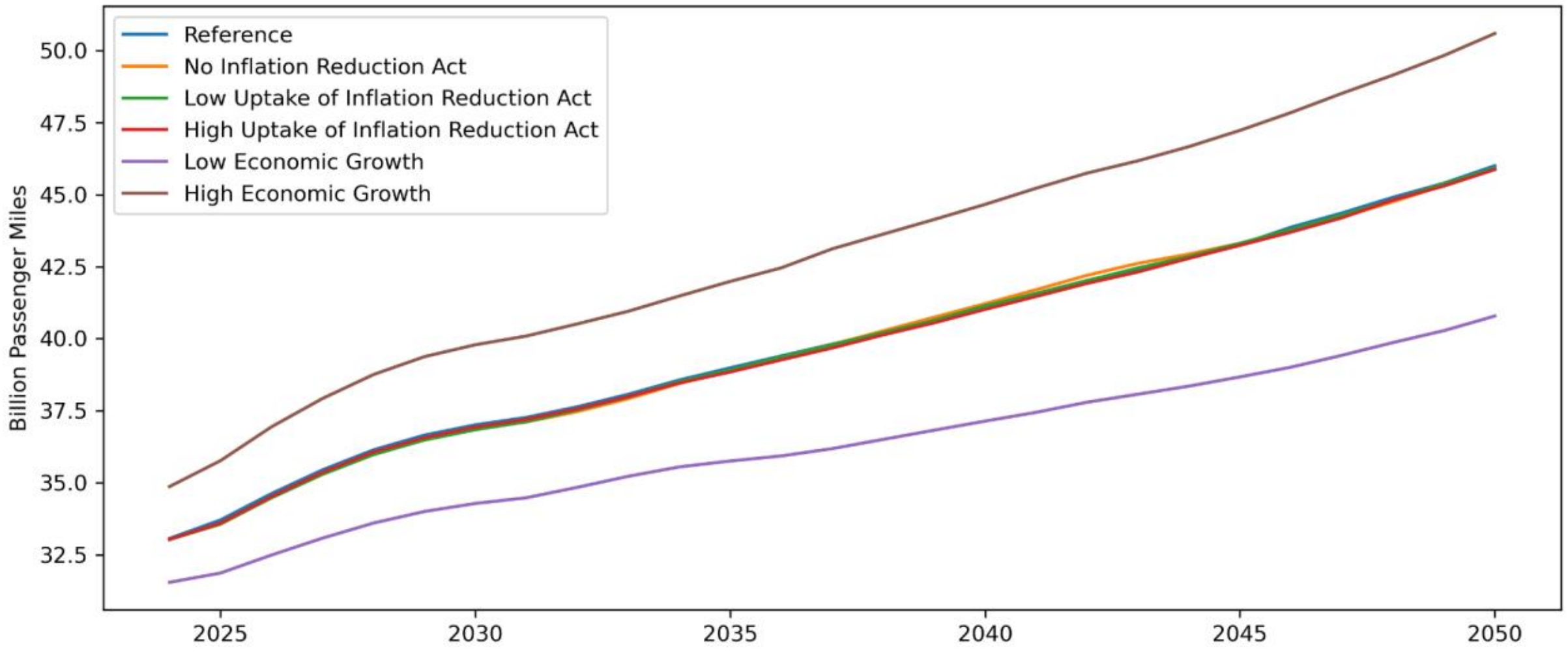

*Figure 28: EIA AEO passenger rail projected demand*

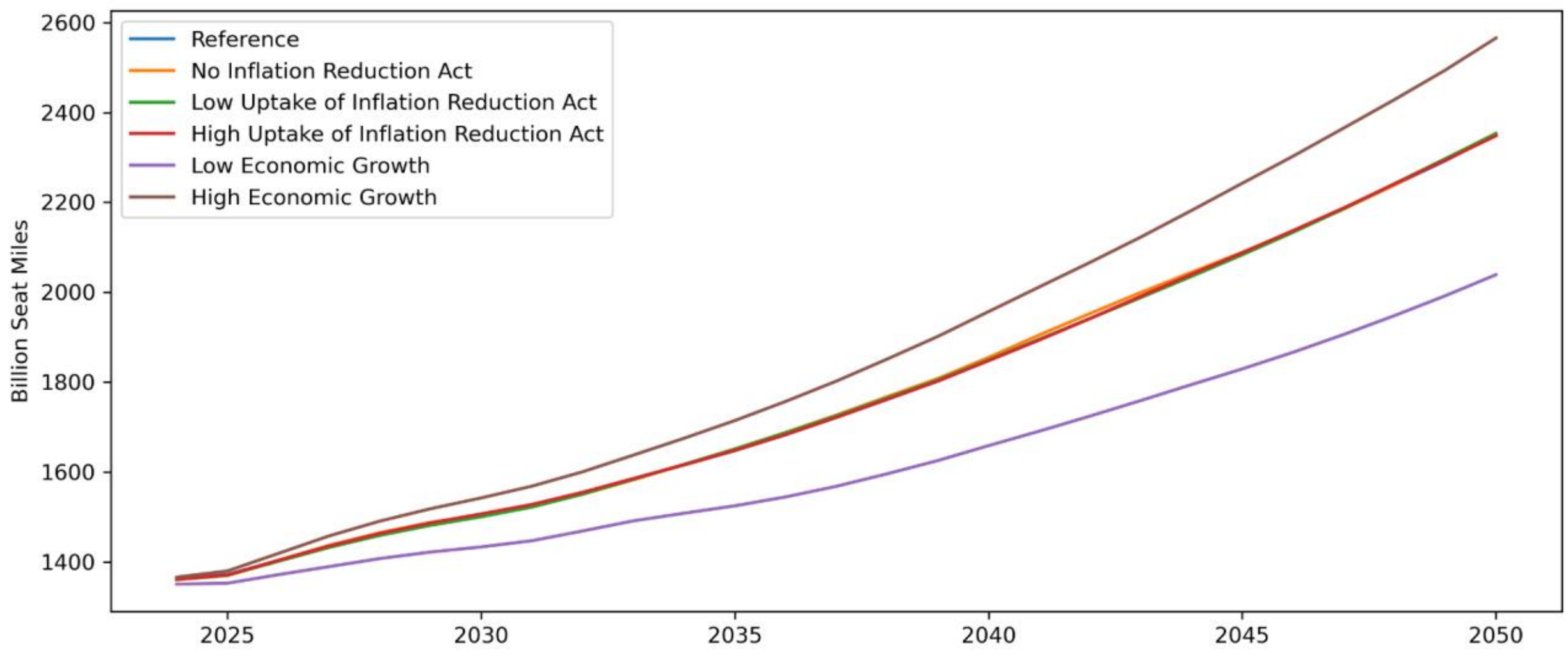

*Figure 29: EIA AEO passenger aircraft projected demand*

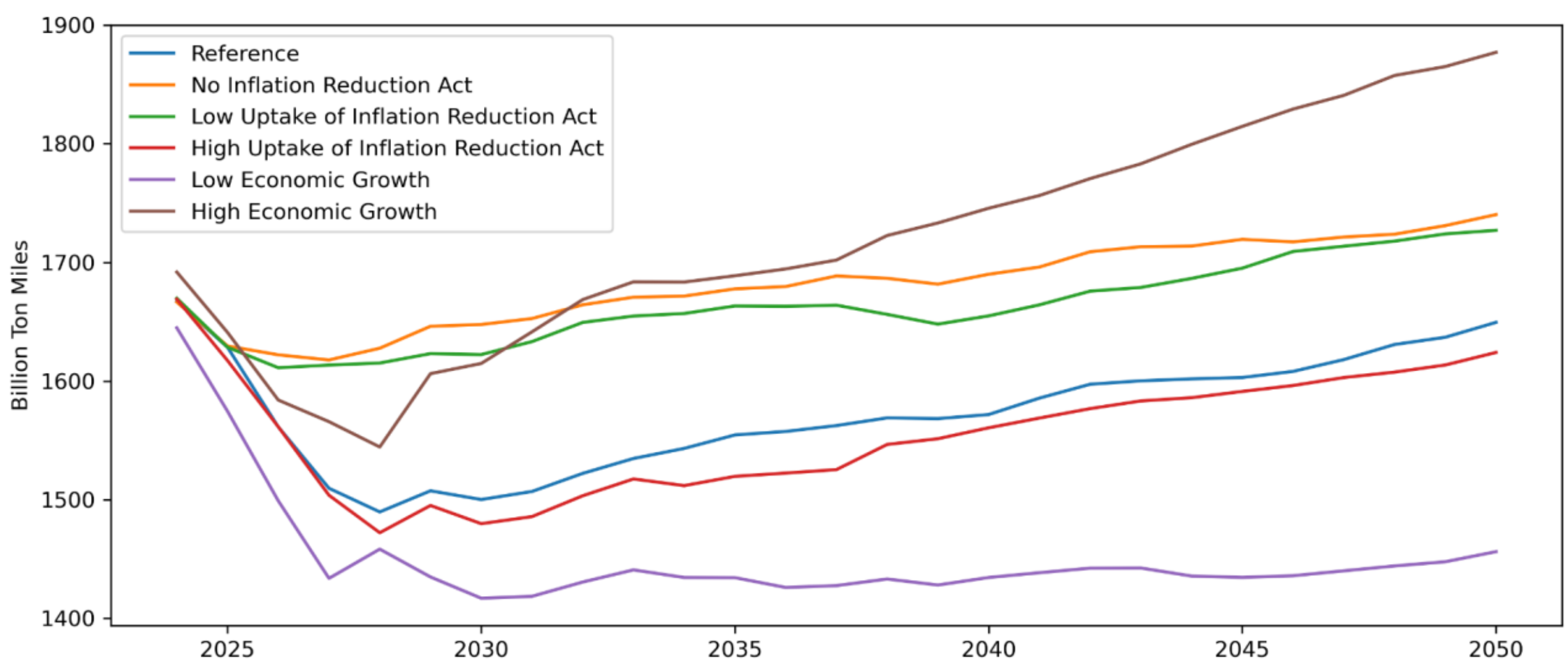

*Figure 30: EIA AEO shipping freight rail projected demand*

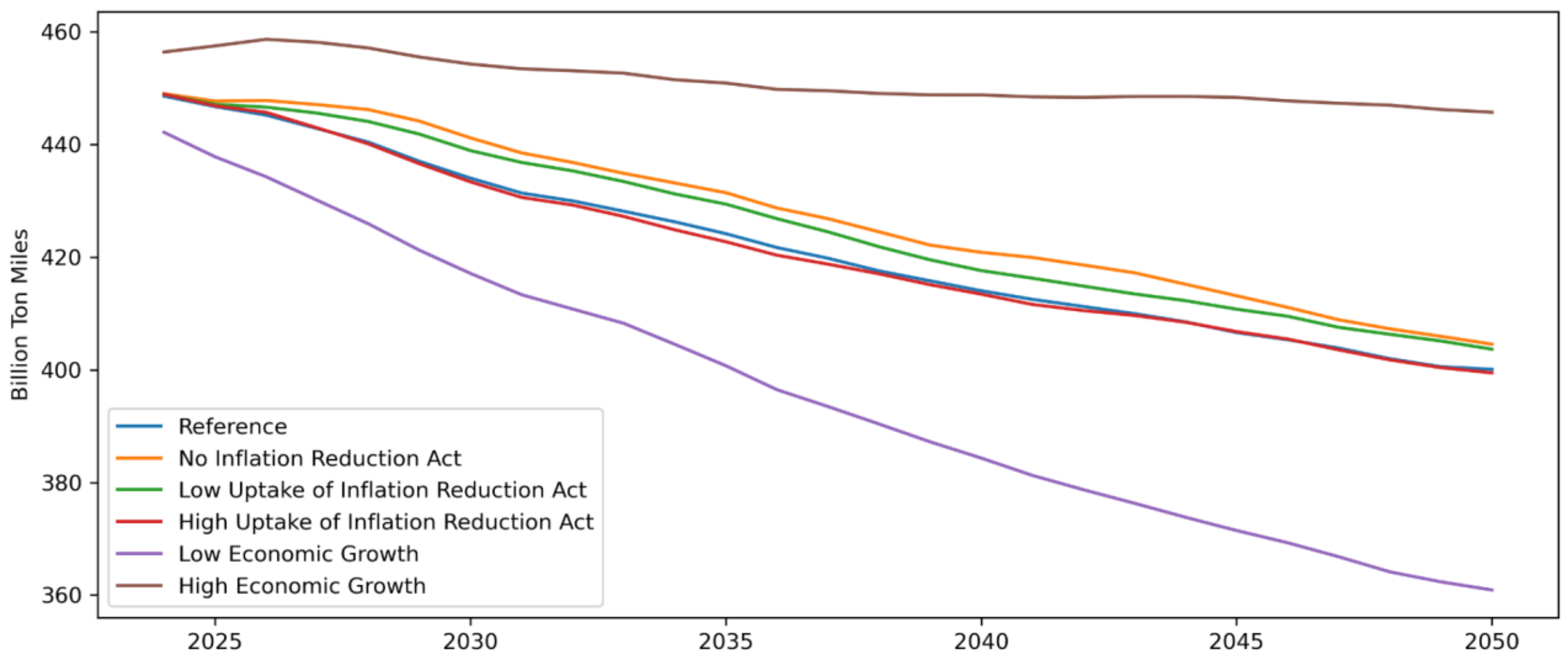

*Figure 31: EIA AEO domestic marine shipping demand*

## S.7 Uncertainty Analysis Methodology

In this section we describe our process to handle model uncertainty. First we describe how we parameterize our uncertainty, second we describe the Global Sensitivity Analysis (GSA) method used, and lastly we describe how we propagate uncertainty through the model.

### S7.1 Uncertainty Parameterization

To perform a sensitivity analysis, all uncertain parameters, and ranges on these parameters, must be correctly parameterized to obtain credible results [37]. We identify ~350 uncertain parameters in our model and estimate uncertainty ranges on them. Primarily, we use research (such as estimating uncertainty on ERA5 [38]) or model outputs (such as from the EIA AEO [24] or NLR EFS [35], [36]) to obtain uncertainty ranges. In cases where uncertainty ranges can not be found, consistent deviations are assumed (ie. $\pm$ 5%). The uncertain parameters are documented on the project GitHub [39] and on the public dashboard (see link at https://github.com/DeltaE/pypsa-gsa). All ranges assume a uniform distribution.

### S7.2 Global Sensitivity Analysis Theory

We follow the process described in detail by Usher et al. [40] on applying GSA to an ESOM. Specifically, we use the Method of Morris [41] (from here on referred to as Morris). Morris computes an average of absolute local derivatives (called the Elementary Effect (EE)) for each parameter, and in turn produces three metrics; $\mu$, $\mu^*$, and $\sigma$. $\mu$ provides the mean EE and gives an indication of how influential a parameter is. However, if effects are both positive an negative, $\mu$ is susceptible to errors. Campolongo et al. [42] introduces $\mu^*$, the absolute value of the mean EE, to address this issue. The higher the $\mu^*$ value, the more influential the parameter is on the result of interest. Finally, $\sigma$ measures the nonlinear effects and/or interactions with other factors.

Saltelli [43] reviews Morris in detail, however, here we highlight the equations used in this paper; equations (27)-(31)[c]. We use the Python package SALib [44], [45] to perform the actual calculations.

$$EE_i^j\left(x^{(l)}\right) = \frac{\left[y\left(x^{l+1}\right) - y(x^l)\right]}{\Delta} \tag{27}$$

$$EE_i^j\left(x^{(l+1)}\right) = \frac{\left[y\left(x^{l}\right) - y(x^{l+1})\right]}{\Delta} \tag{28}$$

$$\mu_i = \frac{1}{r}\sum_{j=1}^{r} EE_i^j \tag{29}$$

$$\mu_i^* = \frac{1}{r}\sum_{j=1}^{r} \left|EE_i^j\right| \tag{30}$$

$$\sigma_i^2 = \frac{1}{1-r}\sum_{j=1}^{r} \left(EE_i^j - \mu\right)^2 \tag{31}$$

Where:

- $l$ is the trajectory (ie. each set of movements within the sample).
- $EE_i^j$ is the EE of the factor i on the j[th] trajectory.
- $r$ is the number of trajectories
- $i$ is the input factor
- $\Delta$ is the step size.

As discussed by Sin and Sin et al. [46] (and applied to an ESOM by Morret et al. [37] and Moksnes et al. [47]), Morris rankings can be dependent on the units of the input factors. If many different units are used, the rankings can lead to incorrect results. As such, Sin et al. [46] introduced the Scaled Elementary Effect (SEE) given in Equations (32) and (33).

$$SEE_i^j\left(x^{(l)}\right) = \left(\frac{\left[y\left(x^{l+1}\right) - y\left(x^l\right)\right]}{\Delta}\right)\left(\frac{\sigma_x^i}{\sigma_y^j}\right) \tag{32}$$

$$SEE_i^j\left(x^{(l+1)}\right) = \left(\frac{\left[y\left(x^{l}\right) - y\left(x^{l+1}\right)\right]}{\Delta}\right)\left(\frac{\sigma_x^i}{\sigma_y^j}\right) \tag{33}$$

The SEE is calculated by scaling the EE using the standard deviation of model inputs $\left(\sigma_y^j\right)$ and input factors $\left(\sigma_x^i\right)$. Therefore, our $\mu$, $\mu^*$, and $\sigma$ Equations (29), (30), (31) are replaced by Equations (34), (35), (36). SEE is a unitless measure that allows comparison between all input factors, regardless of unit or range.

$$\mu_i = \frac{1}{r}\sum_{j=1}^{r} SEE_i^j \tag{34}$$

---

[c] When calculating the EE, Equation (27) is used if the $i^{th}$ component of $x^{(l)}$ is increased by $\Delta$, else Equation (28) is used if the $i^{th}$ component of $x^{(l)}$ is decreased by $\Delta$.

$$\mu_i^* = \frac{1}{r}\sum_{j=1}^{r}\left|SEE_i^j\right| \tag{35}$$

$$\sigma_i^2 = \frac{1}{1-r}\sum_{j=1}^{r}\left(SEE_i^j - \mu\right)^2 \tag{36}$$

### S7.3 Uncertainty Propagation

An advantage of Morris is that it is a computationally efficient GSA strategy, allowing for the identification of the most impactful groups of parameters. For example, capital costs of gas furnaces are not broken out into residential and commercial costs for the GSA, but rater treated as a "service sector furnace capital cost" sensitivity group. The uncertainty propagation step separates these groups and propagates the most influential parameter groups through the model via a Monte Carlo analysis. We use Latin Hyper Cube sampling [48] directly from the SALib library.

## References

[1] T. Brown, J. Hörsch, and D. Schlachtberger, “PyPSA: Python for Power System Analysis,” *J. Open Res. Softw.*, vol. 6, no. 1, Art. no. 1, Jan. 2018, doi: 10.5334/jors.188.

[2] Brown, Tom *et al.*, *PyPSA: Python for Power System Analysis*. (Oct. 15, 2021). Zenodo. doi: 10.5281/ZENODO.3946412.

[3] K. Tehranchi, T. Barnes, M. Frysztacki, F. Hofmann, and I. L. Azevedo, *PyPSA-USA: An Open-Source Energy System Optimization Model for the United States*. (Aug. 25, 2024). Python. doi: 10.5281/zenodo.10815964.

[4] N. Johnson, M. Liebreich, D. M. Kammen, P. Ekins, R. McKenna, and I. Staffell, “Realistic roles for hydrogen in the future energy transition,” *Nat. Rev. Clean Technol.*, pp. 1–21, Apr. 2025, doi: 10.1038/s44359-025-00050-4.

[5] Department of Energy, “Pathways to Commercial Liftoff: Clean Hydrogen,” 2024. Accessed: Mar. 23, 2023. [Online]. Available: https://climateprogramportal.org/wp-content/uploads/2025/02/Pathways-to-Commercial-Liftoff_Clean-Hydrogen_December-2024-Update.pdf

[6] S. Cohen *et al.*, “Regional Energy Deployment System (ReEDS) Model Documentation: Version 2018,” *Renew. Energy*, 2019.

[7] P. Brown *et al.*, *Regional Energy Deployment System Model 2.0 (ReEDS 2.0)*. (Dec. 02, 2025). Python. [Online]. Available: https://www.nrel.gov/analysis/reeds/index.html

[8] pypsa, “PyPSA Documentation,” PyPSA: Python for Power System Analysis. Accessed: Apr. 25, 2026. [Online]. Available: https://docs.pypsa.org/latest/

[9] G. Barbose, “U.S. State Electricity Resource Standards,” 2026.

[10] US EPA, “Energy Attribute Tracking Systems.” Accessed: Apr. 25, 2026. [Online]. Available: https://www.epa.gov/green-power-markets/energy-attribute-tracking-systems

[11] EIA, “Homepage - U.S. Energy Information Administration (EIA).” Accessed: Dec. 28, 2021. [Online]. Available: https://www.eia.gov/index.php

[12] Z. A. Selvans *et al.*, “Public Utility Data Liberation Project (PUDL) Data Release.” Zenodo, Sep. 06, 2025. doi: 10.5281/ZENODO.3653158.

[13] L. Vimmerstedt *et al.*, “2022 Annual Technology Baseline (ATB) Cost and Performance Data for Electricity Generation Technologies.” DOE Open Energy Data Initiative (OEDI); National Renewable Energy Laboratory (NREL), p. 8 files, 2022. doi: 10.25984/1871952.

[14] F. Hofmann, J. Hampp, F. Neumann, T. Brown, and J. Hörsch, “atlite: A Lightweight Python Package for Calculating Renewable Power Potentials and Time Series,” *J. Open Source Softw.*, vol. 6, no. 62, p. 3294, Jun. 2021, doi: 10.21105/joss.03294.

[15] H. Hersbach *et al.*, “ERA5 hourly data on single levels from 1959 to present.” Copernicus Climate Change Service (C3S) Climate Data Store (CDS), 2018. [Online]. Available: 10.24381/cds.adbb2d47

[16] P. R. Brown *et al.*, “A general method for estimating zonal transmission interface limits from nodal network data,” Aug. 2023. [Online]. Available: https://arxiv.org/abs/2308.03612

[17] T. Barnes, K. Tehranchi, B. Reinholz, M. Metcalfe, and T. Niet, “Multi-sector demand response for cost optimal energy transitions,” *PLOS Clim.*, vol. 5, no. 5, p. e0000918, May 2026, doi: 10.1371/journal.pclm.0000918.

[18] HIFLD, “Homeland Infrastructure Foundation-Level Data (HIFLD) | Homeland Security.” Accessed: Apr. 25, 2026. [Online]. Available: https://www.dhs.gov/gmo/hifld

[19] K. Tehranchi, T. Barnes, and C. Chen, “PyPSA-USA Cutouts & Datasets.” Zenodo, Jan. 07, 2025. doi: 10.5281/zenodo.14611937.

[20] Alberta Energy Regulator, “Natural Gas Supply Costs,” Data and Performance Reports. Accessed: Apr. 25, 2026. [Online]. Available: https://www.aer.ca/data-and-performance-reports/statistical-reports/alberta-energy-outlook-st98/natural-gas/natural-gas-supply-costs
[21] S. M. Folga, “Natural Gas Pipeline Technology Overview,” Argonne National Labratory, 2007. Accessed: Apr. 25, 2026. [Online]. Available: https://publications.anl.gov/anlpubs/2008/02/61034.pdf
[22] E. Wilson *et al.*, “End-Use Load Profiles for the U.S. Building Stock: Methodology and Results of Model Calibration, Validation, and Uncertainty Quantification,” NREL/TP-5500-80889, 1854582, MainId:78667, Mar. 2022. doi: 10.2172/1854582.
[23] United States Census Bureau, “U.S. Census Bureau,” Census.gov. [Online]. Available: https://www.census.gov/en.html
[24] EIA, “Annual Energy Outlook 2023,” U.S. Energy Information Administration, 2023, 2023. Accessed: Oct. 29, 2023. [Online]. Available: https://www.eia.gov/outlooks/aeo/
[25] Energy Information Agency, “Updated Buildings Sector Appliance and Equipment Costs and Efficiency,” U.S. Energy Information Agency. Accessed: Oct. 29, 2025. [Online]. Available: https://www.eia.gov/analysis/studies/buildings/equipcosts/
[26] I. Staffell, D. Brett, N. Brandon, and A. Hawkes, “A review of domestic heat pumps,” *Energy Environ. Sci.*, vol. 5, no. 11, pp. 9291–9306, Oct. 2012, doi: 10.1039/C2EE22653G.
[27] Energy Information Agency, “Residential Energy Consumption Survey (RECS),” U.S. Energy Information Agency. Accessed: Oct. 18, 2025. [Online]. Available: https://www.eia.gov/consumption/residential/
[28] Energy Information Agency, “Commercial Buildings Energy Consumption Survey (CBECS),” U.S. Energy Information Agency. Accessed: Oct. 18, 2025. [Online]. Available: https://www.eia.gov/consumption/commercial/
[29] E. Hale *et al.*, “The Demand-Side Grid (dsgrid) Model Documentation,” NREL/TP-6A20-71492, 1465659, MainId:16928, Aug. 2018. doi: 10.2172/1465659.
[30] Electric Power Research Institute, “Load Shape Library.” 2024. Accessed: Jan. 08, 2025. [Online]. Available: https://loadshape.epri.com/
[31] C. McMillan and V. Narwade, “United States County-Level Industrial Energy Use.” National Renewable Energy Laboratory - Data (NREL-DATA), Golden, CO (United States); National Renewable Energy Laboratory, p. 7 files, 2018. doi: 10.7799/1481899.
[32] U.S. Energy Information Agency, “Manufacturing Energy Consumption Survey.” Accessed: Apr. 25, 2026. [Online]. Available: https://www.eia.gov/consumption/manufacturing/
[33] National Labratory of the Rockies, *NatLabRockies/Industry-Energy-Tool*. (Feb. 24, 2025). Python. National Laboratory of the Rockies. Accessed: Apr. 25, 2026. [Online]. Available: https://github.com/NatLabRockies/Industry-Energy-Tool
[34] Danish Energy Agency, “Energy sources | Energistyrelsen.” Accessed: Apr. 25, 2026. [Online]. Available: https://ens.dk/en/energy-sources
[35] P. Jadun, C. McMillan, D. Steinberg, M. Muratori, L. Vimmerstedt, and T. Mai, “Electrification Futures Study: End-Use Electric Technology Cost and Performance Projections through 2050”.
[36] Y. Sun *et al.*, “Electrification Futures Study: Methodological Approaches for Assessing Long-Term Power System Impacts of End-Use Electrification,” NREL/TP-6A20-73336, 1660122, MainId:6124, Jul. 2020. doi: 10.2172/1660122.
[37] S. Moret, V. Codina Gironès, M. Bierlaire, and F. Maréchal, “Characterization of input uncertainties in strategic energy planning models,” *Appl. Energy*, vol. 202, pp. 597–617, Sep. 2017, doi: 10.1016/j.apenergy.2017.05.106.

[38] S. Pfenninger and I. Staffell, “Long-term patterns of European PV output using 30 years of validated hourly reanalysis and satellite data,” *Energy*, vol. 114, pp. 1251–1265, Nov. 2016, doi: 10.1016/j.energy.2016.08.060.

[39] T. Barnes, *DeltaE/pypsa-gsa*. (2026). Jupyter Notebook. ΔE+. Accessed: Apr. 25, 2026. [Online]. Available: https://github.com/DeltaE/pypsa-gsa

[40] W. Usher, T. Barnes, N. Moksnes, and T. Niet, “Global sensitivity analysis to enhance the transparency and rigour of energy system optimisation modelling [version 1; peer review: 1 approved, 2 approved with reservations],” *Open Res. Eur.*, vol. 3, p. 30, Feb. 2023, doi: 10.12688/openreseurope.15461.1.

[41] M. D. Morris, “Factorial Sampling Plans for Preliminary Computational Experiments,” *Technometrics*, vol. 33, no. 2, pp. 161–174, 1991, doi: 10.2307/1269043.

[42] F. Campolongo, J. Cariboni, and A. Saltelli, “An effective screening design for sensitivity analysis of large models,” *Environ. Model. Softw.*, vol. 22, no. 10, pp. 1509–1518, Oct. 2007, doi: 10.1016/j.envsoft.2006.10.004.

[43] A. Saltelli, Ed., *Global sensitivity analysis: the primer*. Chichester, England ; Hoboken, NJ: John Wiley, 2008.

[44] J. Herman and W. Usher, “SALib: An open-source Python library for Sensitivity Analysis,” *J. Open Source Softw.*, vol. 2, no. 9, p. 97, Jan. 2017, doi: 10.21105/joss.00097.

[45] T. Iwanaga, W. Usher, and J. Herman, “Toward SALib 2.0: Advancing the accessibility and interpretability of global sensitivity analyses,” *Socio-Environ. Syst. Model.*, vol. 4, pp. 18155–18155, May 2022, doi: 10.18174/sesmo.18155.

[46] G. Sin and K. V. Gernaey, “Improving the Morris method for sensitivity analysis by scaling the elementary effects,” in *Computer Aided Chemical Engineering*, J. Jeżowski and J. Thullie, Eds., in 19 European Symposium on Computer Aided Process Engineering, vol. 26. Elsevier, Jan. 2009, pp. 925–930. doi: 10.1016/S1570-7946(09)70154-3.

[47] N. Moksnes and W. Usher, “The relative importance of uncertain parameters and structural formulation for electricity systems planning in Kenya and Benin,” *iScience*, vol. 28, no. 2, p. 111792, Feb. 2025, doi: 10.1016/j.isci.2025.111792.

[48] M. D. McKay, R. J. Beckman, and W. J. Conover, “A Comparison of Three Methods for Selecting Values of Input Variables in the Analysis of Output from a Computer Code,” *Technometrics*, vol. 21, no. 2, pp. 239–245, 1979, doi: 10.2307/1268522.

# Additional Results:

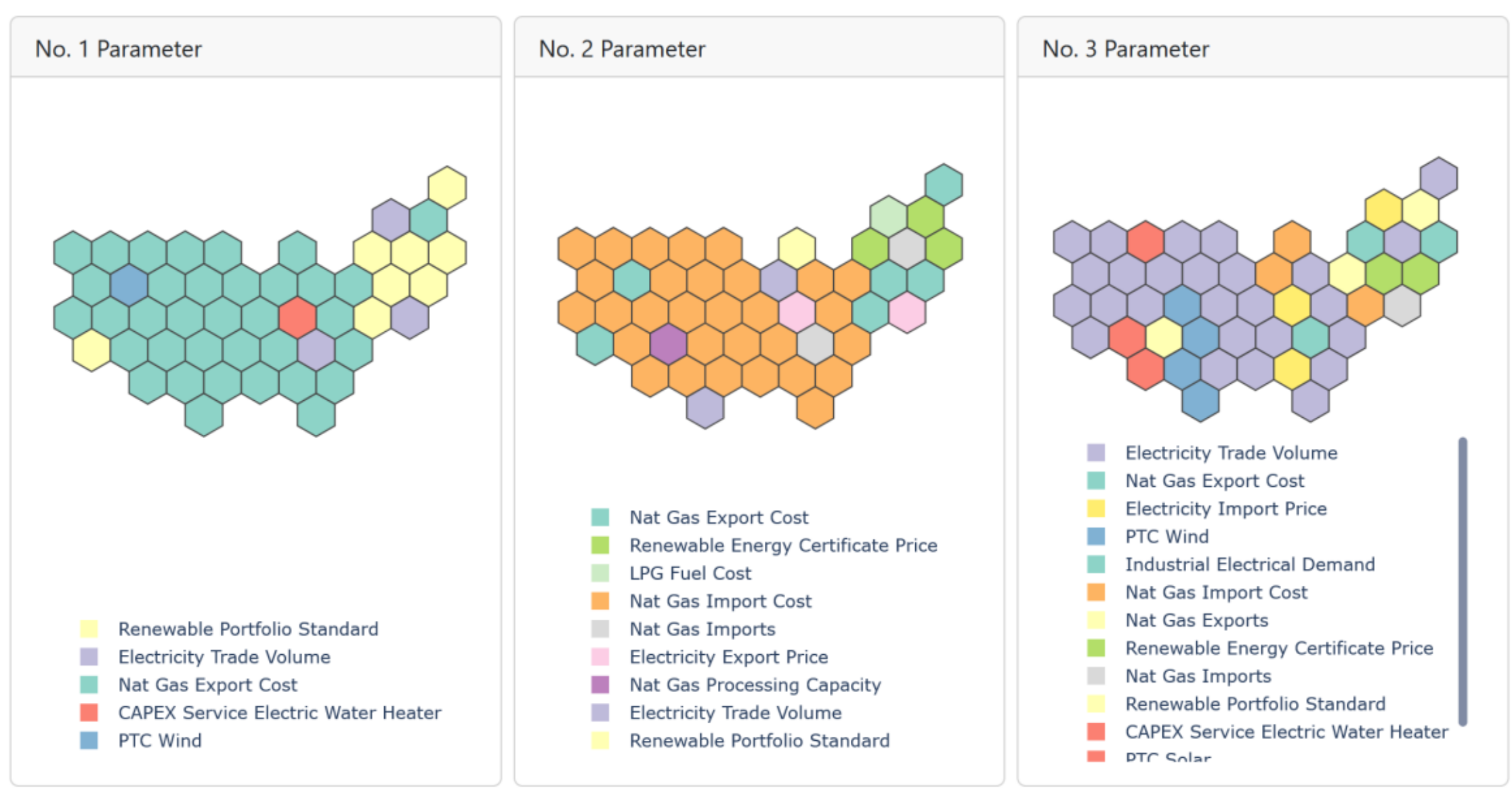

**Figure 1:** ***Average marginal costs of electricity sensitivity.***
The sensitivity to average marginal costs of electricity per state. The import and export costs represent the cost the state pay/earns from trading with the neighboring state. Electricity trade volume represents allowable yearly trade volume (MWh) to/from neighboring states.

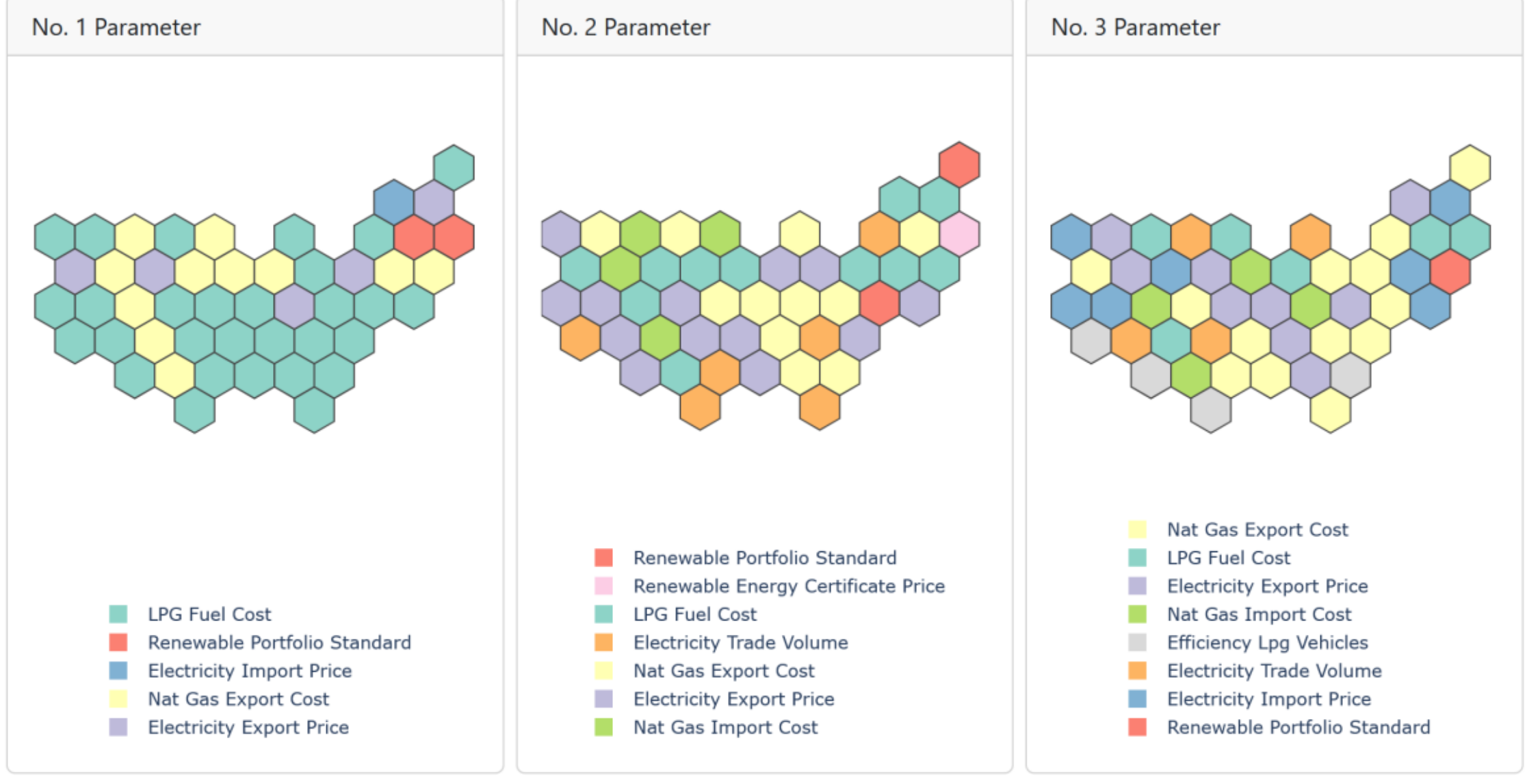

**Figure 2: Objective cost sensitivity.**
The sensitivity to objective costs per state. The import and export costs represent the cost the state pay/earns from trading with the neighboring state. Electricity trade volume represents allowable yearly trade volume (MWh) to/from neighboring states.

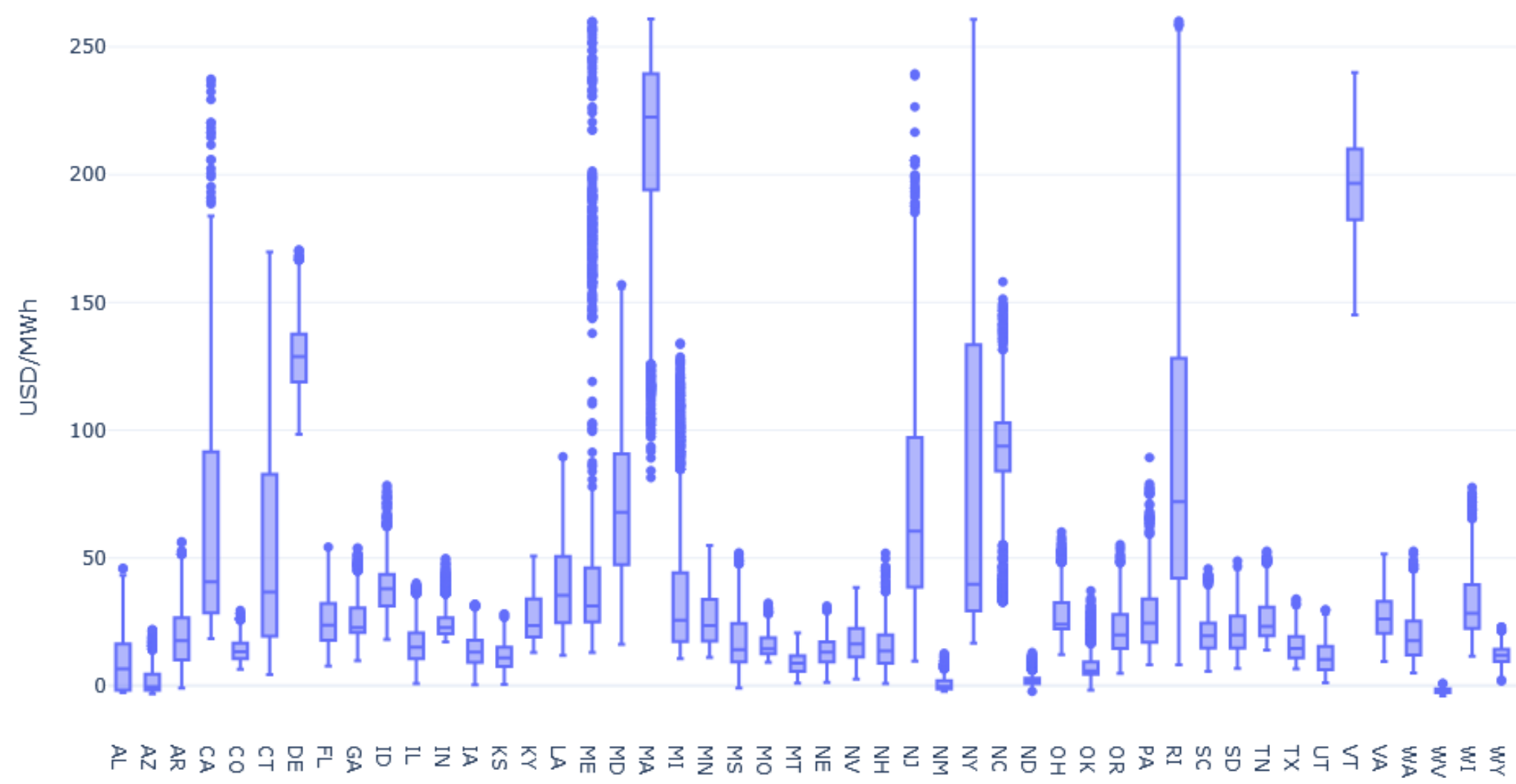

## Figure 3: Marginal costs of energy boxplot.

Boxplot of state level marginal costs of energy. Box plots are constructed by using GSA to find the uncertainty set of the 5 most impactful parameters impacting marginal energy costs, marginal electricity costs, objective costs, carbon emissions, and total emissions. These uncertainty sets, which are different per state, are propagated through the model to perform a Monte Carlo analysis.

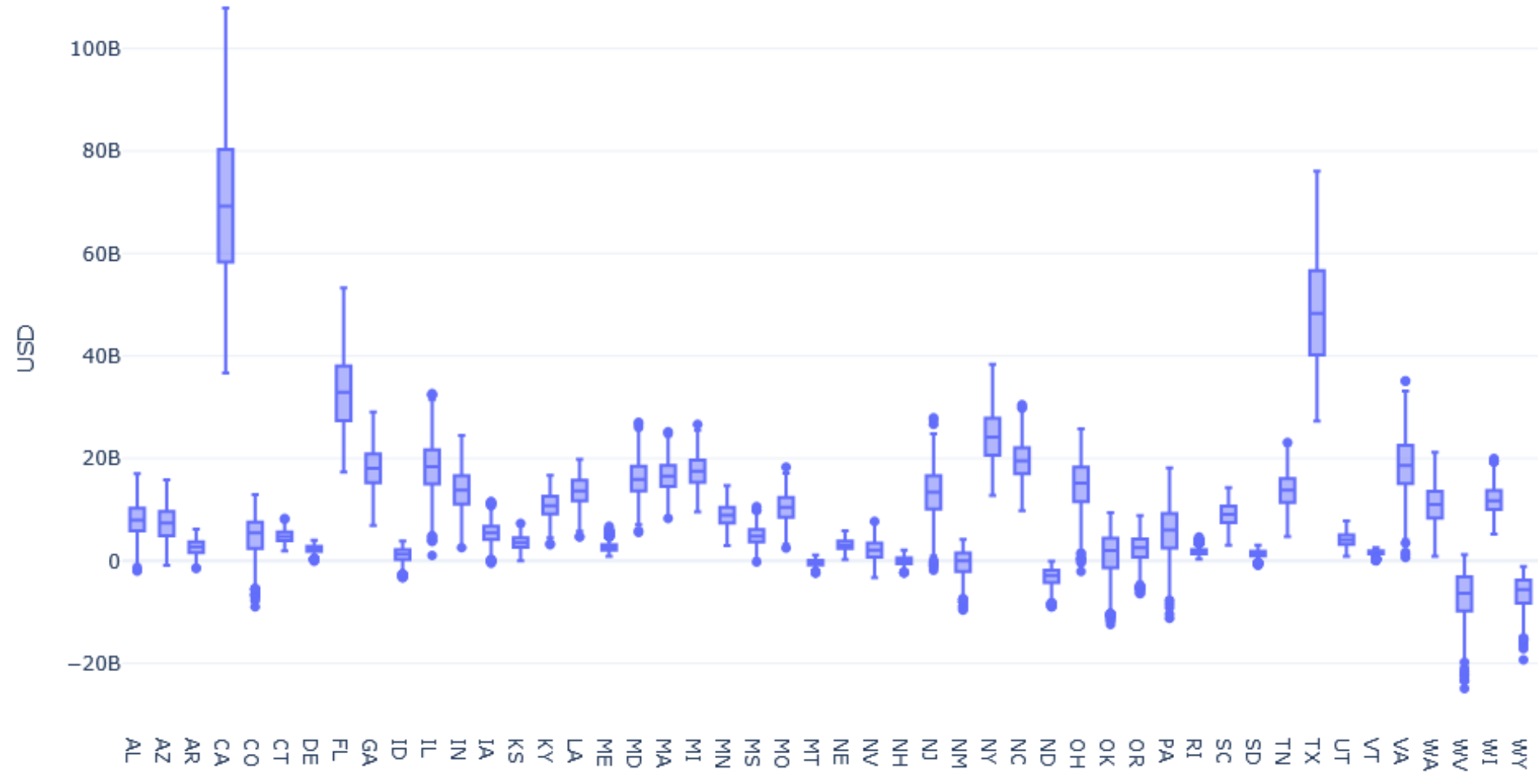

## Figure 4: Objective costs boxplot.

Boxplot of state level marginal costs of energy. Box plots are constructed by using GSA to find the uncertainty set of the 5 most impactful parameters impacting marginal energy costs, marginal electricity costs, objective costs, carbon emissions, and total emissions. These uncertainty sets, which are different per state, are propagated through the model to perform a Monte Carlo analysis.

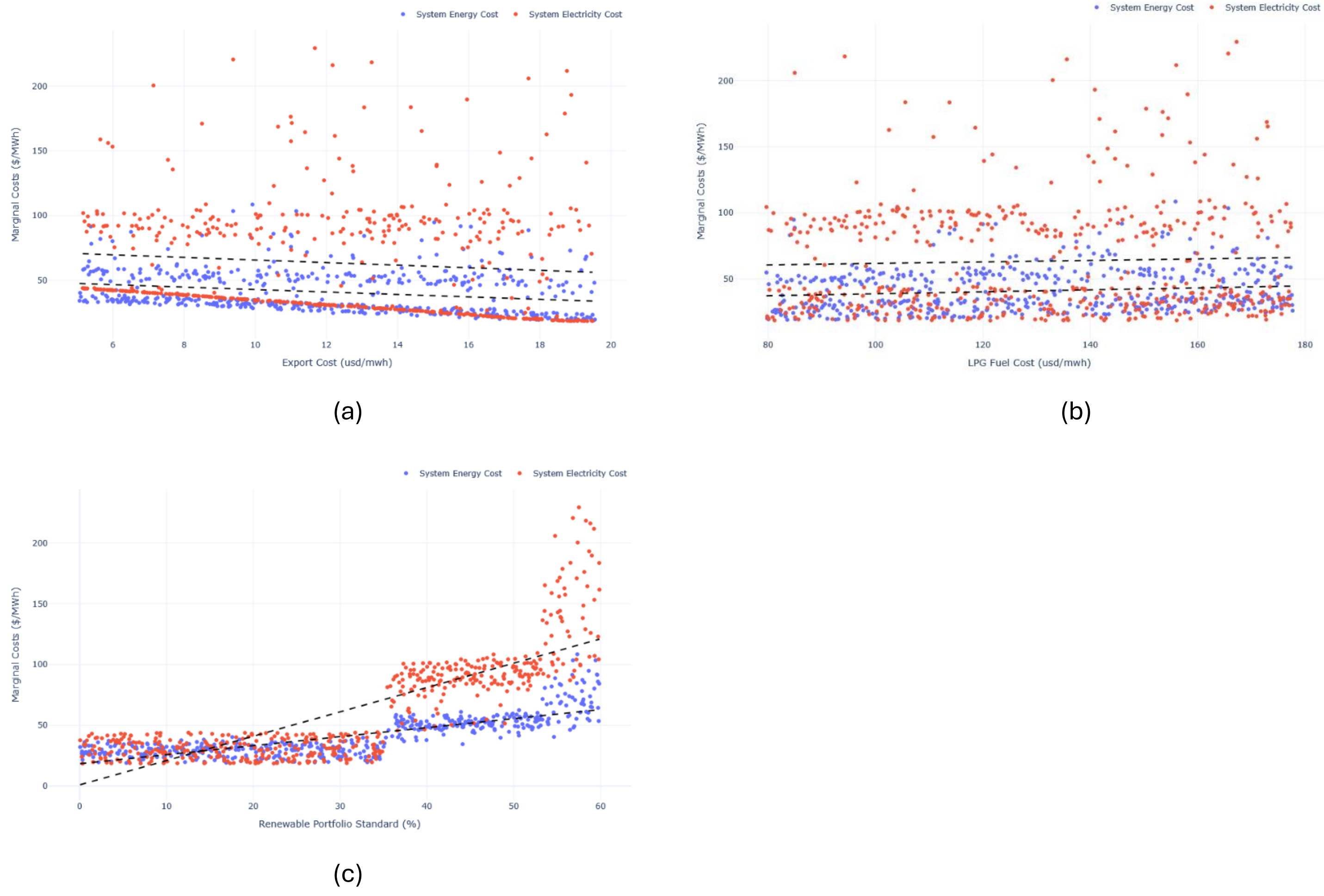

**Figure 5: Uncertainty propagation results for marginal costs in California.**

The results of the uncertainty propagation for California on average marginal energy and electricity costs. (a) Natural gas export price. (b) Gasoline price. (c) Renewable portfolio standards. Note, natural gas import cost was not identified as a sensitive parameter in California.

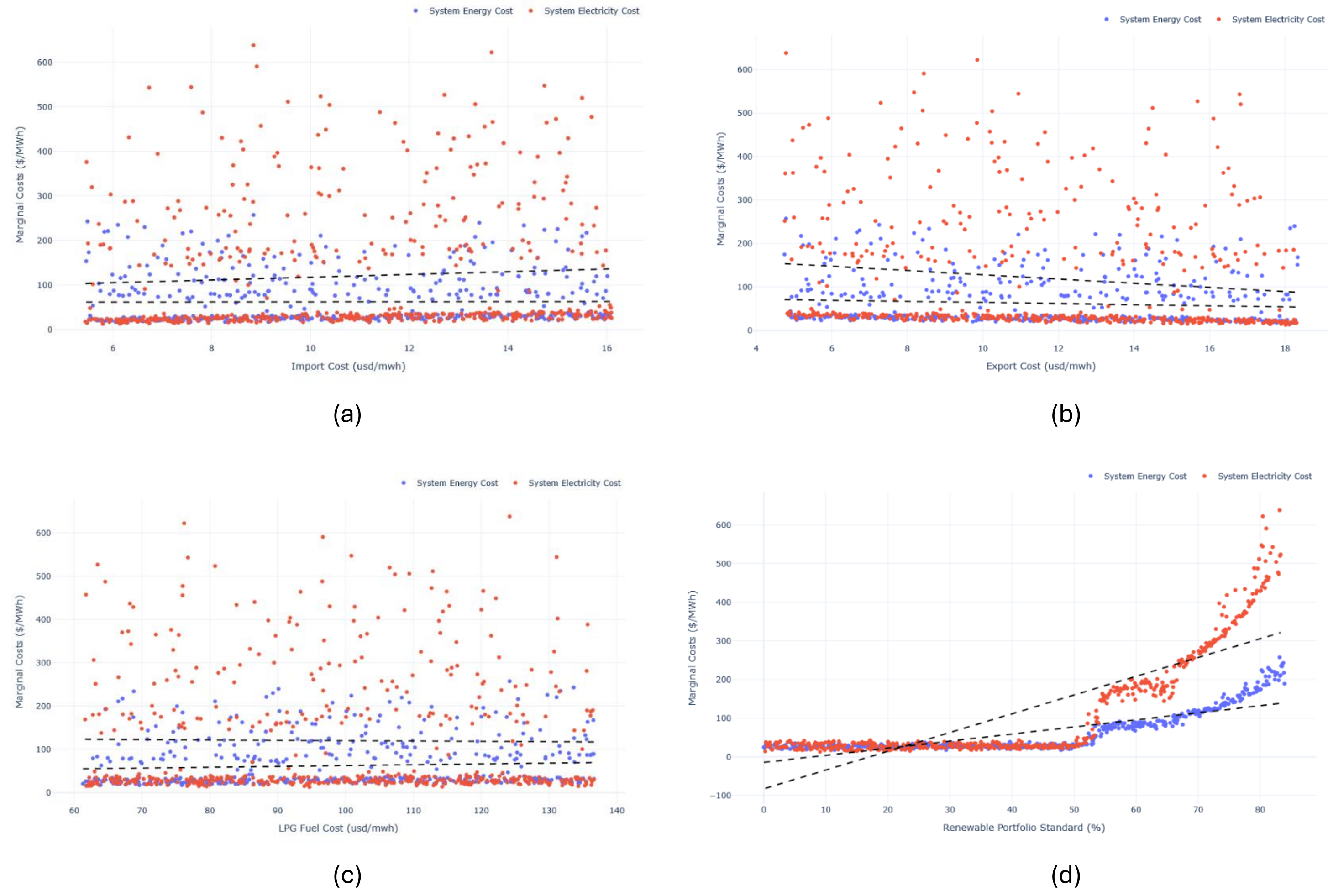

**Figure 6: Uncertainty propagation results for marginal cost in Maine.**

The results of the uncertainty propagation for Main on average marginal energy and electricity costs. (a) Natural gas import price. (b) Natural gas export price. (c) Gasoline price. (d) Renewable portfolio standards.

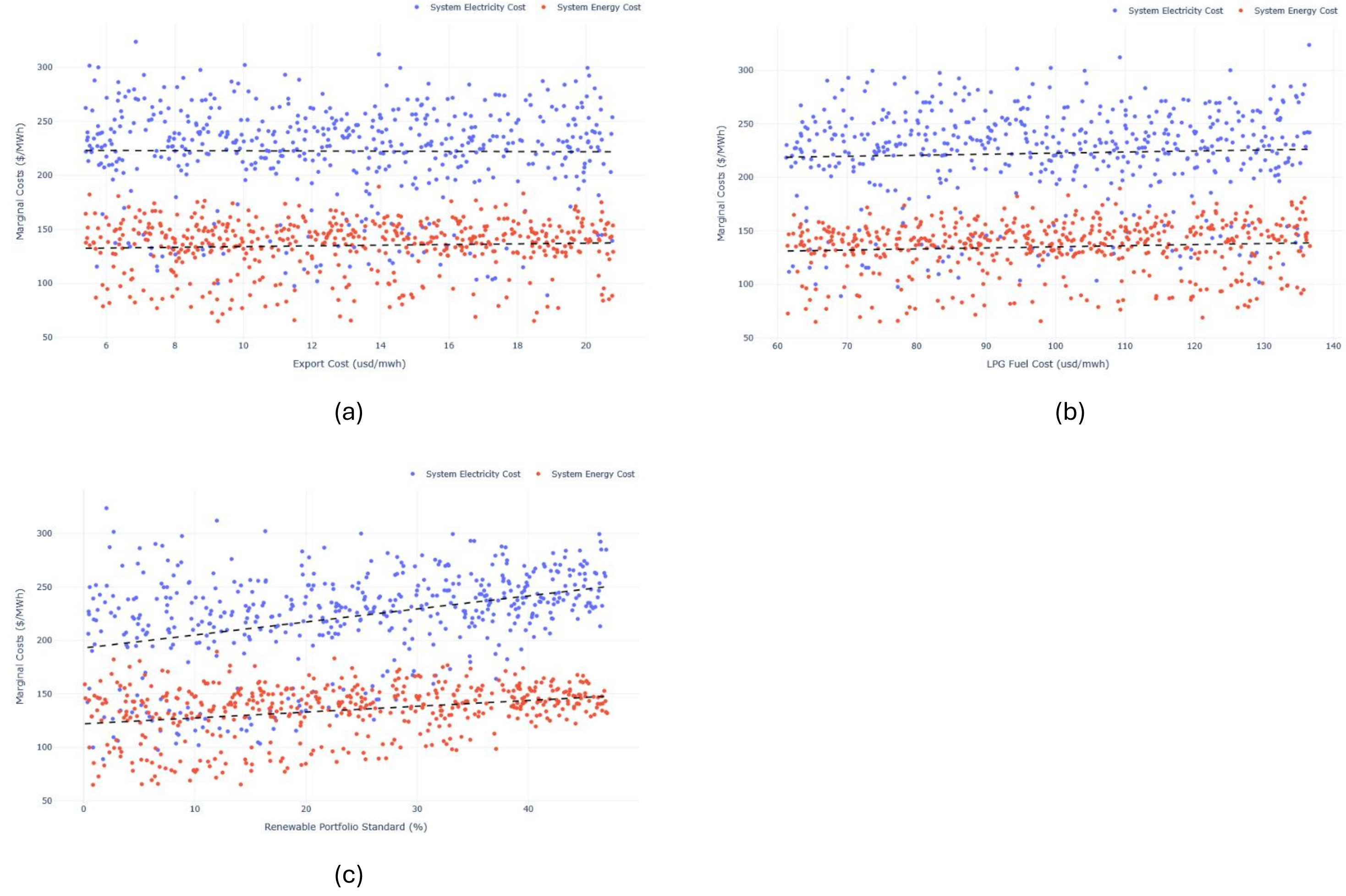

**Figure 7: Uncertainty propagation results for marginal costs in Massachusetts.**
The results of the uncertainty propagation for Massachusetts on average marginal energy and electricity costs. (a) Natural gas export price. (b) Gasoline price. (c) Renewable portfolio standards. Note, natural gas import costs was not identified as a sensitive parameter.

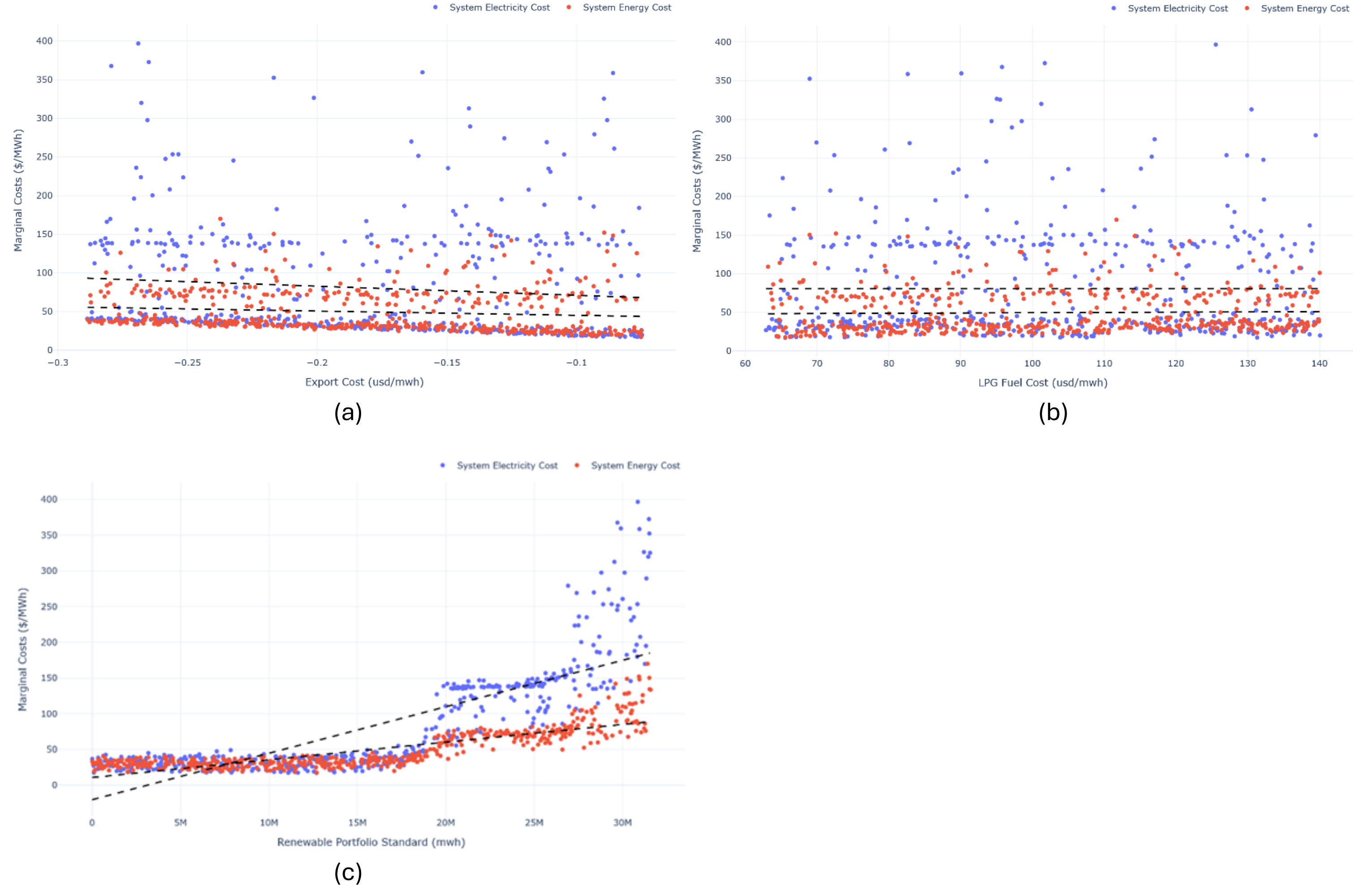

**Figure 8: Uncertainty propagation results for marginal energy costs in New York.**
The results of the uncertainty propagation for California on average marginal energy and electricity costs. (a) Natural gas export price. (b) Gasoline price. (c) Renewable portfolio standards. Note, natural gas import cost was not identified as a sensitive parameter in New York.

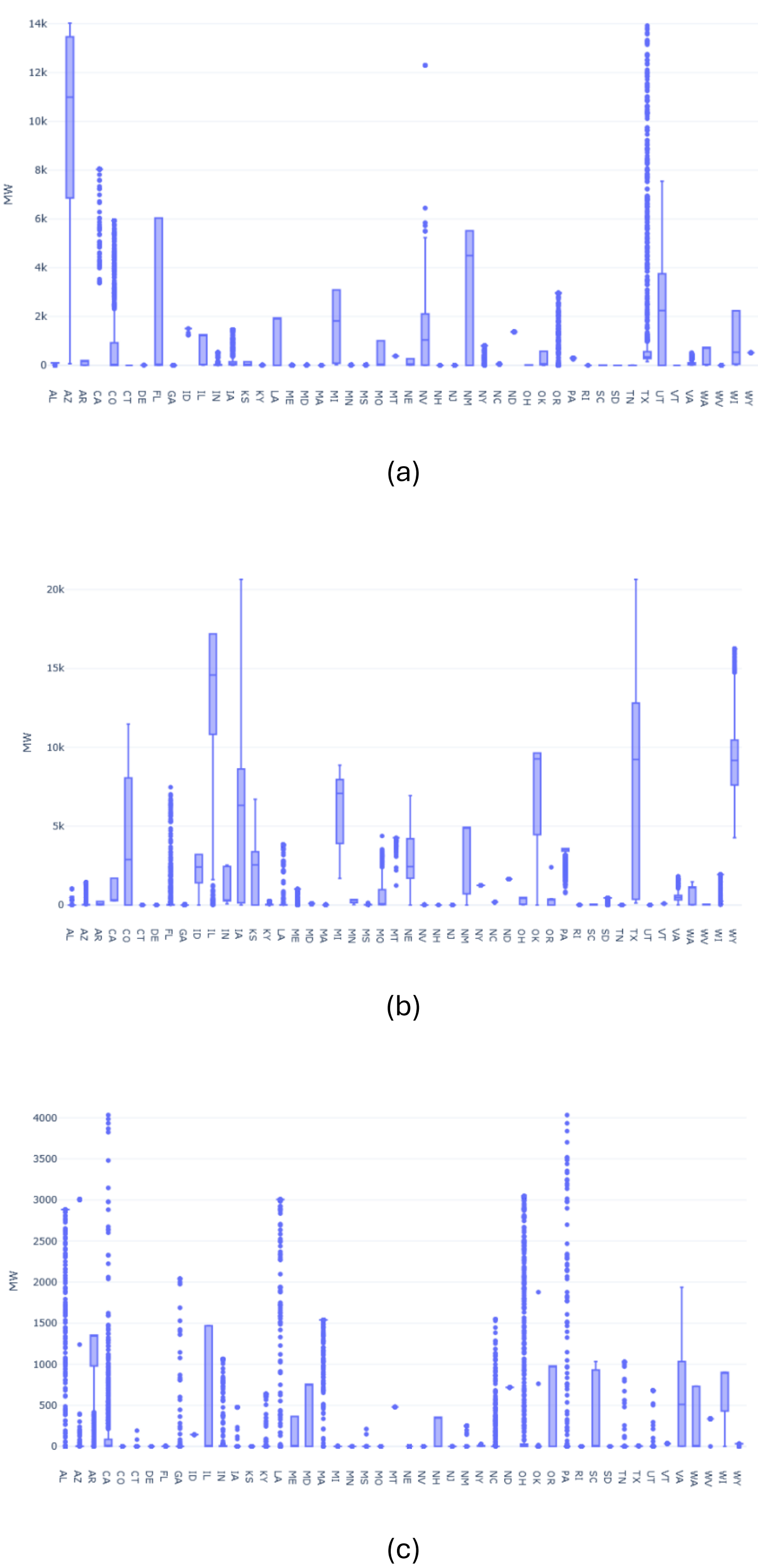

**Figure 9: Uncertainty propagation results for New Capacity.**

The results of the uncertainty propagation on newly installed capacity. (a) Utility scale solar. (b) Utility scale onshore wind. (c) Combined-cycle natural gas. Note, results are shown for 0-99 percentile of results, to avoid extreme outliers skewing the axis.

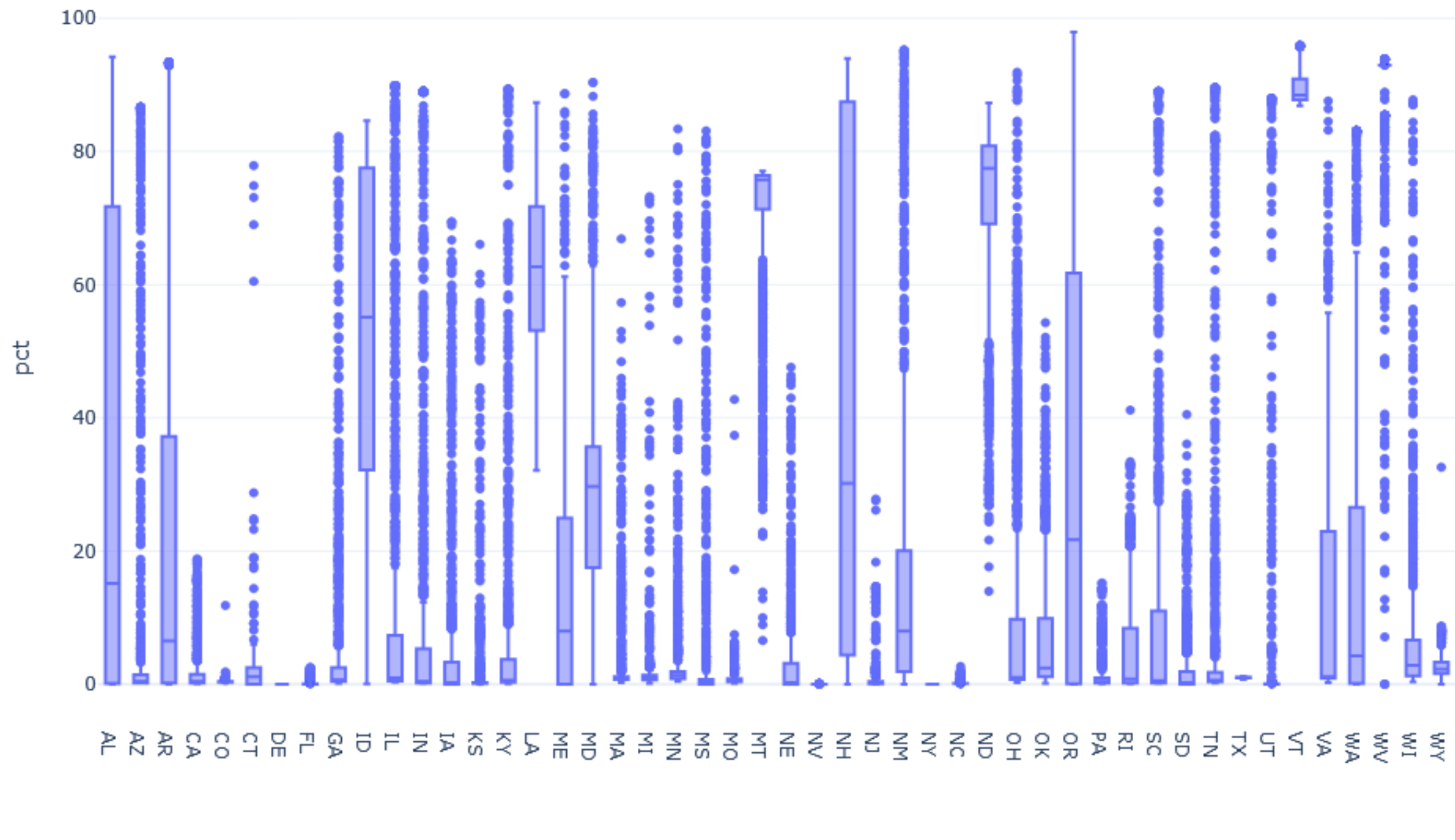

(a)

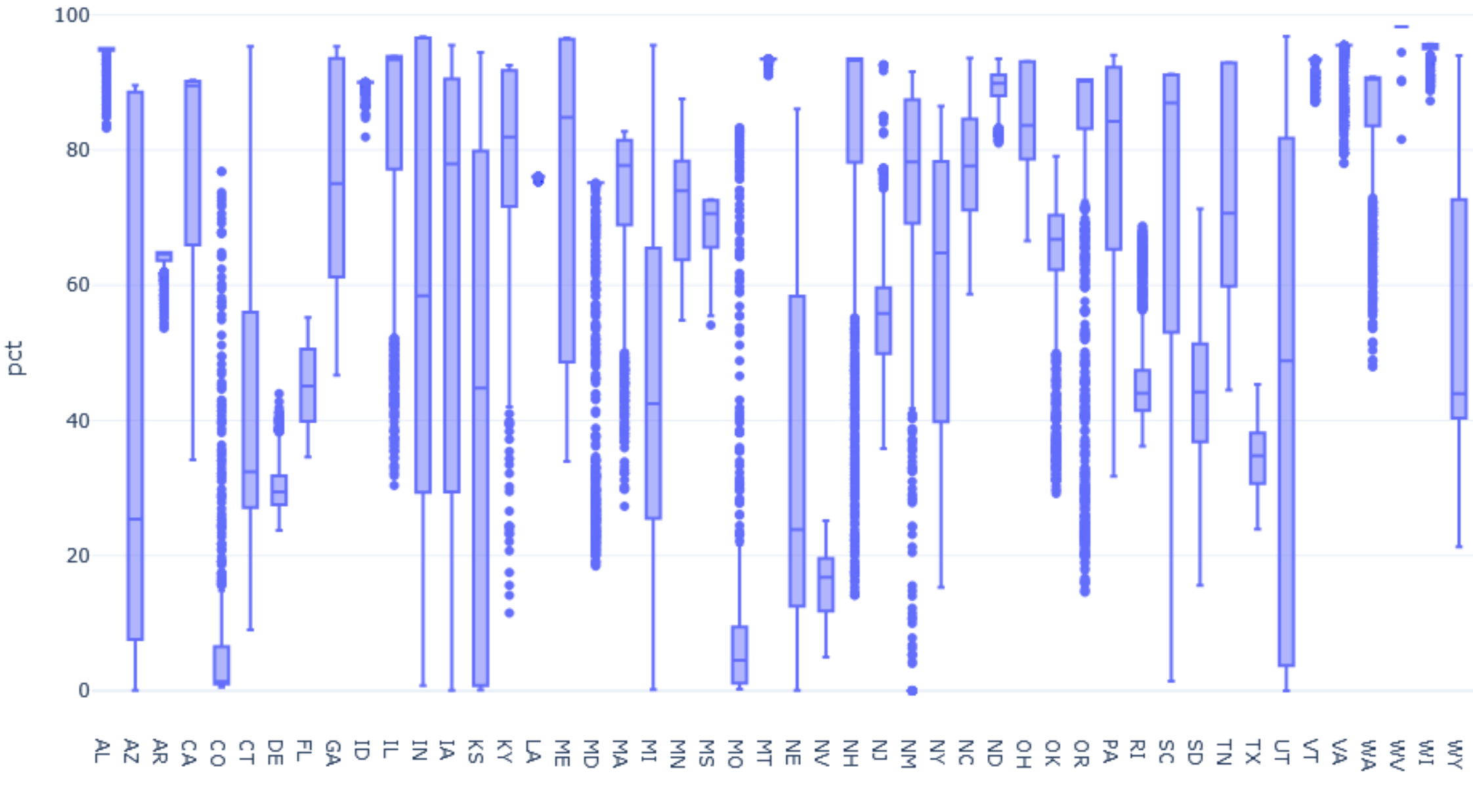

(b)

**Figure 10: Natural Gas Utilization Rates.**

The results of the uncertainty propagation on utilization rates. (a) Open-cycle gas turbines. (b) Combined-cycle gas turbines.

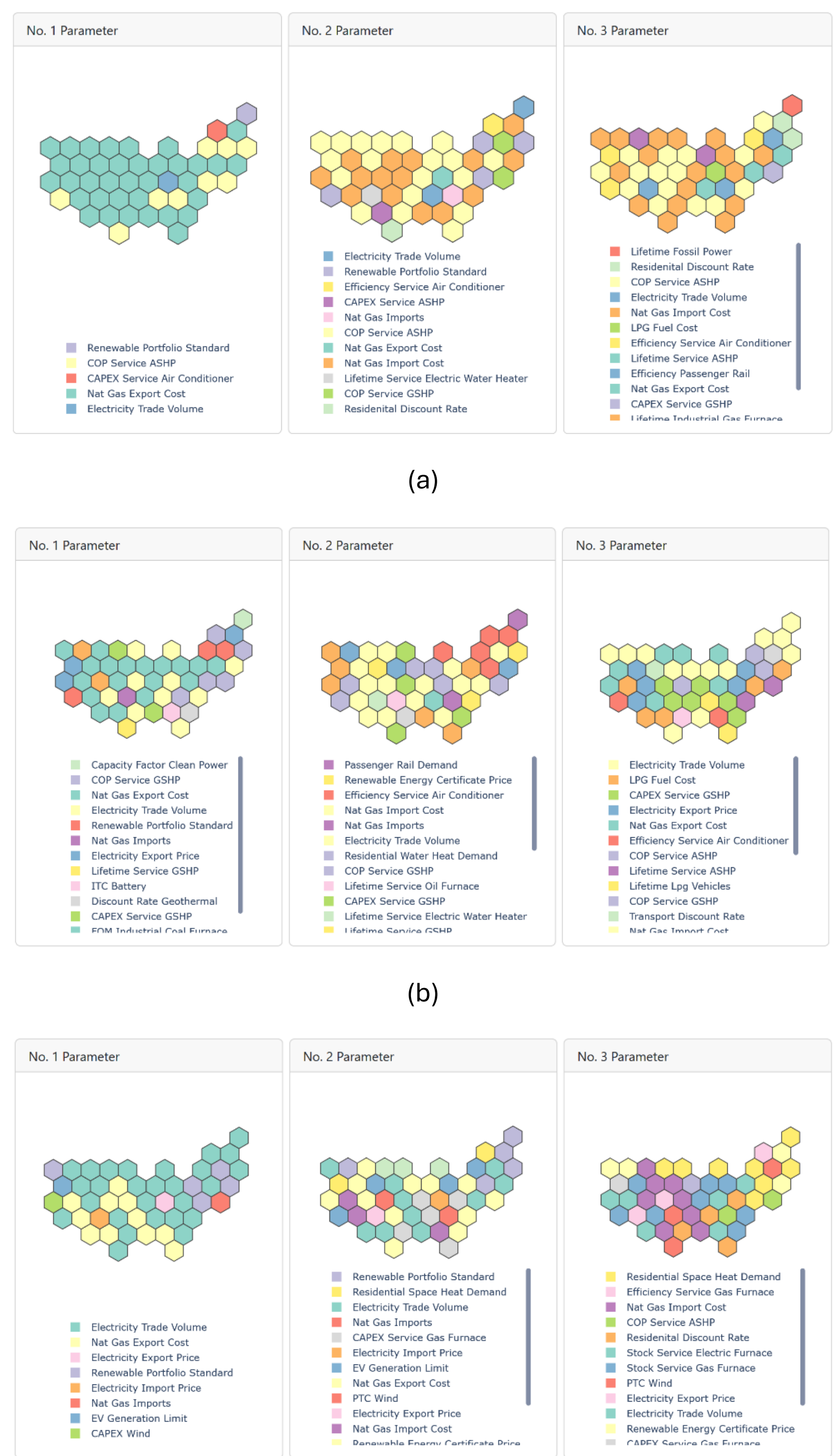

**Figure 11: Service sector heating sensitivity.**

The state level sensitivity on generation from (a) Natural gas furnaces (b) Air-source heat pumps and (c) Ground-source heat pumps.

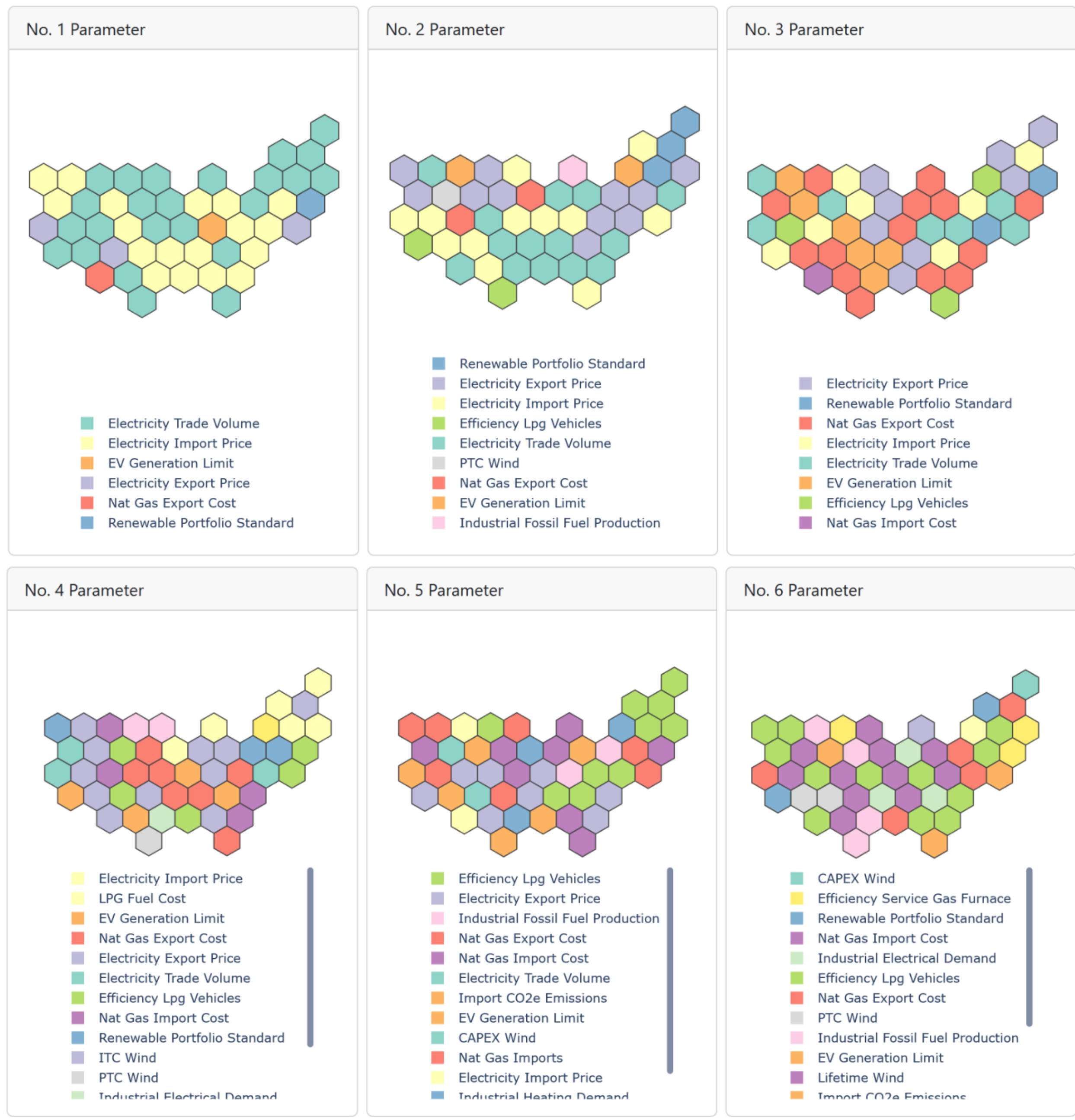

**Figure 12: Carbon sensitivity**

The state level sensitivity on carbon emissions for the top 6 parameters. Notably, EV generation limits and efficiency of petrol cars are a top 5 sensitive parameter in 18 and 20 states respectively.